\documentclass[prd,aps,twocolumn,a4paper,floatfix,showpacs,nofootinbib]{revtex4-1}

\usepackage{graphicx,psfrag}
\usepackage{mathrsfs}
\usepackage{amsmath,amsfonts,amssymb}
\usepackage{multirow,enumerate}
\usepackage{comment,hyperref}
\usepackage{color}
\usepackage{ltablex}
\usepackage{placeins}
\usepackage[utf8]{inputenc}

\newcommand{\be}{\begin{equation}}
\newcommand{\ee}{\end{equation}}
\newcommand{\bea}{\begin{eqnarray}}
\newcommand{\eea}{\end{eqnarray}}
\newcommand{\bel}{\begin{align}}
\newcommand{\eel}{\end{align}}

\newcommand{\da}{\downarrow}
\newcommand{\ua}{\uparrow}

\def\GMc2{{\rm G M_{\odot} c^{-2}}}

\usepackage{color}

\definecolor{cyan}{rgb}{0,0.9,0.9}
\definecolor{orange}{rgb}{0.9,0.5,0}
\definecolor{magenta}{rgb}{1,0,1}
\definecolor{purple}{rgb}{0.8,0.4,0.8}
\definecolor{gray}{rgb}{0.5,0.5,0.5}

\newcommand{\nfinac}{$\lesssim 0.1\%$}
\newcommand{\sfinac}{$\lesssim 6\%$}
\begin{document}

\title{Assessing the Energetics of Spinning Binary Black Hole Systems}

 \author{Serguei \surname{Ossokine}}
 \author{Tim \surname{Dietrich}}
 \affiliation{Max Planck Institute for Gravitational Physics (Albert Einstein Institute), Am M\"uhlenberg 1, Potsdam-Golm, 14476, Germany}
 \author{Evan Foley}
 \author{Reza Katebi}
 \author{Geoffrey Lovelace}
 
  \affiliation{Gravitational Wave Physics and Astronomy Center, California State University, Fullerton, Fullerton, California 92834, USA}
\date{\today}

\begin{abstract}
In this work we study the dynamics of spinning binary black 
hole systems in the strong field regime. 
For this purpose we extract from numerical relativity simulations 
the binding energy, specific orbital angular momentum, 
and gauge-invariant orbital frequency. 
The goal of our work is threefold: 
First, we extract the individual spin contributions to the binding energy, in particular
the spin-orbit, spin-spin, and cubic-in-spin terms.
Second, we compare our results with 
predictions from waveform models and find that while
post-Newtonian approximants are not capable of representing the dynamics 
during the last few orbits before merger, there is good agreement between our data and effective-one-body approximants 
 as well as the numerical relativity surrogate models. 
Finally, we present phenomenological representations for the binding 
energy for non-spinning systems with mass ratios up to $q = 10$ 
and for the spin-orbit interaction for mass ratios up to $q = 8$ 
obtaining accuracies of \nfinac{} and \sfinac, respectively. 
\end{abstract}


\maketitle

\section{Introduction}
\label{sec:intro}

\begin{figure}[t]
  \includegraphics[width=\linewidth]{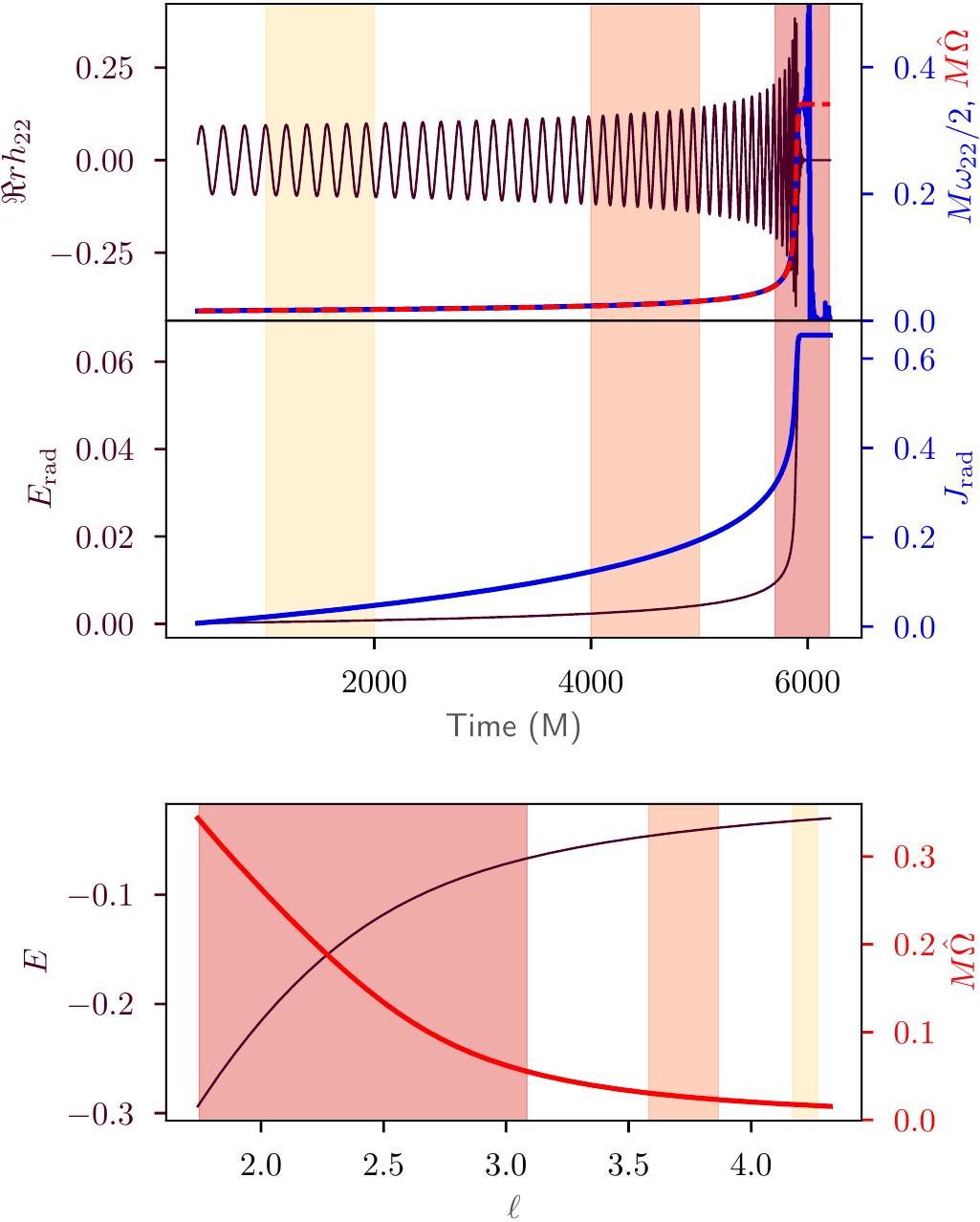}

  \caption{Top: gravitational wave strain $h$ (thin purple line) and
  gravitational wave frequency (thick blue line) as a function of time.
  The dashed red line is the gauge invariant frequency $M\hat{\Omega}$.
  Middle: emitted energy (thin purple lines) and angular momentum (thick blue line) 
  Bottom: the binding energy $E$ and the orbital 
  frequency $M \hat{\Omega}$ as a function of the specific 
  orbital angular momentum $\ell$. 
  All quantities are plotted for an equal mass BBH
  system with dimensionless spin $\chi_1=\chi_2=0.6$.
  }
  \label{fig:sketch}
\end{figure}

With the direct detection of gravitational waves (GWs) from a binary black hole (BBH)
system in 2015, Advanced LIGO ushered in the era of gravitational 
wave astronomy~\cite{Abbott:2016blz}.
Since then, several detections have been 
made~\cite{Abbott:2016nmj,Abbott:2017vtc,Abbott:2017oio,Abbott:2017gyy}, 
including the spectacular observation of both GWs 
and electromagnetic radiation from 
a binary neutron star merger~\cite{TheLIGOScientific:2017qsa}. 
Due to the increasing sensitivity of the advanced detectors, 
an increasing number of detections is expected in the 
third observing run~\cite{TheLIGOScientific:2016pea}. 
To interpret the large number of detections and extract 
astrophysical properties of the systems
it is more important than ever to construct
accurate models of merging compact binary systems.

In recent years, a large effort has been made to 
improve the modeling of the GW signal,
leading to rapid progress in the field. 
Currently, there are at least three routes for constructing 
state-of-art waveform models: by 
directly using numerical relativity (NR) simulations 
and interpolating between those to obtain generic 
waveforms~\cite{Blackman:2015pia,Blackman:2017pcm};
by calibrating to NR waveforms following a 
phenomenological approach,
e.g.~\cite{Ajith:2007qp,Ajith:2009bn,Hannam:2013oca}; or
by combining analytical information from post-Newtonian (PN)
theory(see ~\cite{Blanchet:2013haa} and references therein), with NR results
in form of the effective-one-body (EOB) 
description~\cite{Buonanno:1998gg,Damour:2000we,Damour:2001tu,Buonanno:2005xu,
Damour:2008qf,Damour:2008gu,Damour:2009kr,Barausse:2009aa,Barausse:2009xi,Pan:2010hz,
Barausse:2011dq,Damour:2012ky,Pan:2011gk,Pan:2013rra,Taracchini:2013rva,Damour:2014sva,Nagar:2015xqa,Nagar:2016ayt,Bini:2016cje,Babak:2016tgq,
Bohe:2016gbl}. 

In this paper we do not focus on the gravitational 
waveform; instead, we examine the global dynamics in terms of 
gauge-invariant quantities. In particular we 
study 
the binding energy $E$, 
specific orbital angular momentum $\ell$,
and the gauge-invariant orbital frequency $\hat{\Omega}$, 
see~\cite{Damour:2011fu,Nagar:2015xqa}.
Those quantities can be constructed from the
strain by computing the radiated energy 
and angular momentum. 
An example of this calculation is given in Fig.~\ref{fig:sketch}. 
From the GW strain (top panel), we compute the energy and angular momentum 
emitted by the system (middle panel). 
From these, we construct gauge-invariant quantities: $E$ and $\hat{\Omega}$ 
as a function of $\ell$ (bottom panel). 
Considering the marked times, one finds that due to 
the emission of energy and angular momentum, $E$ and $\ell$ decrease over time, while 
the orbital frequency $\hat{\Omega}$ increases. The top panel shows that the dimensionless frequency $M \hat{\Omega}$ is half the 
frequency of the GW. Consequently, during the ringdown, $\hat{\Omega}$ 
can be interpreted as the main emission frequency of the quasi-normal 
modes.

Although the evolution of $E,\ell,\hat{\Omega}$
incorporates nonadiabatic effects, those are practically 
negligible during the inspiral until the LSO~\cite{Damour:2011fu}. 
Thus, investigation of $E,\ell,\hat{\Omega}$ permits a direct probe of the  
conservative dynamics of a BBH in a highly relativistic regime. 
Not only does this improve our understanding of the 
inspiral dynamics, but it also allows us to study the 
influence of the mass ratio separately from those of the spin-orbit, spin-spin, and cubic-in-spin terms.

In this article, we focus primarily on three tasks: 
(i) extracting (up to our knowledge) for the first time different contributions 
to the binding energy from NR simulations of BBHs; 
(ii) comparing our results to state-of-the-art waveform models 
to test their accuracy; and 
(iii) proposing phenomenological fits, including unknown higher order PN coefficients, to constrain the binding energy for aligned spin systems.

The structure of the paper is as follows.
In Sec.~\ref{sec:methods} we describe the employed numerical methods, 
the construction of gauge-invariant quantities, 
and numerical errors. 
In Sec.~\ref{sec:Ej:nonspinning} we present
the results for non-spinning configurations
accessing the effect of the mass-ratio on the dynamics. 
In Sec.~\ref{sec:Ej:constant_spin} we study spinning-equal mass configurations and extract the individual spin components up to cubic-in-spin terms.
Sec.~\ref{sec:Ej:generic} presents results for generic non-precessing systems. 
We conclude in Sec.~\ref{sec:conclusion}. 
In Appendix~\ref{sec:Ej:nonprecBBH:chiflexible} we extract spin orbit and spin-spin 
contribution with an alternative approach to 
Sec.~\ref{sec:Ej:constant_spin}, using only two different simulations per spin magnitude. 
In Appendix~\ref{sec:config} we list the studied NR simulations.

Throughout the paper we employ geometric units with $G=c=1$.

%%%%%%%%%%%%%%%%%%%%%%%%%%%%%%%%%%%%%%%%%%%%%%%%%%%%%%%%%%%%%%%%%%%%%%%%%%%%%%%%%%%%%
%%%%%%%%%%%%%%%%%%%%%%%%%%%%%%%%%%%%%%%%%%%%%%%%%%%%%%%%%%%%%%%%%%%%%%%%%%%%%%%%%%%%%
\section{Methods}
\label{sec:methods}

\subsection{Simulation method}
\label{sec:methods:simulation}

In this article, we restrict ourselves to non-precessing
configurations leaving precession for a future work. The studied  
NR simulations are produced with the
Spectral Einstein Code ({\tt
  SpEC})~\cite{SpECwebsite,Scheel:2006gg,Szilagyi:2009qz,Buchman:2012dw}
  and are publicly available in the SXS catalogue~\cite{SXSCatalog,Mroue:2013xna}.

{\tt SpEC} is a multi-domain, pseudospectral collocation code for the simulation of
compact object binary spacetimes. Conformally curved
initial data~\cite{Lovelace:2008tw} are constructed in the
extended-conformal-thin-sandwich formalism~\cite{Pfeiffer:2002iy} using a
pseudo-spectral elliptic solver~\cite{Pfeiffer:2002wt}. Dynamical evolutions use
a first order formulation~\cite{Lindblom:2005qh} of the generalized harmonic
formulation of Einstein's equations~\cite{Pretorius:2005gq, Friedrich1985} in
the damped harmonic gauge~\cite{Lindblom:2009tu}. {\tt SpEC} uses a dual frame
method~\cite{Scheel:2006gg,Hemberger:2012jz} to solve Einstein's equations based
on explicit coordinate transformations. These map a set of inertial
(physical) coordinates, in which the BHs orbit and approach each other, to a set
of grid coordinates, in which the BHs remain at fixed coordinate locations. To
handle the BH singularity, dynamical excision is
used~\cite{Hemberger:2012jz,Scheel:2014ina}. An adaptive mesh refinement,
which adjusts both the order of the spectral basis functions as well as the
number of subdomains, is employed to achieve the desired truncation error,
see~\cite{Szilagyi:2014fna}. {\tt SpEC} has been successfully employed to
simulate a large variety of physical systems; see, among
others~\cite{Blackman:2017pcm,Szilagyi:2015rwa,Barkett:2015wia,Lovelace:2014twa,Mroue:2013xna,Buchman:2012dw}).

\subsection{Constructing binding energy curves}
\label{sec:methods:Ej}

To extract the reduced binding energy and the specific angular momentum
from our BBH simulations, we compute the energy and
angular momentum flux as described in~\cite{Ruiz:2007yx}.
The binding energy itself is defined as
\be
E_{b}=\frac{E_{\rm ADM}(t=0)-E_{\rm rad}-M}{\mu}, \ee
with $\mu=m_1 m_2/M$, where $E_{\rm ADM}$ is extracted at 
the beginning of the simulation $t=0$.
The dimensionless angular momentum is
\be
j=\frac{|{\bf{J}}_{\rm ADM} (t=0) -\bf{J}_{\rm rad}|}{M\mu}
\ee
from which we can construct the dimensionless orbital angular momentum
\be
\ell =\frac{|\bf{J}_{\rm ADM}- \bf{S}_1 -\bf{S}_2 -\bf{J}_{\rm rad}|}{M\mu}.
\ee
For spin-aligned/anti-aligned cases,
$\bf{J}_{\rm ADM}, \bf{S}_1, \bf{S}_2, \bf{J}_{\rm rad}$
reduce to $J_{\rm ADM}^z, S_1^z, S_2^z, J^z_{\rm rad}$ respectively, since the
orbital plane is identical to the $x$-$y$-plane. \\

The presence of junk radiation in our simulations causes an ambiguous shift of
the curves; thus, as outlined in Ref.~\cite{Taracchini:2013rva,Nagar:2015xqa}, the binding energy
curves must be further processed. The first step involves a vectorial shift in
the $E-j$ plane, such that the curve passes through the final values of mass and
spin of the merger remnant. This can be done in a way that eliminates the need
of ADM quantities entirely. Indeed, letting $(\tilde{E},
\tilde{j})\equiv (E+\Delta E, j+\Delta j)$, and requiring that at the final time
$t_f$, the curve passes through the point given by the mass $M_f$ and spin $S_f$
of the final black hole, $(E(t_{f})+\Delta E,j(t_{f})+\Delta
j)=(M_{f},S_{f})$, it follows readily that $\Delta E =
M_{f}-E(t_{f})=M_{f}-E_{ADM}+E_{rad}(t_f)$. Consequently, $\tilde{E}(t)=E_{\rm
  rad}(t_f)-E_{\rm rad}(t)+M_{f}$, and similarly for $j$. For convenience we work
exclusively with $\ell$ instead of $j$, since all the cases considered in this paper 
are non-precessing and difference between the two quantities is equivalent to a shift 
by a constant under the assumption that the spins of the individual black holes 
stay constant.

\begin{figure}[t]
  \includegraphics[width=\linewidth]{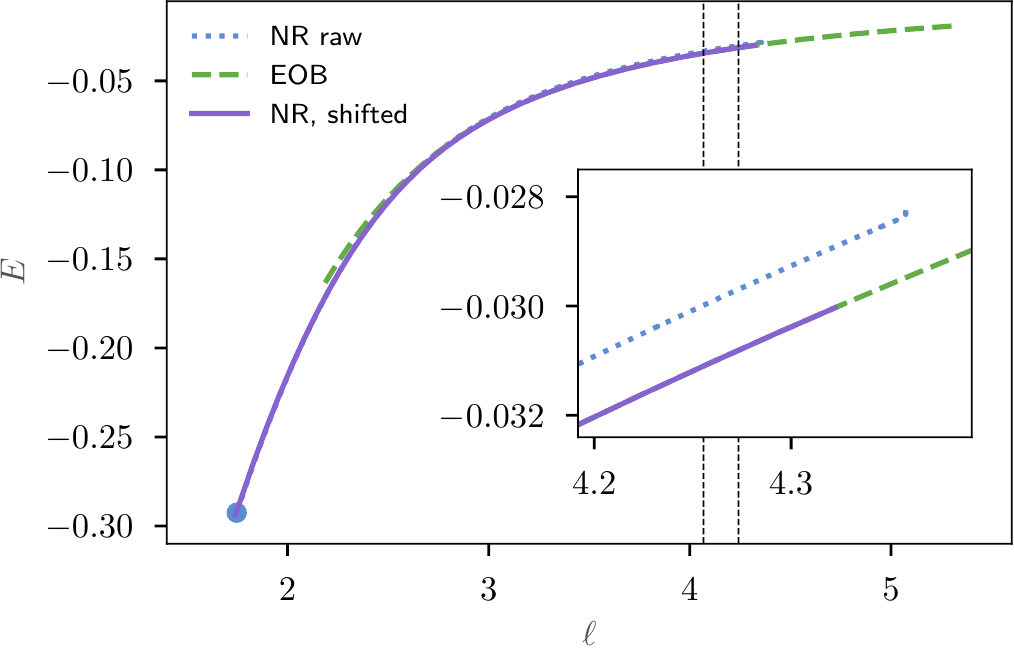}
  \caption{The binding energy $E$ as a function of the reduced orbital angular
    momentum $\ell$ for an equal mass binary with dimensionless spins $\chi_{1}=\chi_{2}=0.6$.
    The dashed green curve corresponds to EOB data, while the
    dotted blue and solid purple curves are the raw and shifted NR data,
    respectively. The curves were aligned in the $\ell$-interval $(4.07,
    4.24)$ shown as dashed vertical lines. The dot indicates the state of the black hole
    as measured from quasi-local quantities. The inset highlights the
    requirement for the shifted NR curve in the early inspiral behavior (note that this shifted NR
    curve is also truncated to eliminate junk radiation). }
  \label{fig:e_vs_j_q1_downzero}
\end{figure}

As an example, Figure~\ref{fig:e_vs_j_q1_downzero} shows $E-\ell$ curves for an
equal mass binary, with spins $\chi_{1}=0.6,\ \chi_{2}=0.6$ from NR
(blue dotted line), and the non-precessing EOB
model~\cite{Bohe:2016gbl} (green dashed line), 
which we hereafter denote as SEOBNR~\footnote{This model is known as SEOBNRv4 in the LIGO Algorithm Library.}. Although the overall agreement is good, it is obvious that an offset between the EOB and the NR curves exists even in the early inspiral, at a time where the EOB description is expected to give accurate results. We assume that there are at least two reasons for this offset: 
(i) parts of the unresolved junk radiation imprint on the $E-\ell$ curves and can not be eliminated with a single
shift $(\Delta E, \Delta \ell)$; 
(ii) there are extrapolation errors in the ringdown of higher modes. 
To resolve these issues, we perform an additional 
shift of the NR curve, aligning it with the EOB curve early in the inspiral. 
This is done by minimizing the difference
in the binding energy over the interval $[0.94,0.98]\ell_{\rm max}$, where
$\ell_{\rm max}$ is the maximum value of $\ell$ in the NR data, 
after junk radiation has been removed. 
The shifted NR curve (purple solid line in Fig.~\ref{fig:e_vs_j_q1_downzero}) now agrees better with the EOB curve. 
A similar approach to construct binding energy
curves from \texttt{SpEC} data was used in~\cite{Nagar:2015xqa}. 

From $E(\ell)$, we compute the gauge-invariant frequency 
\be
M \hat{\Omega}\equiv\frac{\partial
  E}{\partial \ell}.  
\ee
which corresponds to the orbital frequency for circular orbits in the adiabatic regime.
We apply a Savitzky-Golay filter to
eliminate oscillations due to residual eccentricity and numerical noise. From
$M \hat{\Omega}$ it is possible to define the post-Newtonian parameter
\be
x\equiv (M \hat{\Omega})^{2/3} .
\ee

\subsection{Extracting individual contributions to the binding energy}
\label{sec:methods:components}

In order to gain insight into the role of different spin interactions during
the BBH coalescence, we make the assumption that
\be
\label{eq:Epieces}
E \approx E_0 + E_{SO} + E_{S^2} + E_{SS} + E_{S^3} + E_{S^2S} +
\mathcal{O}(S^{4})\ ;
\ee 
i.e., that the binding energy of a spinning BBH
configuration can be approximated by the \emph{sum} of separate contributions,
 as predicted by PN theory (see e.g. \cite{Blanchet:2013haa}) and applied successfully for NR
simulations of neutron stars~\cite{Bernuzzi:2013rza,Dietrich:2016lyp,Dietrich:2017xqb}. 
When the spins are oriented 
only along the orbital angular momentum, the dependence of the individual 
contributions in Eq.~\eqref{eq:Epieces} on the spin orientation can be given 
explicitly. Here, for equal-mass cases, 
the nonspinning point-mass term $E_0$ is independent of spin;
the spin-orbit term $E_{SO} \propto (S_1 +S_2) $; the self-spin ($S^2$) term
$E_{S^2} \propto S_1^2 + S_2^2$; the spin-spin ($SS$) interaction term $E_{SS}
\propto S_1 S_2$; the cubical self-spin term $E_{S^3} \propto S_1^3 + S_2^3$ and
lastly, the additional cubical spin term $E_{S^2S} \propto S_1^2 S_2 + S_1 S_2^2$. In 
PN theory, the different terms in Eq.~\eqref{eq:Epieces} enter
at different orders: SO terms start at 1.5 PN, spin-squared terms start at
2PN, and cubic-in-spin terms start at 3.5PN. 
Quartic-in-spin contributions are not included in this paper, 
but PN results have been presented in~\cite{Levi:2016ofk}. 

All individual contributions in Eq.~\eqref{eq:Epieces} can be calculated by
combining simulation with different spin orientations. The
overall sign of the individual pieces and their magnitude depend on the spin
magnitude, the spin's orientation, and the size/sign of the prefactor of every
binding energy contribution.

\subsection{Error estimates}
\label{sec:methods:errors}

The presented gauge invariant quantities have several possible 
uncertainties: truncation errors due to numerical discretization, errors due to finite radii extraction, filtering errors, and alignment errors. 
All uncertainties are included in our work and a detailed description
is given below. \\

\textit{Numerical discretization:}
Every NR simulation is only an approximate representation of the 
continuum solution of the general relativistic field equations; 
thus, truncation errors ($\epsilon_{\rm trunc}$) affect all results. 
In general, {\tt SpEC} shows small absolute errors for the
waveforms phasing ~\cite{Lovelace:2016uwp,Kumar:2015tha,Scheel:2014ina,Mroue:2013xna}. Consequently, 
we expect the same for the binding energy, orbital angular momentum,
and orbital frequency.
To quantify the uncertainty caused by resolution effects, we perform our
analysis for the same physical configuration using the two highest resolutions
available. The truncation error is estimated as the difference between the
resolutions. For most cases, the relative truncation error is typically of the order of 
$0.1\%$ .

\textit{Finite radii extraction:} Another possible source of error is caused by the finite radii extraction 
of the waves and the need for extrapolation of the waveform to infinity
to eliminate near-field effects. In {\tt SpEC} the
extrapolation is done in the co-rotating frame~\cite{Boyle:2013nka}, in terms of
a corrected radial coordinate \cite{Boyle:2009vi} using several extrapolation
orders $N=2,3,4$. It has been shown for many {\tt SpEC} waveforms that GW
extraction errors ($\epsilon_{\rm extrap}$) are comparable with truncation 
errors~\cite{Chu:2015kft}. For the computation of the
fluxes, which depend on the amplitude of the gravitational waves and their time
derivates, extrapolation errors can be significant.
In~\cite{Taylor:2013zia} it was found that higher-order
extrapolation produces more accurate results in the inspiral but can amplify
high frequency noise during merger-ringdown, leading to non-convergence,
especially for subdominant modes. To avoid amplification of noise during merger, we
choose not to use $N=4$ waveforms in this work. Comparing $N=2$ and $N=3$
waveforms reveals that the latter is less sensitive to the effect of omitting
data at small radii when performing extrapolation. Although $N=3$
introduces more noise during ringdown, the effect on the binding energy curves
is rather small. Thus, we adopt $N=3$ for all the results in the
paper. To estimate $\epsilon_{\rm extrap}$ we consider the difference in $E-\ell$ and $E-x$
curves between $N=2$ and $N=3$. For the majority of cases considered here, this
gives a relative error of order $0.5\%$, which is approximately a factor of 5
larger than the truncation error and is the dominant source of uncertainty for most cases.

Aside from errors inherent to the NR waveforms themselves,
other sources of error will enter the computation of $E-\ell$ and $E-x$ curves.

\textit{Alignment procedure:} 
The alignment procedure, which is needed to compensate for the junk radiation,
relies on the accuracy of the EOB approach and a properly chosen alignment
window. We tested the effect of using different EOB approximants (the
uncalibrated/calibrated SEOBNRv2 \cite{Taracchini:2013rva}
and SEOBNRv4~\cite{Bohe:2016gbl} and the EOB model
of~\cite{Nagar:2015xqa}). While we found best agreement with
\cite{Nagar:2015xqa}, the effect was small. Hence, because of convenience, the calibrated SEOBNRv4 model is used throughout the paper. 
The effect of the alignment error ($\epsilon_{\rm
  al}$) is studied by: i) changing the alignment window from
$[0.94,0.98]\ell_{\rm max}$ to $[0.9,0.98]\ell_{\rm max}$ and taking the
difference between the results ($\epsilon_{int}$) and ii) aligning to SEOBNRv2 instead of SEOBNRv4
and again taking the difference between the results ($\epsilon_{\rm v2}$). 
We find the relative errors to be of order $0.1\%$ in
$E-x$ and $E-\ell$ curves.

\textit{Filtering:}
For computation of $E-x$ curves, another possible source of error
($\epsilon_{\rm filt}$) is the use of a Savitzky-Golay filter for the
computation of the PN parameter $x$. To gauge the impact of the chosen stencil size, we
redid the computation using half as many points for filtering and taking the difference with
the standard procedure. We find that these errors are on the order of $0.1\%$ in $E-x$ curves.\\

We use the $L_1$ norm to combine the alignment errors and $L_2$ norm to compute the total error estimate:
\begin{align}
  \epsilon_{\rm al} &= |\epsilon_{\rm v2}|+|\epsilon_{\rm int}|, \\
  \epsilon_{\rm total} &= \sqrt{\epsilon_{\rm al}^{2}+\epsilon_{\rm trunc}^{2}+\epsilon_{\rm extrap}^{2}+\epsilon_{\rm filt}^{2}}.
\end{align}
We compute this error estimate for every $E-j$ and $E-x$ curve, and use standard
error propagation to compute errors in linear combinations giving the various
contributions to the total binding energy. Where multiple resolutions are not available,
the truncation error is taken to be the maximum error among all other configurations.

%%%%%%%%%%%%%%%%%%%%%%%%%%%%%%%%%%%%%%%%%%%%%%%%%%%%%%%%%%%%%%%%%%%%%%%%%%%%%%%%%%%%%
%%%%%%%%%%%%%%%%%%%%%%%%%%%%%%%%%%%%%%%%%%%%%%%%%%%%%%%%%%%%%%%%%%%%%%%%%%%%%%%%%%%%%
\section{Non-spinning Binaries}
\label{sec:Ej:nonspinning}

\subsection{Imprint of the mass ratio}

\begin{figure}[t]
  \includegraphics[width=\linewidth]{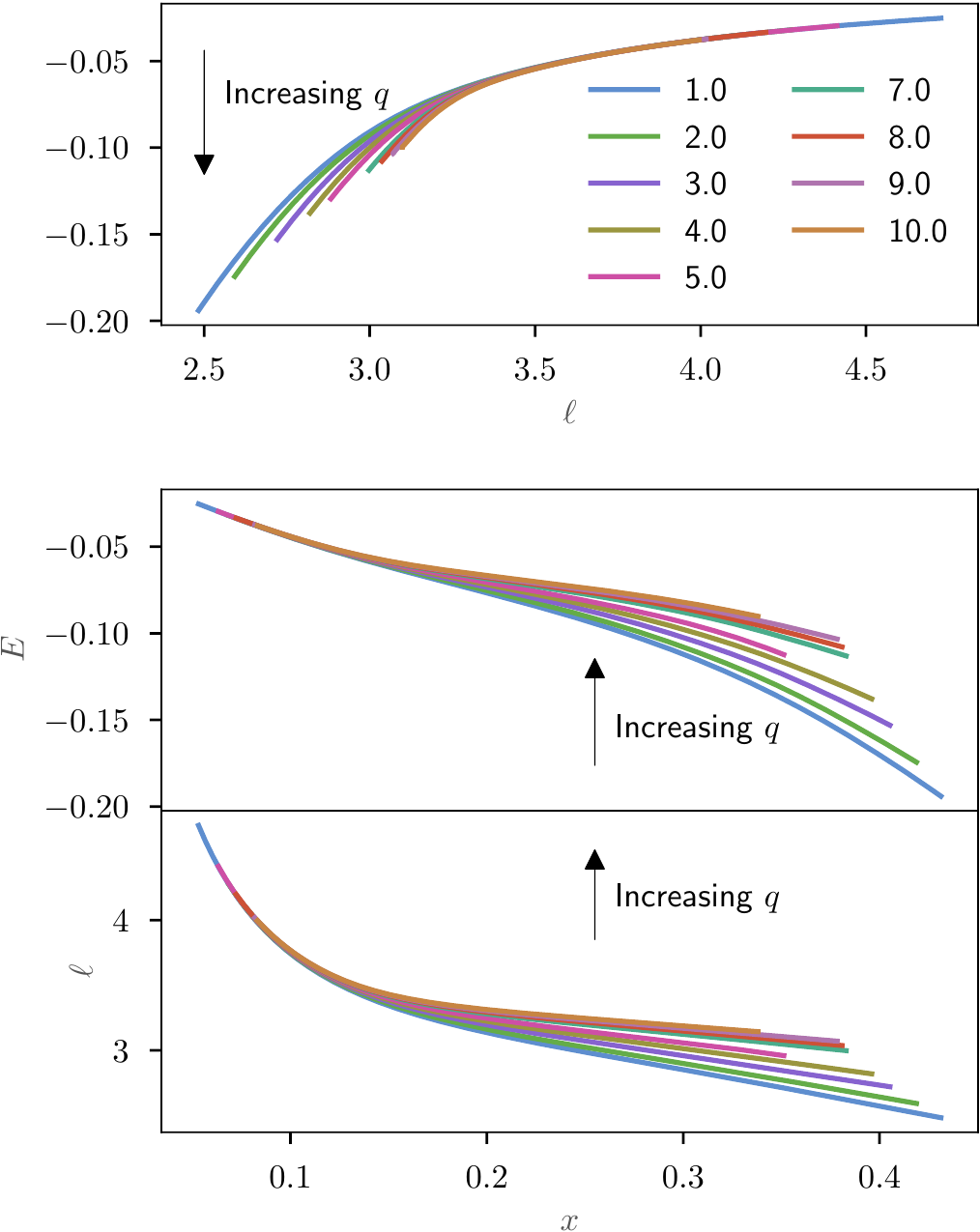}
  \caption{
    Top panel: Binding energy $E$ as a function of the reduced orbital angular
    momentum $\ell$. Consistent with the Newtonian limit all $E(\ell)$ curves
    approach the same value for large $\ell$. The curves show a hierarchy in which 
    for fixed $\ell$ the most unequal mass systems are
    the most bound, where we define $q\geq1$. 
    Middle panel: 
    Binding energy $E$ as a function of $x$, where for a fixed $x$, higher 
    mass ratio systems are less bound. 
    Bottom panel: 
    Reduced orbital angular
    momentum $\ell$ as a function of $x$. For fixed $x$ the angular
    momentum increases for increasing $q$. }
  \label{fig:e_x_ell_x_nospin}
\end{figure}

To isolate point-particle contributions from the binary dynamics, we start by
considering non-spinning binaries. Figure~\ref{fig:e_x_ell_x_nospin} (top panel) shows the
binding energy as a function of the reduced orbital angular momentum for a subset of
simulations in Table~\ref{tbl:non_spinning}. There is a clear trend in the
data: for higher mass ratios and for given value of $\ell$, the system becomes more bound and merges at smaller values of $\ell$. This observation is in
agreement with PN predictions, once higher PN orders are included. 

The opposite trend in mass ratio is demonstrated in the middle and bottom panel of Fig.~\ref{fig:e_x_ell_x_nospin}, in which the binding energy and angular
momentum as a function of the PN parameter $x$ are presented. 
For a fixed value of $x$ (i.e. orbital frequency), the most unequal binaries are the least bound and have the lowest values of angular momentum.

\subsection{Comparison with waveform approximants}

\begin{figure*}[t]
  \includegraphics[width=\linewidth]{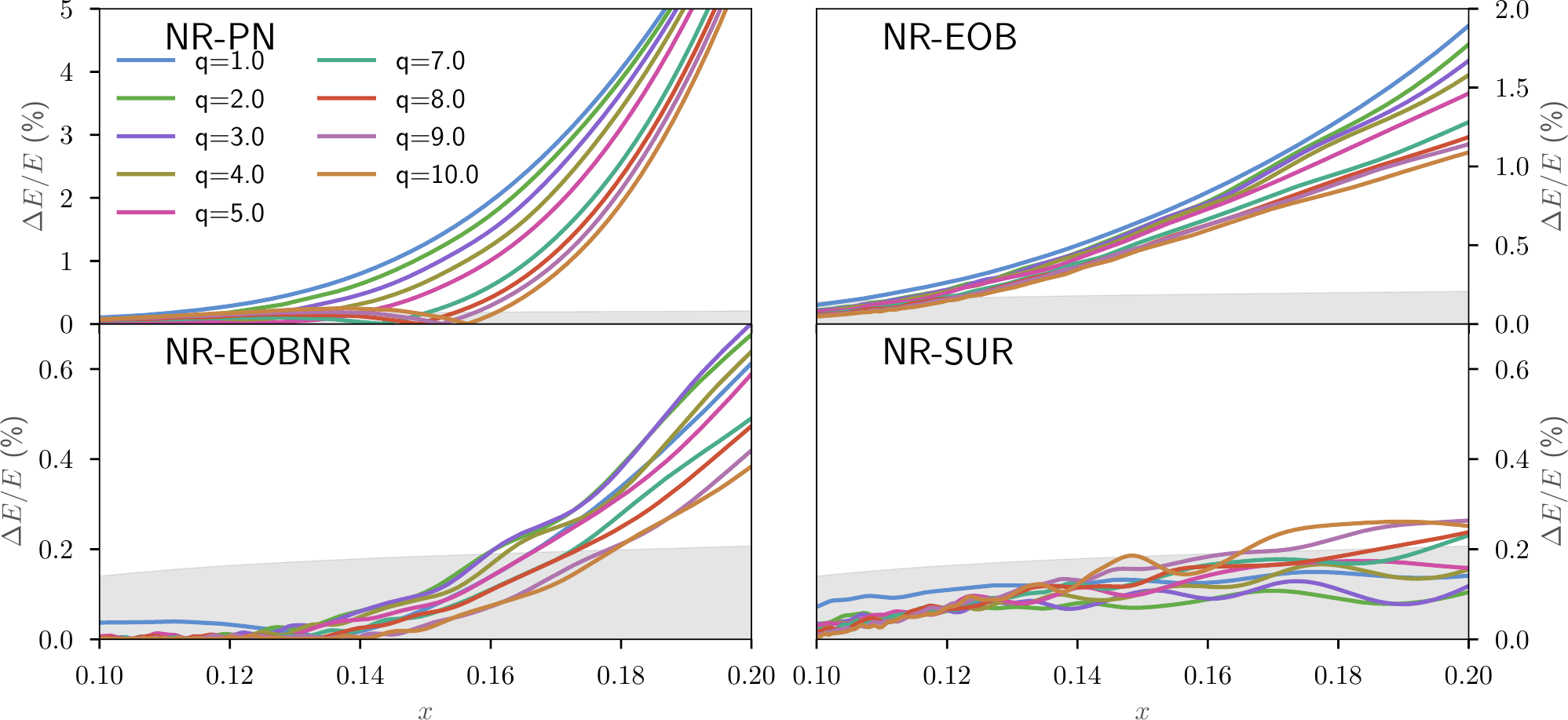}
  \caption{Difference of NR binding energy $E(x)$ with respect to PN (upper left panel), 
    uncalibrated EOB (upper right panel), calibrated EOBNR (lower left panel), and
    the non-spinning surrogate~\cite{Blackman:2015pia} (lower right panel). 
    The gray region represents an estimate of the NR
    error, given as the maximum of all errors in all the curves.
    Notice the improvement in the agreement of the calibrated EOBNR model
    compared to the uncalibrated EOB model.}
  \label{fig:delta_e_x_ell_x_nospin_all}
\end{figure*}

Next we compare the NR results to PN predictions, calibrated/uncalibrated
EOB models~\cite{Bohe:2016gbl}\footnote{The uncalibrated model is obtained from
SOEBNRv4 by setting all calibration and NQC parameters to zero.}, as well as the non-spinning NR surrogate
from~\cite{Blackman:2015pia}. Previous comparisons between PN, EOB, and NR curves
have been presented in e.g.~\cite{Damour:2011fu,LeTiec:2011dp,Nagar:2015xqa}.

Unsurprisingly, the PN predictions show the largest difference with NR data as seen in the top left panel of
Figure~\ref{fig:delta_e_x_ell_x_nospin_all}. The error grows quickly as a
function of $x$, where equal mass systems show the largest fractional error. 
It should be noted, however, that during the inspiral the error is still small $\lesssim 6 \%$ until
$x=0.2$ which is close to merger. The top right panel shows the difference
with the uncalibrated SEOBNR model. The difference with NR data is smaller and
the agreement persists to higher frequencies showcasing the effect of the EOB
approach. In the bottom left panel, the difference with respect to the full
calibrated model (SEOBNR) is shown. Here, the differences are even smaller,
which indicates that the calibration of the model to match NR waveforms also
improves the agreement of the conservative dynamics. Finally, the bottom right panel contrasts
the NR data with a surrogate model. We find the smallest differences here, as
expected from a model that is entirely based on NR information. Although the surrogate provides a very powerful way to obtain NR information (dynamics and waveforms), it is limited in scope of application in two crucial ways: i) the information is only available about the last $\simeq15$ orbits prior to merger, and ii) the results are only accurate inside the domain where the surrogate is constructed (i.e. $1\leq q\leq 10$).

\subsection{Phenomenological $E(x)$ - fit}

Generically, $E(x)$ depends sensitively on the
symmetric mass ratio $\nu= \mu/M$. An examination of the 4PN expression for the
binding energy \cite{Damour:2014jta} shows that progressively higher powers of
$\nu$ appear at each PN order. Schematically, the expansion is
\be E(x) =
-\frac{x}{2}(1+c_1x+c_2x^2+c_3x^3+c_4x^4),
  \ee
where the coefficients $c_{i}$ are polynomial functions of $\nu$. 
Starting at 4PN, terms logarithmic in $x$ also appear.
We construct a fit for the binding energy using the following procedure. First we
introduce unknown higher order coefficients $c_{5},\ c_{5\rm log},\ c_{55}$ corresponding
to the 5PN non-log and log corrections and the 5.5 PN correction, respectively.
We then consider a log-resummed expression for the binding energy

\begin{align}
E(x) =
-\frac{x}{2}\left(1+\log [1+a_1
  x+a_2x^2+a_3x^3+a_4x^4  \right. \nonumber \\
\left.  + a_5 x^5+a_{55}x^{11/2}] \right) 
  \label{eq:fit_nospin}
\end{align}  
following~\cite{Barausse:2009xi}. The coefficients
$a_{i}$ for $i=1,...,4$ are completely constrained by post-Newtonian theory,
while $c_{5}$ and $c_{5\rm log}$ enter $a_{5}$ and trivially $a_{55}=c_{55}$.
We fit this log-resummed expression separately to each dataset for all the
different mass ratios.
Then we fit each
coefficient to a low order polynomial in the symmetric mass ratio. This allows
us to assemble the full mass ratio dependence of $E(x)$. The coefficients are:

\begin{widetext}
\begin{eqnarray}
 c_5 & =  \quad  -605.96 +   20901\nu -76662\nu^2 + 84623\nu^{3} \\
 c_{5log} & = \quad  454.06 +  14360\nu -60400.0\nu^{2} + 56875\nu^{3}\\
 c_{55} & = \quad  1271.2  -81986\nu  + 320930\nu^{2} -327270\nu^{3}
\end{eqnarray}
 \end{widetext}

\begin{figure}[t]
  \includegraphics[width=\linewidth]{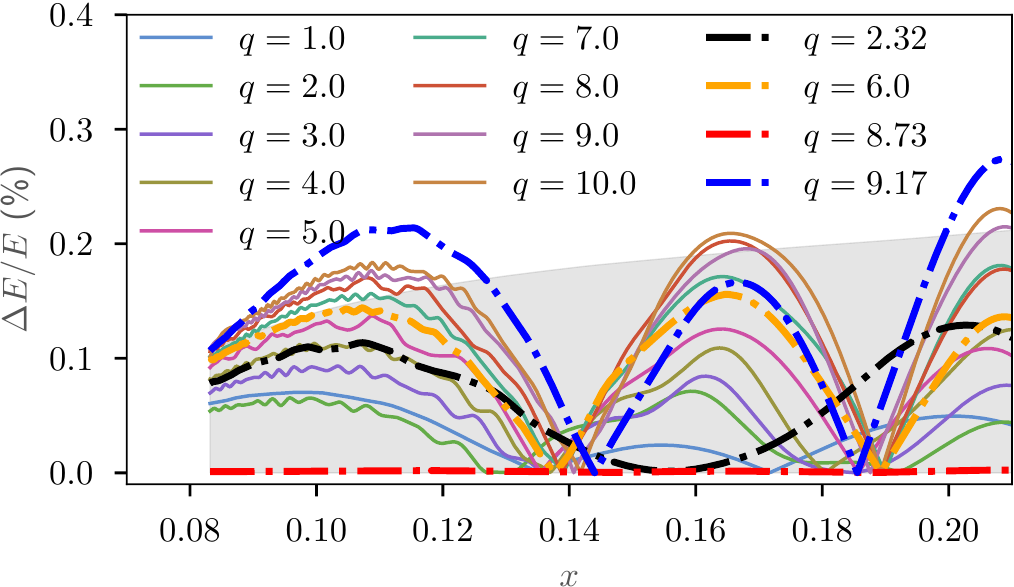}
  \caption{The difference between the binding energy of our NR data and 
    the phenomenological fit, Eq.~\eqref{eq:fit_nospin}. The thin
    solid lines correspond to the same curves shown in 
    Figure~\ref{fig:delta_e_x_ell_x_nospin_all}. The dashed-dotted lines
    showcase the agreement with configurations that were \emph{not} used in the
    construction of the fit. All differences are essentially within the NR error estimate,
    shown in gray.}
  \label{fig:fit_agreement_non_spinning}
\end{figure}

Figure~\ref{fig:fit_agreement_non_spinning} demonstrates the excellent agreement
of the fit with the original data, giving even slightly better performance 
than the NR surrogate. We
also test the fit by comparing it with NR data that were not used for construction. 
These curves are shown as bold, dashed-dotted lines in
Figure~\ref{fig:fit_agreement_non_spinning}. As one can see, the agreement
remains remarkable up to the very late inspiral. 
We note that the fit is only applicable in
the region $x\le0.2$ and nominally limited in the range of mass ratios
to $q\in [1,10]$. However, we have verified that 
reasonable results are even obtained outside of this range, 
e.g.~for $q=100$, by comparing results to SEOBNR predictions.

There are several possible applications for the phenomenological fit 
obtained with Eq.~\eqref{eq:fit_nospin}. 
While it is obvious that the fit can be used to directly test the dynamics 
inferred from waveform models, it might also be used to constrain higher
order PN terms, see e.g.~\cite{Kapadia:2015yra}.
Additionally, an alternative calibration of the 
EOB $A$-potential is possible.
Restricting to circular orbits, i.e., to a vanishing radial conjugate momentum, 
one can compare Eq.~\eqref{eq:fit_nospin} with 
$E=(H_{\rm EOB}-M)/\mu$ and obtain the EOB-Hamiltonian $H_{\rm EOB}$ 
and thus the unknown high order PN coefficients in the $A$-potential.  
This allows a direct calibration of the conservative dynamics of EOB without
the direct use of the waveform.

%%%%%%%%%%%%%%%%%%%%%%%%%%%%%%%%%%%%%%%%%%%%%%%%%%%%%%%%%%%%%%%%%%%%%%%%%%%%%%%%%%%%%
%%%%%%%%%%%%%%%%%%%%%%%%%%%%%%%%%%%%%%%%%%%%%%%%%%%%%%%%%%%%%%%%%%%%%%%%%%%%%%%%%%%%%
\section{Equal mass, aligned spin binaries}
\label{sec:Ej:constant_spin}

\subsection{Extracting spin contributions}
\label{sec:Ej:constant_spin:extraction}

\begin{table}
\begin{ruledtabular}\begin{tabular}{l|cccccc}
Name       & (00)            &($\ua0$)       &($\da0$)   &($\ua\da)$     &($\ua\ua$)&$(\frac{\uparrow}{2}0)$ \\
SXS ID     & 0180    & 0227  & 0585 &0217  & 0152         & 0223                       \\
\hline
$\chi_{1}$ & 0.0 &   0.6  &  -0.6  & 0.6      & 0.6      &   0.3        \\
$\chi_{2}$ & 0.0 &   0.0  &   0.0  & -0.6     & 0.6      &   0.0
\end{tabular}\end{ruledtabular}
\caption{Configurations used to separate out contributions to the binding energy. }
\label{tbl:special}
\end{table}

\begin{figure*}[t]
  \includegraphics[width=\linewidth]{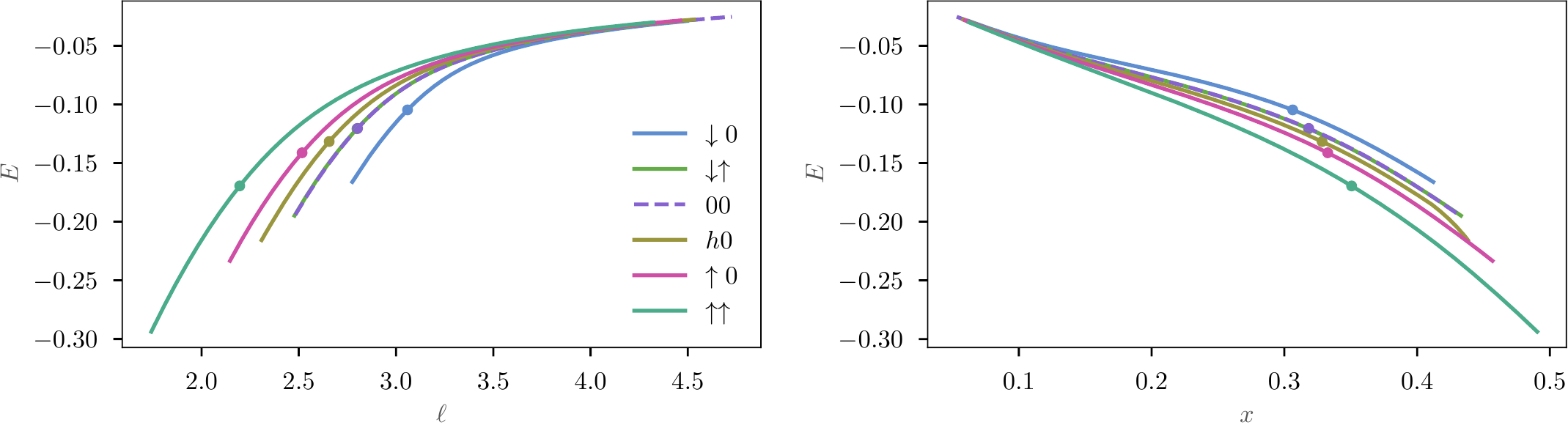}

  \caption{The binding energy $E$ as a function of reduced orbital angular
    momentum $\ell$ (left) and the post-Newtonian parameter $x$ (right) for configurations from Table~\ref{tbl:special}. The
    scatter points indicate the location of the merger.
  }
  \label{fig:e_vs_ell_q1_all}
\end{figure*}

As a starting point for extracting spin-dependent contributions to the binding
energy, we begin by considering a sequence of equal-mass, non-precessing
simulations. By neglecting quartic-in-spin corrections, all the individual
components identified in Eq.~\eqref{eq:Epieces} can be extracted from the simulations in
Table~\ref{tbl:special} using the following linear combinations 
(see also~\cite{Bernuzzi:2013rza,Dietrich:2016lyp}):
\begin{eqnarray}
E_{SO} &=& \frac{1}{6}\left[ -(\da 0)+ 16 \left(\frac{\uparrow}{2}0\right) - 12 (00) -3(\ua 0)\right], \label{eq:e_comps0}\\
E_{S^{2}} &=& \frac{1}{2}\left[(\da0) -2(00)+(\ua0)\right],\\
E_{SS} &=& (\da0) - (00) - (\ua\da) + (\ua0), \\
E_{S^{3}} &=& \frac{1}{3}\left[-(\da0)-8\left(\frac{\uparrow}{2}0\right)+6(00)+3(\ua0)\right], \\
E_{S^{2}S} &=& \frac{1}{2}\left[-(\da0)+2(00)+(\ua\da)-3(\ua0)+(\ua\ua)\right].
\label{eq:e_comps}
\end{eqnarray}

Figure~\ref{fig:e_vs_ell_q1_all} shows the $E-\ell$ and $E-x$ curves for all
configurations after applying the procedure described in
Sec.~\ref{sec:methods:Ej}. 
Furthermore, it shows clearly that the larger the total
angular momentum of the binary, the higher the 
frequency of merger and the lower the
orbital angular momentum.
This effect is sometimes referred to as the hang-up effect~\cite{Campanelli:2006uy}.
We further find that the curves corresponding to the non-spinning $(00)$ configuration and to the $(\ua\da)$ configuration lie essentially on top of each other,
since the leading order difference is the quadratic-in-spin term.

\begin{figure*}[t]
  \includegraphics[width=\linewidth]{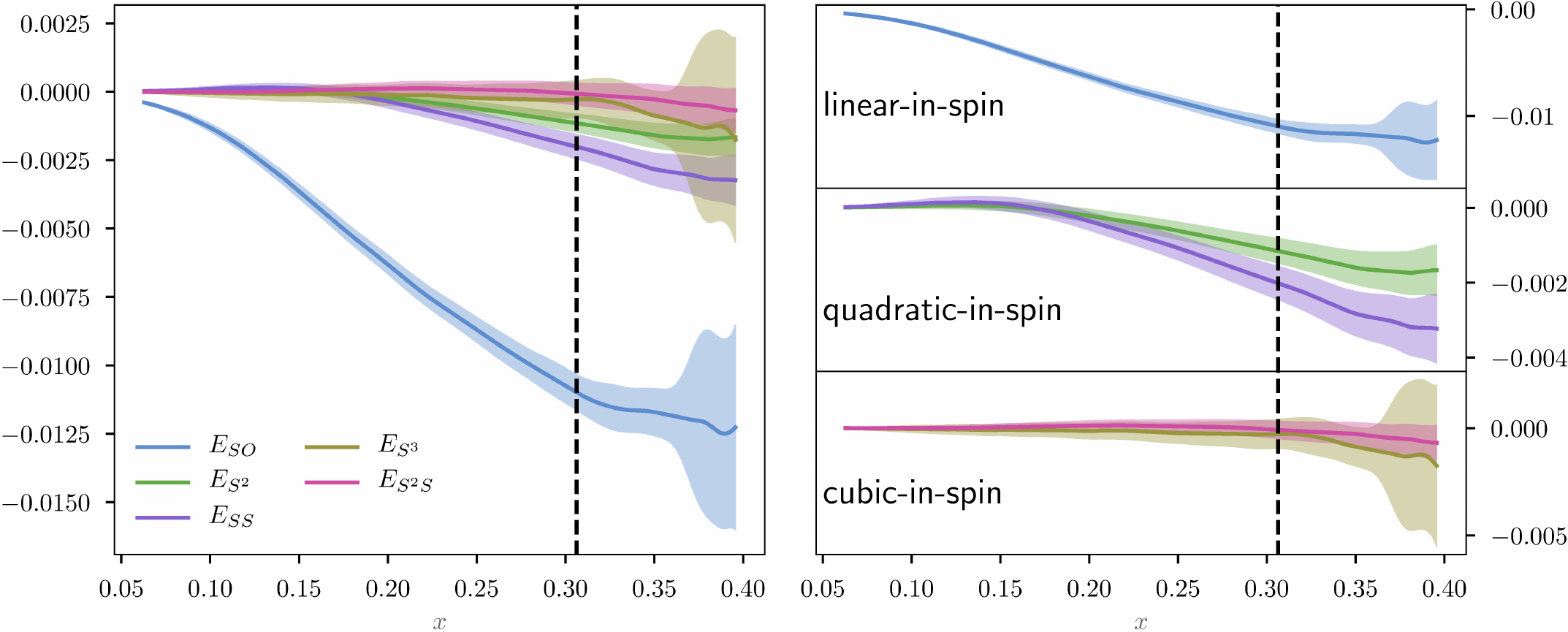}

  \caption{Spin contributions to the binding energy as a function of $x$. The dashed
    vertical line represents the point of merger for the configuration that
    merges earliest (i.e., at the lowest frequency). The shaded regions represent
    the NR error. Note that the spin-orbit contribution is about one order of magnitude
    larger than the other spin contributions. 
    The spin-squared and spin-cubed terms are consistent with zero early in the inspiral and are monotonically
    decreasing as one approaches merger.    }
  \label{fig:e_components_q1}
\end{figure*}

Using Eqs.\eqref{eq:e_comps0}-\eqref{eq:e_comps} we obtain the various contributions shown in
Fig.~\ref{fig:e_components_q1}. The right panels highlight the
linear, quadratic and cubic-in-spin terms. Throughout the inspiral, the
spin-orbit term dominates the others by an order of magnitude. On the
other hand, the quadratic and cubic-in-spin terms have comparable magnitudes,
with the quadratic terms growing larger near the merger.

\begin{figure}[t]
 \includegraphics[width=\linewidth]{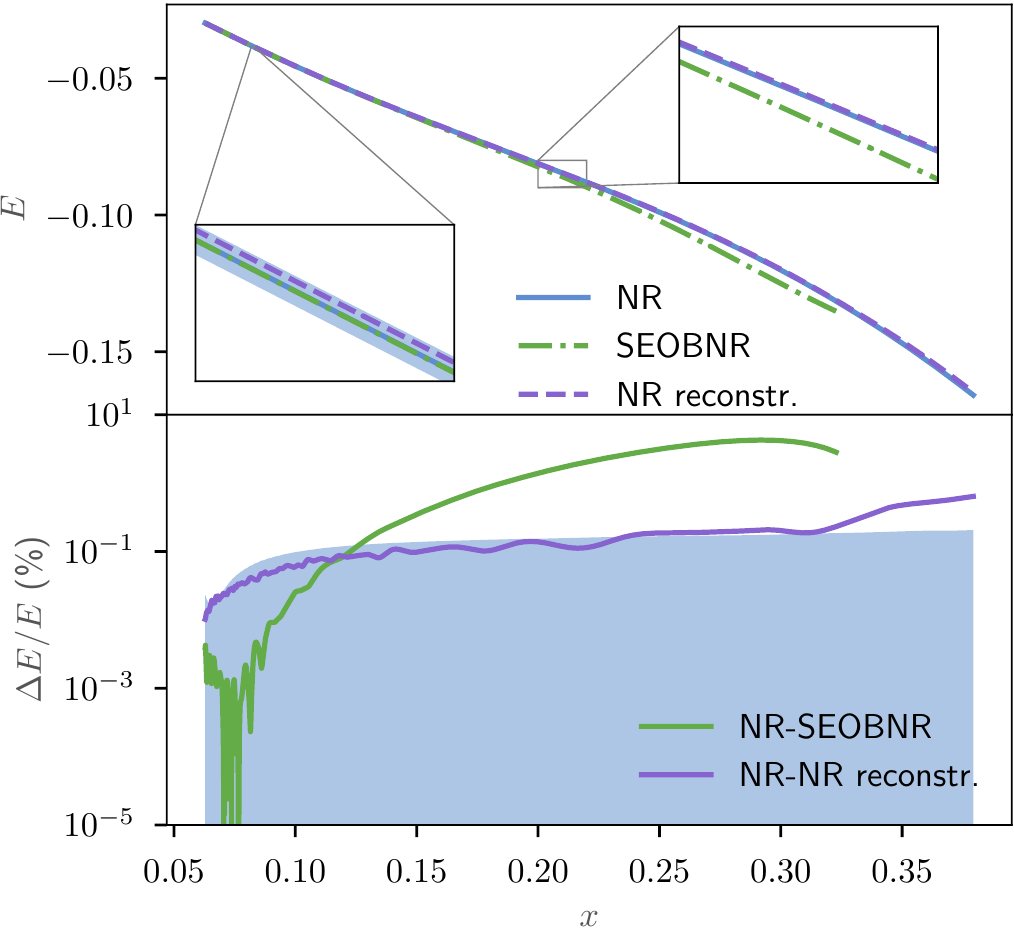}
 \caption{Detailed analysis of the  equal mass $\chi_{1}=-0.4,\ \chi_{2}=0.8$ system. 
 The top panel shows the binding energy while
 the bottom shows the fractional error. 
 The NR error is shown as the shaded region.}
 \label{fig:e_reconstruction_detail}
\end{figure}

\begin{figure}[t]
\includegraphics[width=\linewidth]{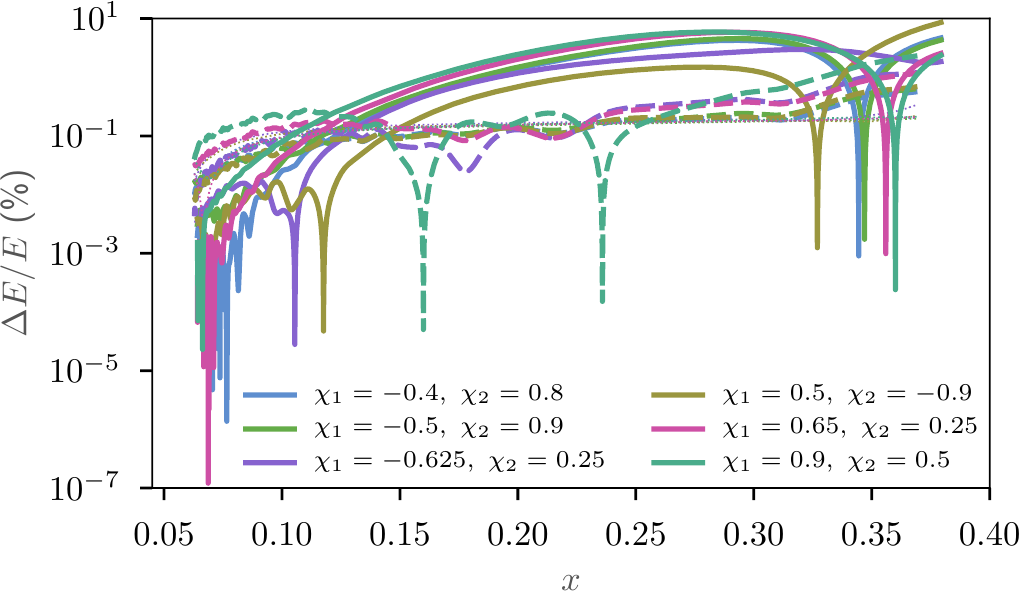}
\includegraphics[width=\linewidth]{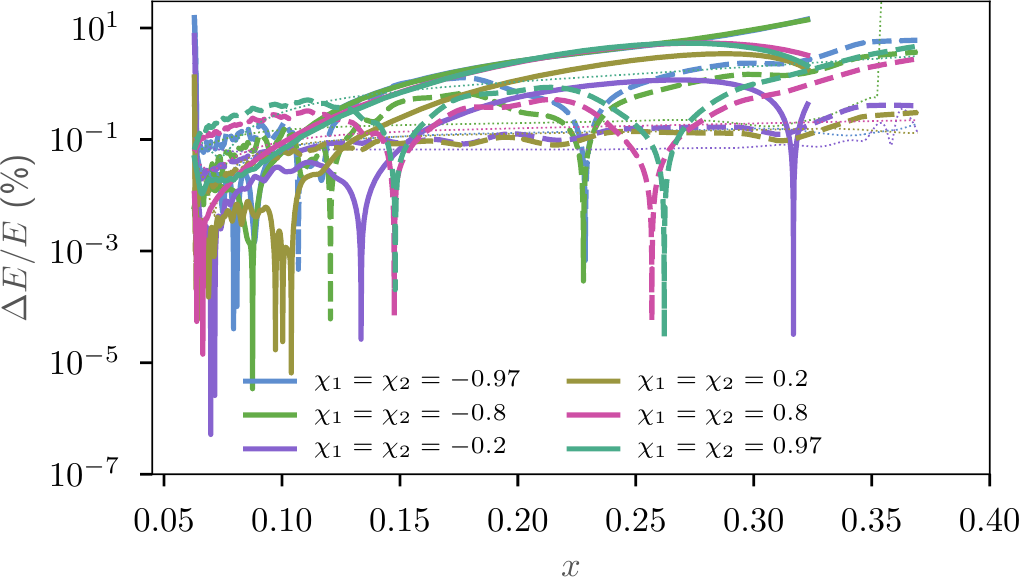}
 \caption{Top panel: residuals of all the unequal spin cases from Table~\ref{tbl:equal_mass} between NR and SEOBNR (solid), NR and 
 reconstructed (dashed).  Bottom panel: same, but for equal spin cases. Both models show low residuals in the early inspiral,
 and the reconstructed curves generally have smaller residuals at higher frequencies and stay within the NR errors (indicated by dotted curves) longer.} 
 \label{fig:e_reconstructions}
\end{figure}

To evaluate how well the energy components are extracted and to test our ansatz, 
Eq.~\eqref{eq:Epieces}, we check whether we can
reconstruct the full $E-x$ curves for simulations which were not used in the
computation above. In particular, we choose six equal-spin cases and six 
unequal spin cases, as described in Table~\ref{tbl:equal_mass}.
The $E-x$ curves are reconstructed as
\begin{equation}
 E=E_{0}+E_{SO}\frac{\chi_{1}}{\chi_{0}}+E_{SO}\frac{\chi_{2}}{\chi_{0}}+E_{S^{2}} \frac{\chi_{1}^{2}}{\chi_{0}^{2}}+E_{S^{2}} \frac{\chi_{2}^{2}}{\chi_{0}^{2}}
 +E_{SS}\frac{\chi_{1}\chi_{2}}{\chi_{0}^{2}},
\end{equation}
where $\chi_{0}=0.6$. Note that we omit the cubic-in-spin terms 
as they are not measured with sufficient accuracy.
A detailed analysis of the case $\chi_{1}=-0.4,\ \chi_{2}=0.8$
is shown in Fig.~\ref{fig:e_reconstruction_detail}, 
in which we also include SEOBNR results~\cite{Bohe:2016gbl} for comparison. 
Early in the inspiral, both the EOB approximant and the 
NR reconstructed curves stay close to the NR data. As the frequency increases, the EOB approximant departs from the NR curve, 
while the reconstructed curve remains close throughout the inspiral, 
although it leaves the region of NR error near merger. 
The behavior presented in Fig.~\ref{fig:e_reconstruction_detail} 
is typical for the cases we have considered.
In Fig.~\ref{fig:e_reconstructions}, we present the residuals between NR, 
EOB and the reconstructed results. 
For unequal spins, the difference between the reconstructed curves and NR result
remains small throughout the inspiral (typically below $0.3\%$) and throughout 
the entire simulation $\lesssim 3\%$. Only 
for the two extreme cases with $\chi_{1}=\chi_{2}=\pm 0.97$, 
the curves show more disagreement $\lesssim 0.7\%$ during the inspiral
and around merger $\simeq6\%$. 
The larger disagreement for higher spin magnitudes 
indicates the importance of cubic-in-spin contributions.

The calibrated EOB models also perform well, with errors
below $0.5\%$ during the early inspiral and maximum errors 
throughout the simulation $\lesssim 10\%$. 

Having extracted the various components of energy, we proceed to 
a detailed comparison with different models.

\subsection{Comparison with waveform approximants}
\label{sec:Ej:constant_spin:comparison}

\begin{figure*}[t]
  \includegraphics[width=0.42\linewidth]{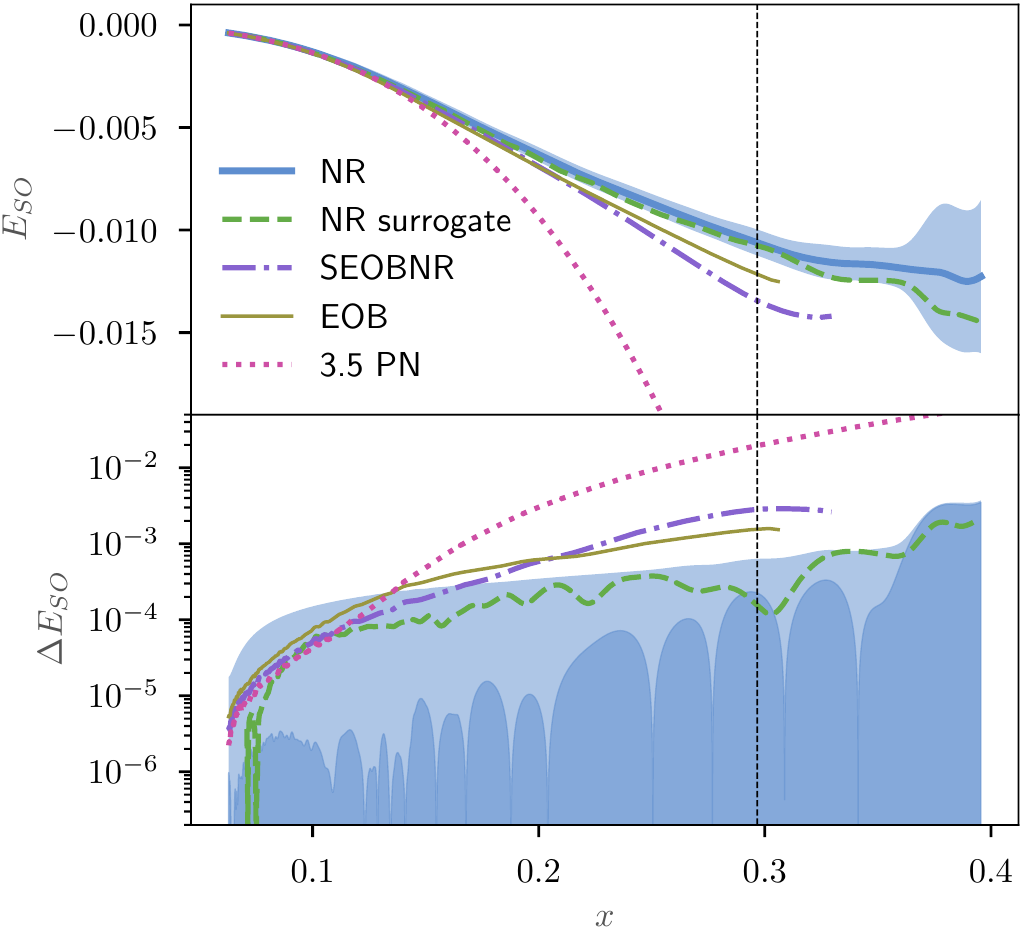}
  \includegraphics[width=0.42\linewidth]{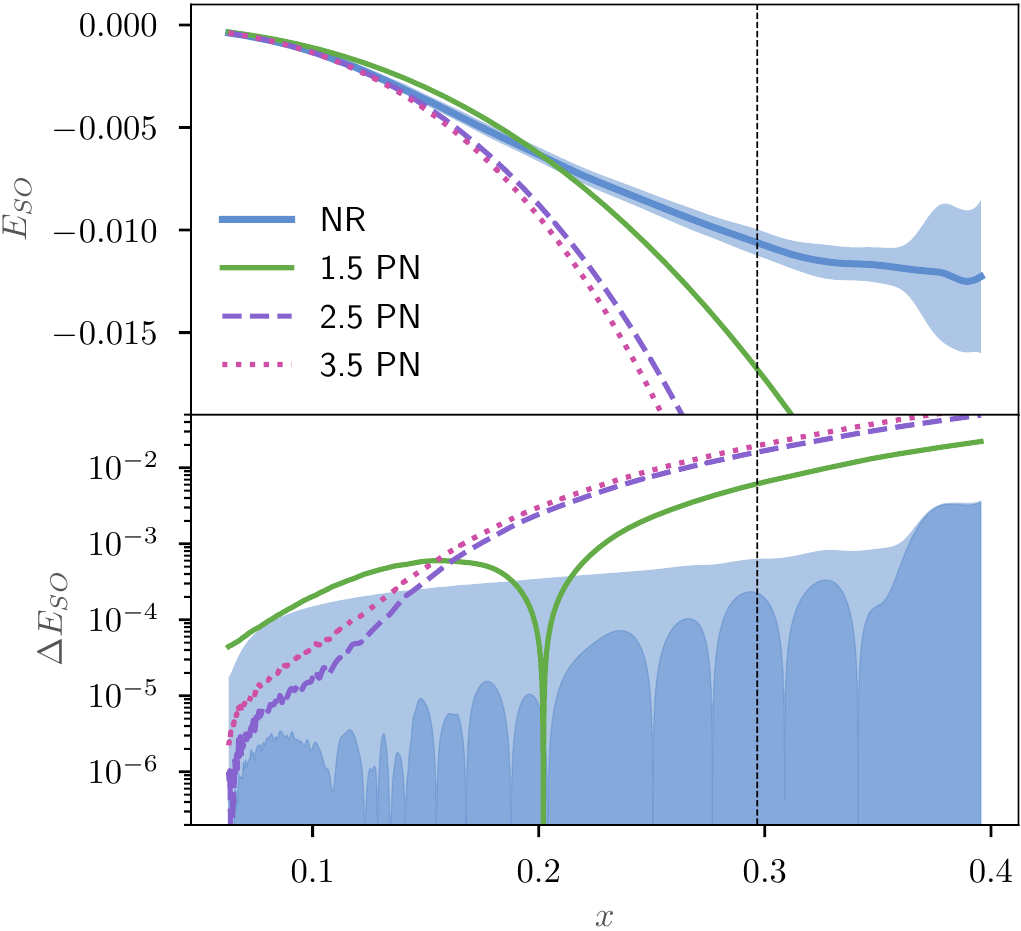}
  \caption{Left: the spin-orbit contribution to the binding energy $E$
    as a function of $x$ from NR (blue thick), the NR surrogate 
    (dashed), SEOBNR (dashed-dotted), EOB (thin) as well as the PN result 
    up to 3.5 PN order (dotted). 
    Note the good agreement of PN in the inspiral, which quickly deteriorates at $x\approx 0.12$ where the
    binary is only a few orbits to merger. The surrogate and EOB curves
    track the NR results substantially longer. The EOB curves terminate at EOB merger.
    The bottom panel shows the residuals between the approximants and NR.
    Until late inspiral, the surrogate and SEOBNR remain within the NR error.  
    The light shaded region gives the estimate of NR error as defined in Sec.~\ref{sec:methods:errors}. The dark shaded
    region shows instead the difference in the result between two different NR resolutions, a common
    metric used in literature. It is evident that our error estimate is indeed conservative.
    Right: comparison of binding energy in NR and PN at various
    orders. Throughout the inspiral, the results with at 2.5 PN order are 
    closest to NR. The dashed vertical line indicates the merger
    of the NR simulation that merges at the lowest frequency ($\chi_1=\chi_2=-0.6$).}
  \label{fig:ESO_PN_vs_NR_q1}
\end{figure*}

\subsubsection{Spin-orbit effects}

We begin by considering the spin-orbit effects, which enter at 1.5 PN order. In
the left panel of Fig.~\ref{fig:ESO_PN_vs_NR_q1}, we compare NR data to PN
expression for the spin-orbit contributions to the binding energy up to 3.5 PN
order, as well as to predictions from EOB with (SEOBNR) and without NR calibration and the NR surrogate model~\cite{Blackman:2017pcm}. Early in the
inspiral, the agreement between NR and PN (and all other models) is very good,
with differences of about $1\%$. However, as the inspiral proceeds, the PN
curves deviate sharply from NR data, while the other curves continue to track
the NR data much longer. The surrogate has the smallest differences but once
more suffers from length and parameter space constraints that limit its
application. Considering the difference between the uncalibrated EOB model and 
SEOBNR, we find that during most of the inspiral ($x\lesssim 0.2$), the calibrated model performs better than the uncalibrated one, with the uncalibrated 
models doing slightly better closer to merger.

The effect of PN order is shown in the right panel of
Fig.~\ref{fig:ESO_PN_vs_NR_q1}. For the early inspiral, increasing the
PN order helps to capture better the NR data, however the 2.5 PN results are marginally
better than the 3.5 PN. Once again, when the evolution approaches several orbits before the merger, the differences rise
sharply.

\subsubsection{Spin-squared effects}

\begin{figure*}[t]
  \includegraphics[width=0.42\linewidth]{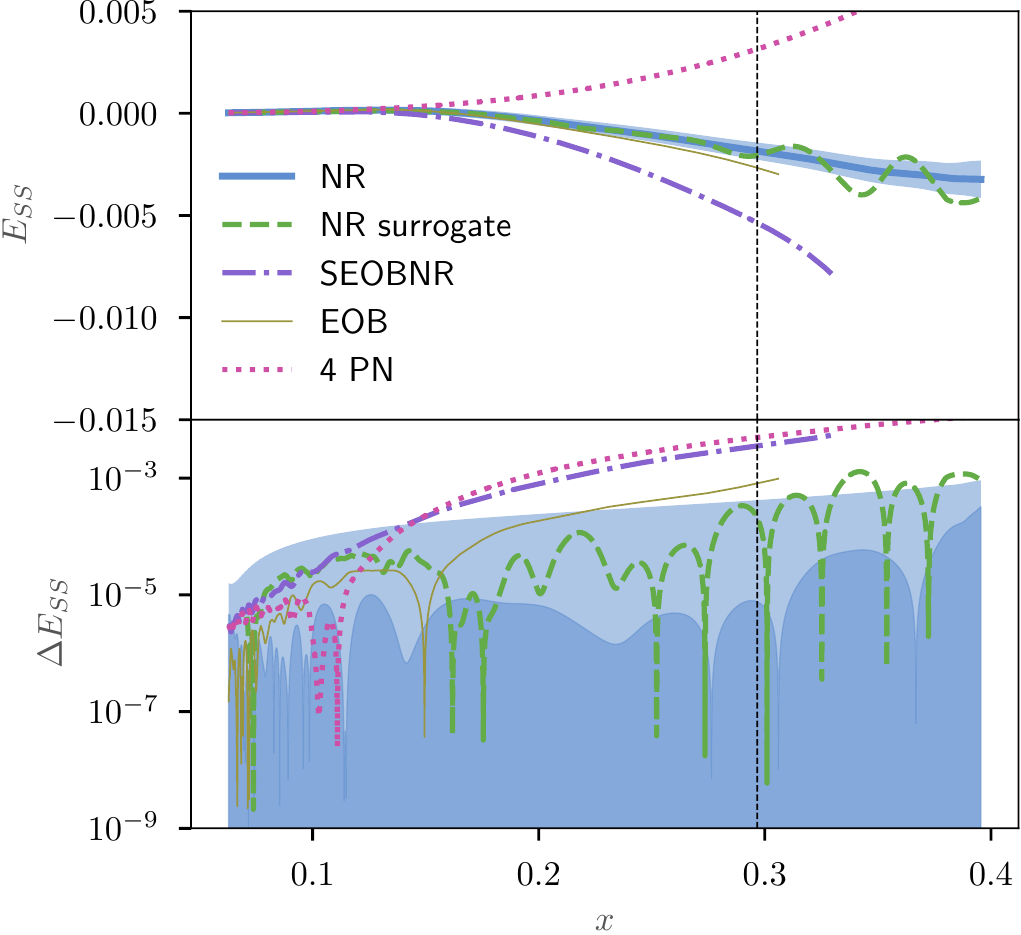}
  \includegraphics[width=0.42\linewidth]{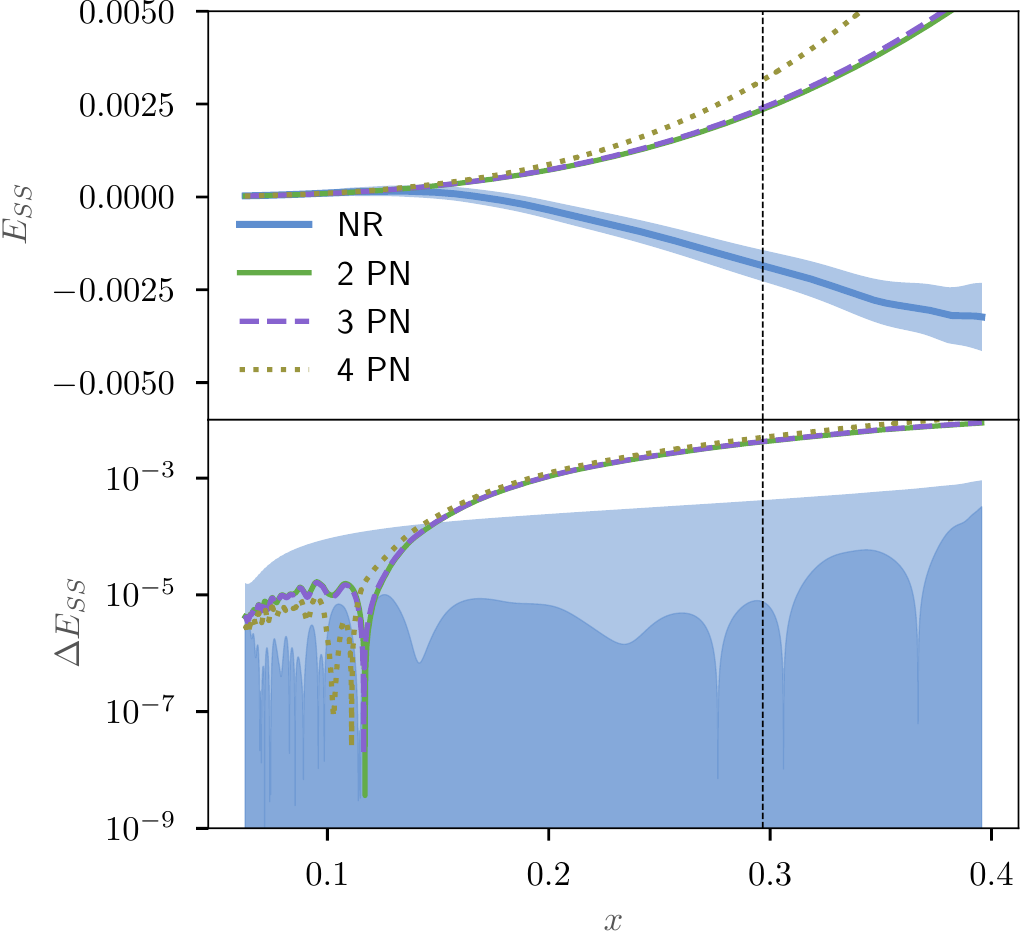}
  \caption{
    As Fig.~\ref{fig:ESO_PN_vs_NR_q1} but showing the spin-spin interaction. 
    Left: the spin-spin contribution to the binding energy $E$
    as a function of $x$ from NR, PN expression up to 4 PN order as well
    as SEOBNR, EOB and surrogate. Note
    the sharp contrast of the behavior between PN and the other models
    close to merger: the PN prediction grows monotonically, while
    the rest decrease. }
  \label{fig:ESS_PN_vs_NR_q1}
\end{figure*}

The results for the spin-spin term are shown in Fig.~\ref{fig:ESS_PN_vs_NR_q1}
and exhibit an interesting effect: during most of the inspiral, the PN and and
NR curves agree, but after about $x\approx 0.12$  the SS contribution to the binding energy decreases, in contrast to
the PN prediction which continues to grow monotonically. Meanwhile, using the
same expression as for NR but with the surrogate and the EOB models, we reproduce the NR behavior. 
More pronounced than for the spin-orbit contribution, we find a 
smaller difference between the NR result and the uncalibrated EOB model 
than to the calibrated SEOBNR model.
This shows that the calibration of the SEOBNR model, which 
focuses on minimizing unfaithfulness and difference in the time to merger~\cite{Bohe:2016gbl}, can result 
in a worse description of some aspects of the conservative dynamics in the strong field regime. 
One possible reason might be that the calibration parameters, which were introduced in the SEOBNR model, correspond to coefficients of higher unknown PN 
orders. When modeled as polynomials in $\nu,\chi$, these parameters break the symmetry underlying the extraction of the terms presented here. 

Considering the effect of PN order, we find no difference in the behavior of the different PN expressions. All PN orders show a monotonic growth with $x$ in contrast to NR and EOB,
and virtually no improvement with increasing PN order is found. Exactly
the same effects are present in the self-spin contribution, with the only
difference being that the magnitude of the self-spin effect is slightly smaller.

\subsubsection{Cubic-spin effects}

\begin{figure}[t]
  \includegraphics[width=\linewidth]{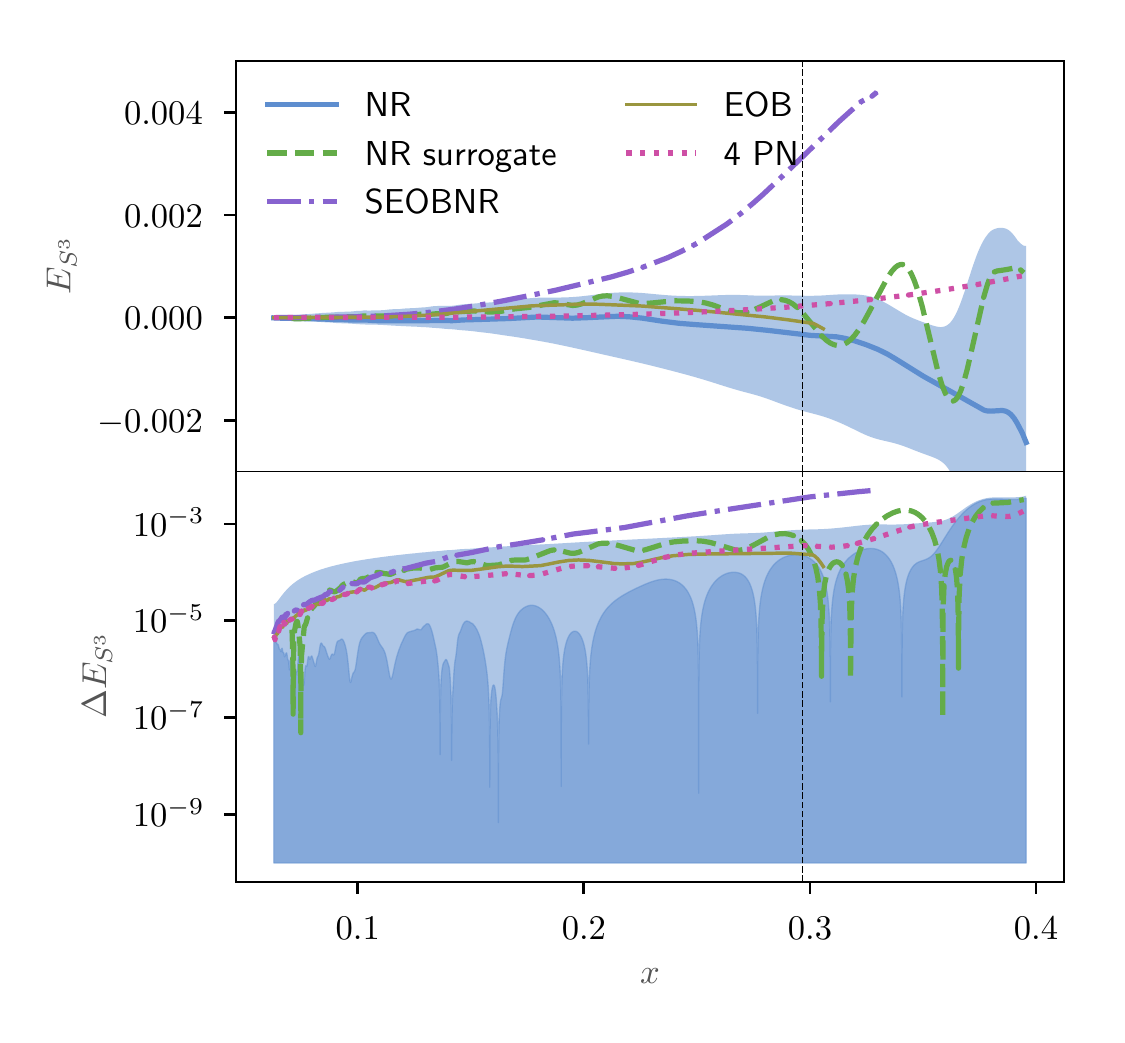}
  \caption{The cubic-in-spin contribution to the binding energy $E$
    as a function of $x$ from NR, PN
    as SEOBNR, EOB and surrogate. Since the uncertainties in the NR results are large,
    we are in effect putting an upper bound on the cubic terms. Note
    the sharp contrast of the behavior between  calibrated EOB versus the other models. 
    We do not show a comparison
  between various PN orders since spin-cubed terms are only known to leading order.}
  \label{fig:ES3_PN_vs_NR_q1}
\end{figure}

Finally we consider cubic-in-spin contributions to the binding energy. To enhance
the effect we consider the combination of both types, namely
$E_{S^{3}}+E_{S^{2}S}$. Figure~\ref{fig:ES3_PN_vs_NR_q1} shows the results. The
spin-cubed effects from the NR data display residual oscillations due to
eccentricity and are relatively noisy. For this reason, for
cubic-in-spin terms we focus on the qualitative behavior and put an upper bound on them. 
For most of the inspiral, all the models except the calibrated EOB stay within the NR error. This is likely
due to the symmetry breaking terms mentioned earlier. It should be noted, however, that these effects are 
small and contribute little to the overall disagreement. We leave it to future work to explore the cubic-in-spin
terms in more detail with the aid of additional NR simulations. 

%%%%%%%%%%%%%%%%%%%%%%%%%%%%%%%%%%%%%%%%%%%%%%%%%%%%%%%%%%%%%%%%%%%%%%%%%%%%%%%%%%%%%
%%%%%%%%%%%%%%%%%%%%%%%%%%%%%%%%%%%%%%%%%%%%%%%%%%%%%%%%%%%%%%%%%%%%%%%%%%%%%%%%%%%%%

\section{Generic spin aligned BBHs}
\label{sec:Ej:generic}

In this section we extend our investigation to aligned-spin binaries
with various masses ratios and spin magnitudes. 
We compare the obtained binding energies with state-of-the-art waveform approximants 
and present a method to phenomenologically describe 
generic spin-aligned systems. 

\subsection{Comparison with waveform approximants}

\subsubsection{Total binding energy}

\begin{figure*}[t]
  \includegraphics[width=0.49\linewidth]{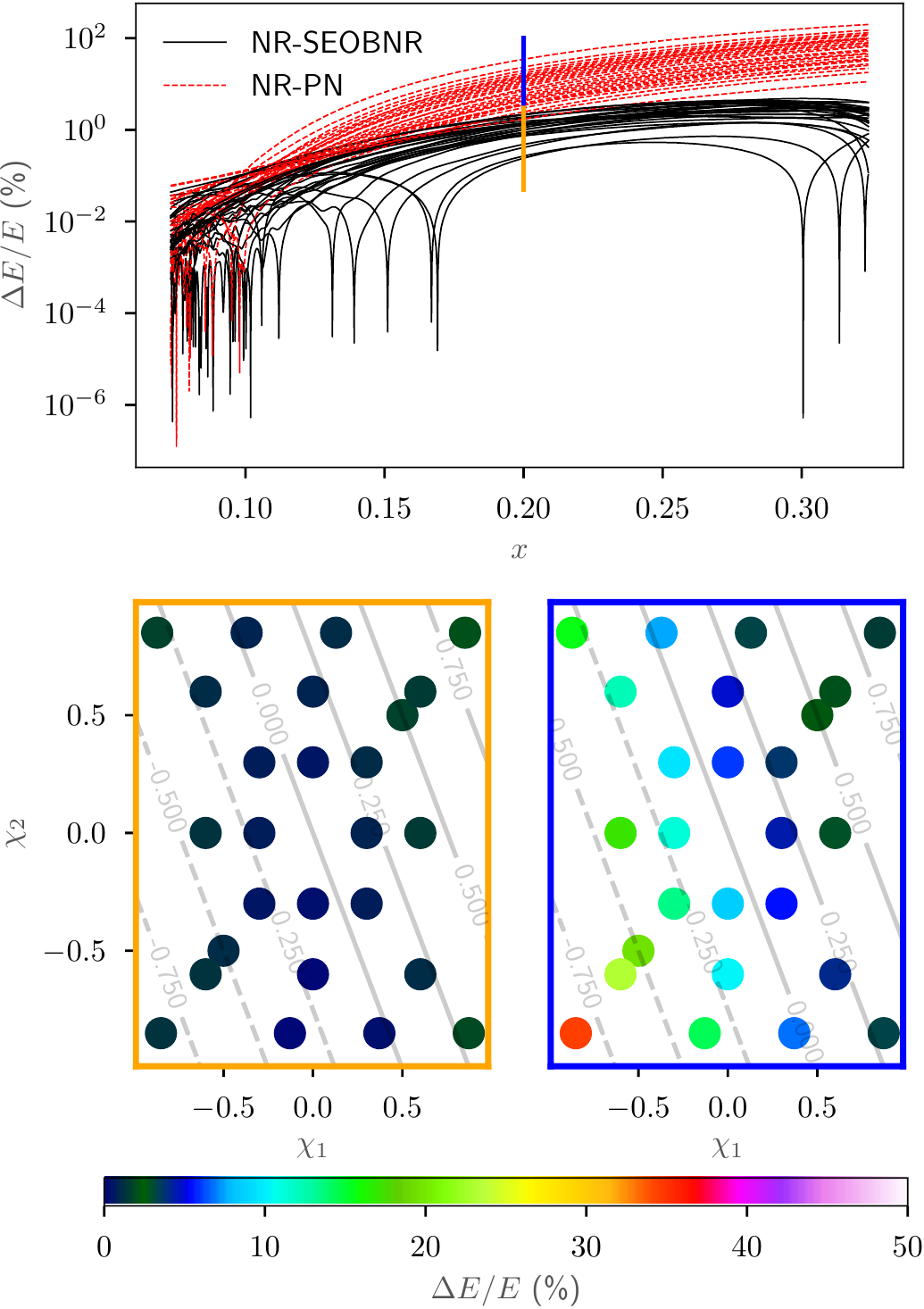}
  \includegraphics[width=0.49\linewidth]{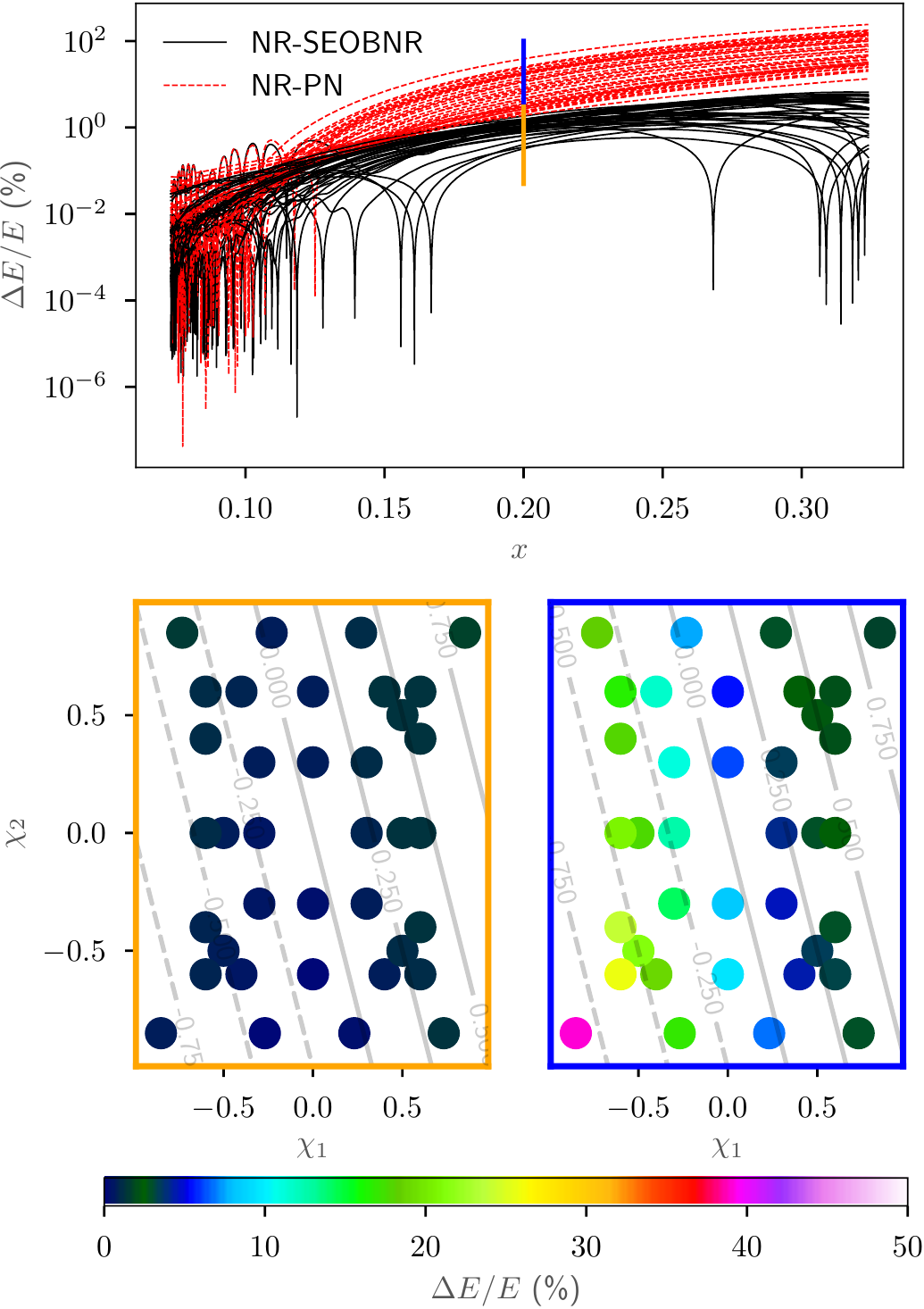}
  \caption{Fractional differences between NR, SEOBNR (black) and PN (red, dashed) $E-x$ curves for $q=2$ (left) and $q=3$ (right). The top panels give an overview
  demonstrating the errors as a function of frequency. 
  Note the excellent agreement of PN and SEOBNR with NR in the early inspiral. The bottom panels show the differences evaluated at $x=0.2$ as functions
  of the two aligned-spin components for SEOBNR (left) and PN (right). 
  In gray are contours of constant $\chi_{eff}$, with negative values dashed.
}

  \label{fig:comparison_q2q3}
\end{figure*}

In the following we construct the $E-x$ curves for a set of 70 simulations of 
unequal mass, non-precessing binaries described in Table~\ref{tbl:unequal_mass} and compare 
those to PN and SEOBNR predictions. The PN curves include 
non-spinning terms up to 4PN order, 
spin-orbit terms up to 3.5 PN order, 
spin-spin terms up to 4PN order, and
cubic-in-spin terms up to 3.5 PN order. 
Figure \ref{fig:comparison_q2q3} shows the difference between NR, 
SEOBNR and PN curves separately for 
mass ratios $q=2$ (left panels) and $q=3$ (right panels). 
The top panels show all considered $E(x)$ curves for the given mass ratios. 
In the inspiral ($x\lesssim 0.1$)
both PN and SEOBNR agree with NR remarkably well, 
with errors of order $0.1\%$. Closer to merger, the PN errors grow quickly to 
about $\sim 10\%$ and are consequently several times larger than EOB errors, which are typically $\sim 1 \%$. 
The bottom panels show the error for both SEOBNR (left)
and PN (right) at $x=0.2$ as functions of the two spins. From the PN results, one notes that for cases with large negative spins on the primary
black hole, the error is larger. This pattern is explained by 
considering contours of the effective spin, shown as labeled gray lines: 
the error is smallest for large positive $\chi_{\rm eff}$ and grows monotonically as the effective spin decreases. This is another manifestation
of the hang-up effect: cases with larger $\chi_{\rm eff}$ merge at higher frequencies than those with large negative effective spins. Since
we are computing the error at a fixed frequency,  for some cases it is evaluated ``closer'' to merger than for others. In contrast, such a correlation is not present for the SEOBNR model which shows small errors for all values of the effective spin. The same picture applies to the $q=3$ case,
with the main difference being the larger values of error at 
fixed frequency, as seen in the right panel of Fig.~\ref{fig:comparison_q2q3}.

Considering cases with $q\in(5,7,8)$, we find qualitative agreement with results with lower mass ratios. For a given effective spin, the cases with
higher mass ratios have higher errors for PN, while the errors are essentially unchanged for SEOBNR.

When repeating this procedure using the SEOBNRv2 models we find that for different cases one model may have lower
errors than the other, but no distinct pattern emerges. This is because SEOBNRv4 was constructed to attain better unfaithfulness
with NR in a larger region of parameter space, and includes no direct information about the binding energy.

\subsubsection{Spin-orbit contribution}
\label{sec:SO:all}

\begin{figure}[t!]
 \includegraphics[width=\linewidth]{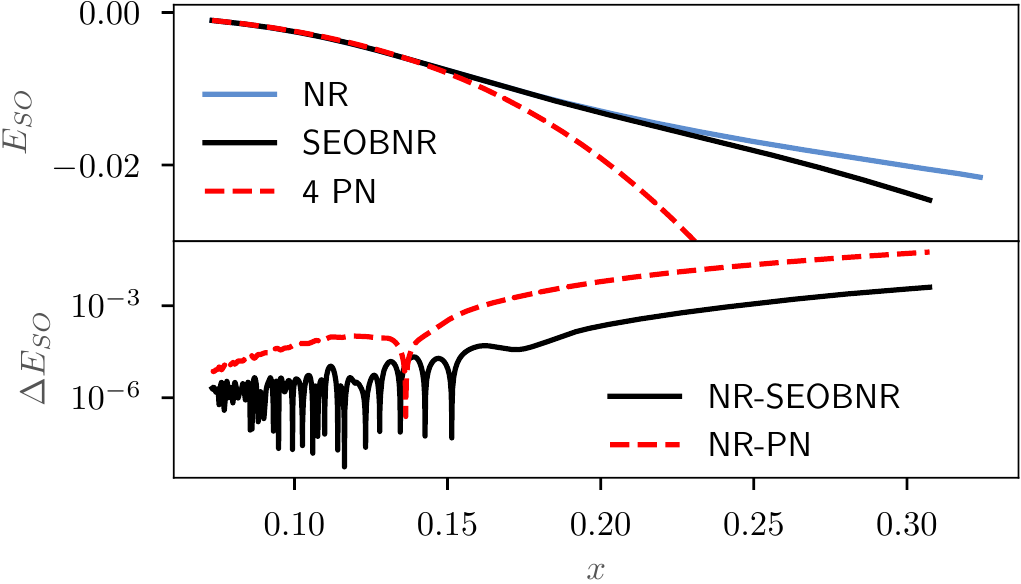}\\
  \includegraphics[width=\linewidth]{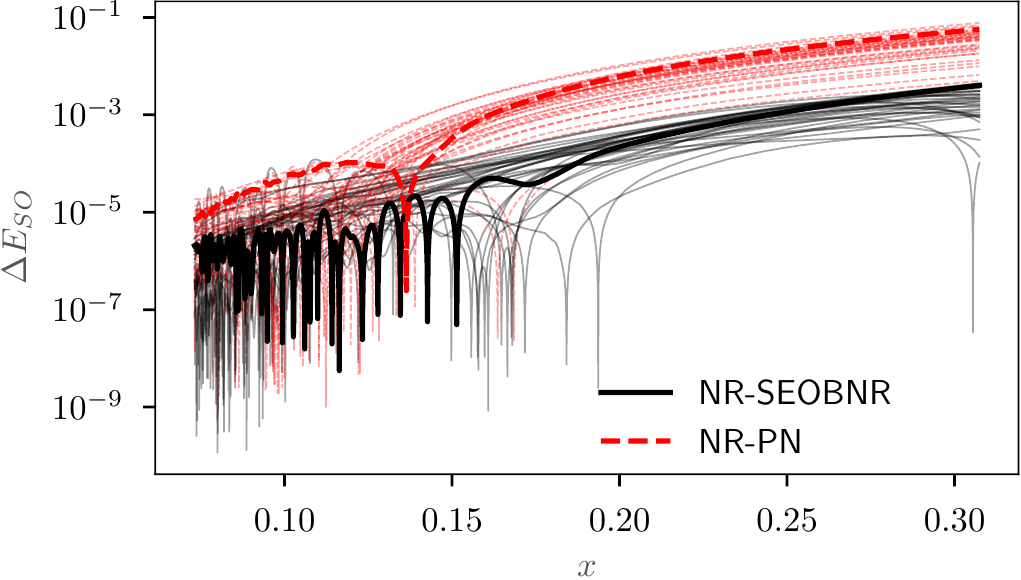}
  \caption{Absolute differences between NR, SEOBNR (black, solid) and NR, PN (red, dashed) 
  for the spin-orbit contribution. The top panel shows
  detailed behavior for the $q=8.0,\chi_1=0.5,\chi_2=0$ case and the bottom shows all cases. Note that 
  SEOBNR has fractional errors $\lesssim 8\%$ even in late inspiral ($x=0.2$), 
  while PN differences are a factor of $5$ larger.}
  \label{fig:ESO_comparison_all}
\end{figure}

\begin{figure}[t]
  \includegraphics[width=\linewidth]{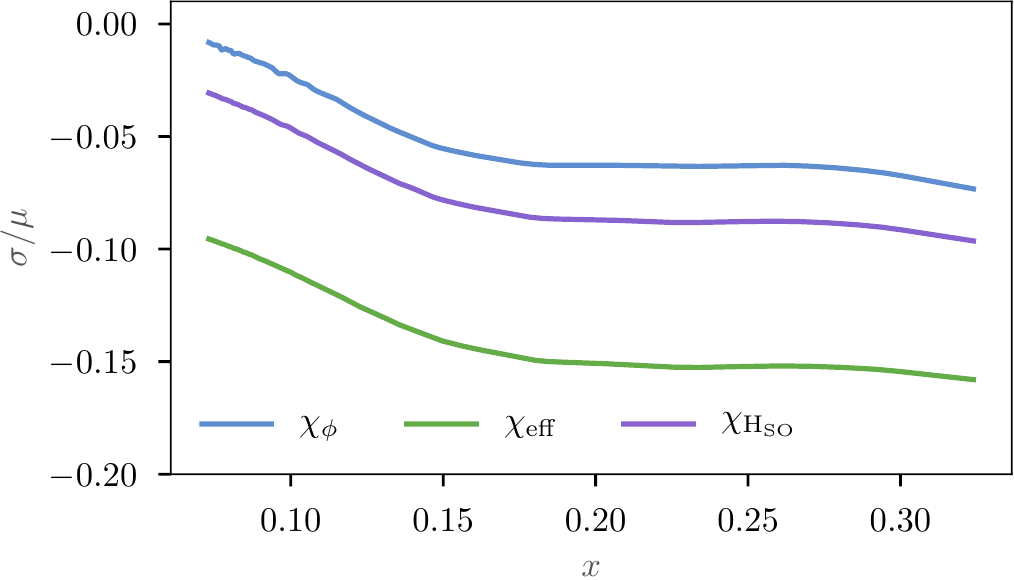}
  \caption{The coefficient of variation as a function of $x$ computed from all 35 (70) 
  cases that we possess with (without) symmetry. It is clear
  that the effective spin that arises in phase evolution, $\chi_{\phi}$ has the smallest spread. }
  \label{fig:chi_eff_comparison}
\end{figure}

After comparison of the total energies, we proceed to extract the various energy components for the cases 
where this is possible as described in Appendix~\ref{sec:Ej:nonprecBBH:chiflexible}.
The top panel of Figure~\ref{fig:ESO_comparison_all} demonstrates the typical behavior of 
the spin-orbit term for a $q=8,\chi_1=0.5,\chi_2=0$ system, the bottom panel shows the same behavior for all other cases.
At low frequency both models agree well, but as the inspiral proceeds, 
the SEOBNR model has differences smaller by an order of magnitude.

The \emph{absolute} errors in the spin-orbit term
for SEOBNR and PN are shown in bottom panels of 
Fig.~\ref{fig:ESO_comparison_all} for the $q=8,\chi_1=0.5,\chi_2=0$ system 
as well as all cases. 
At low frequency the errors are $\sim 10^{-5}$ for both models, 
corresponding to fractional errors of $\sim1\%$.  As the frequency increases, the errors
grow faster in PN than SEOBNR, such that in the late inspiral the 
SEOBNR fractional errors are $\sim 8\%$, a factor of $5$ smaller than PN.

Previously we saw that for an equal-mass binary we were able to rescale the spin-orbit contribution to the energy from a given configuration
to one with different spins, simply by rescaling by the magnitude of the spin. It is interesting to examine whether the various definitions of
effective spins proposed in the literature can serve the same task. To do so we rescale all the NR results by different effective spins
and compute the \emph{coefficient of variation}, $CV\equiv \frac{\sigma}{\mu}$
where $\mu$ and $\sigma$  are the mean and standard deviations respectively. 
We use the following definitions: 
\begin{align}
 \chi_{\rm mw} &= \frac{m_{1}\chi_{1}+m_{2}\chi_{2}}{m_1+m_2},\\
 \chi_{\phi}&=\chi_{\rm mw}-\frac{38\nu}{133}(\chi_{1}+\chi_{2}),\\
 \chi_{H_{SO}}&=\nu\left[\left(\frac{3}{4}+\frac{m_1}{m_2}\right)\chi_{1}+\left(\frac{3}{4}+\frac{m_2}{m_1}\right)\chi_{2}\right].
\end{align}
Surprisingly, Figure~\ref{fig:chi_eff_comparison} shows that it is the effective spin that corresponds to the phase evolution $\chi_{\phi}$ 
results in the smallest overall spread, followed
by the Hamiltonian inspired spin $\chi_{H_{SO}}$ and 
the standard mass-weighted spin $\chi_{\rm mw}$. Thus, $\chi_{\phi}$ should be used
when trying to reduce dimensionality of the problem.

\subsubsection{Spin-squared contribution}

\begin{figure}[t]
 \includegraphics[width=\linewidth]{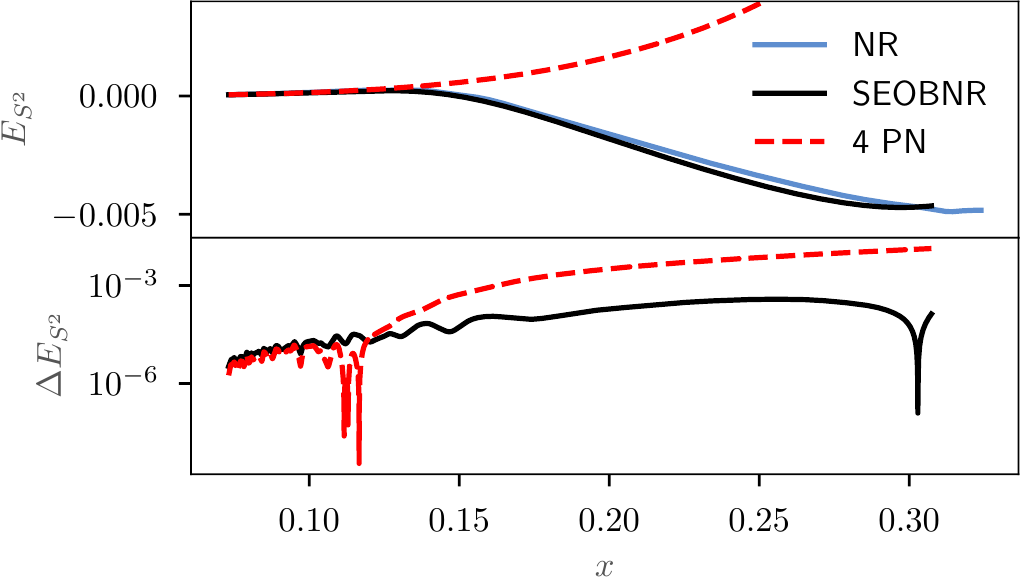}\\
  \includegraphics[width=\linewidth]{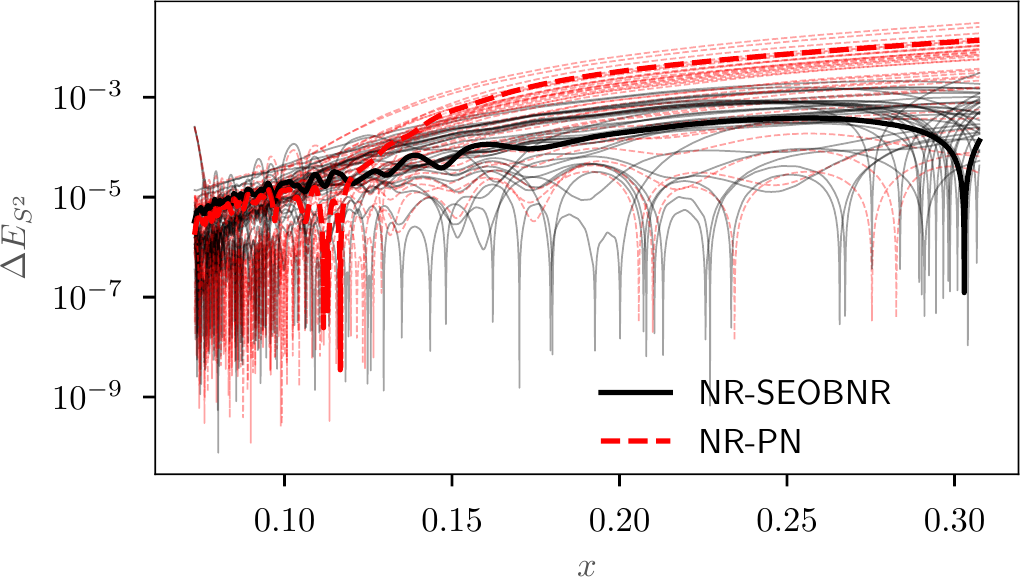}
  \caption{Absolute differences between NR, SEOBNR (black, solid) and NR, PN (red, dashed) for the spin-squared contribution. 
  The top panel shows detailed behavior for the $q=8.0,\chi_1=0.5,\chi_2=0$ case and the bottom shows all cases. Note that 
  SEOBNR has fractional errors $\lesssim 20\%$ even in late inspiral ($x=0.2$), 
  while PN no longer represents the contribution even qualitatively.}
  \label{fig:ES2_comparison_all}
\end{figure}

We finish the binding energy comparison by considering the spin-squared contributions. 
As before, the generic behavior we saw for equal masses carries over to this
case: PN generically predicts a monotonic spin-squared contribution, 
while SEOBNR correctly follows the NR results that have a local maximum,
as shown in Fig.~\ref{fig:ES2_comparison_all}.
The absolute errors for all cases are summarized in the bottom panel 
of Fig.~\ref{fig:ES2_comparison_all} and show a similar behavior as the spin-orbit contribution. 
In the early inspiral, the errors are of order $10^{-5}$, which corresponds to a fractional error of $\sim10\%$. However, they increase very quickly with PN errors growing faster than EOB errors. 

\subsection{Phenomenological E-$x$ curves}
\label{sec:unequal:phen}

Next we investigate if we can extend the phenomenological discussion of the binding 
energy for non-spinning and equal mass aligned-spin cases, with the aim of finding a closed-form
expression for binding energy contributions for generic spin-aligned systems. 
Assuming that the \emph{form} of the binding energy in NR is similar to that in PN theory, 
the spin-orbit contribution can be expressed as 
\begin{equation}
E_{SO}=x^{5/2}\nu\left[(S_{1}+S_{2})A(\nu,x)+(S_{1}/q+qS_{2})B(\nu,x)\right].
\end{equation}
Assuming that, for any fixed mass ratio the functions $A,B$ are independent of spin, then given any two
configurations $E_{SO},\tilde{E}_{SO}$ with given mass ratio and spins that are not multiples of each other, one 
can trivially obtain $A,B$ as functions of $x$. After fitting these functions for different mass
ratios the spin-orbit term for all mass ratios and spins is determined. In principle, spin-squared coefficients may be extracted using the same approach; however, this is hampered by numerical uncertainties.

The accuracy of determining $A,B$ depends on the accuracy of extraction 
of the spin-orbit terms, both in terms of numerical error and the contribution of higher spin terms.
As proof of principle, we applied this approach using SEOBNR data as the underlying model for a set of 15 configurations
between mass ratios $1.1-50$ and using spin magnitudes of $0.6$. We find the differences between the reconstructed curves
and the EOB data from Tables~\ref{tbl:equal_mass} and \ref{tbl:unequal_mass}, which were not used for the fit, to be $\lesssim 10^{-3}$ at $x=0.3$, 
comparable to the expected magnitude contribution of the cubic-in-spin terms.  This difficulty can in 
principle be overcome by creating data at more spin configurations to completely decouple the linear
and cubic-in-spin terms. For SEOBNR this is feasible, but would require $10$ configurations per mass 
ratio and thus is too computationally expensive for full NR simulations.
Furthermore, combining an increasing number of datasets leads to a growth of uncertainty caused by 
the individual errors of the configurations. 

Instead, we adopt a completely phenomenological model for the odd-spin terms. In particular, we take
\begin{equation}
E_{SO} \approx \left(1+\frac{a_3x^3+a_4x^4+a_5x^5}{1+b_1x+b_2x^2}\right) E_{SO}^{PN}
\label{eq:ESO_Q}
\end{equation}
where $E_{SO}^{PN}$ is 3.5PN expression, and $a_{i},b_{i}$ are polynomial functions of the
symmetric mass ratio $\nu$ and the effective spin $\chi_{\phi}$. 
The particular form of Eq.~\eqref{eq:ESO_Q} enforces that at small frequencies the PN prediction 
is obtained and correction only enter higher order PN terms. 
Furthermore, to preserve the correct symmetry under the transformation 
$(\chi_{1},\chi_{2})\rightarrow (-\chi_{1},-\chi_{2})$ the polynomials $a_i,b_i$ are restricted
to the form
\begin{equation}
 p +s_{0}\chi_{\phi}^{2}+s_{1}\chi_{\phi}^{4} + n_{0}\nu.
\end{equation}

\begin{figure}[t]
 \includegraphics[width=\linewidth]{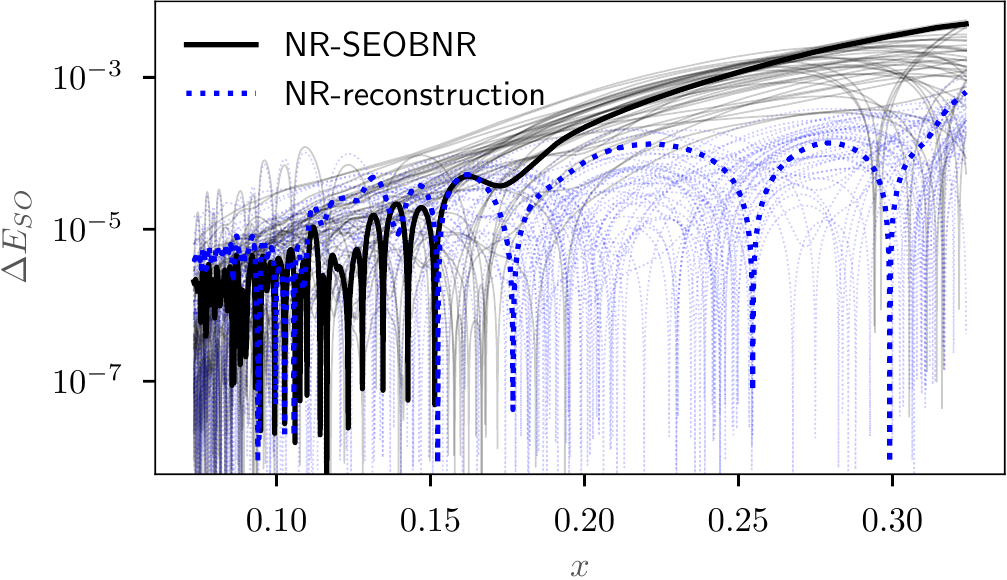}
  \caption{Absolute differences NR and reconstruction (blue, dotted) for the spin-orbit contribution. For comparison
  the difference between NR and SEOBNR is shown in black thin lines. The thick curves correspond to the 
  configuration $q=8,\ \chi_{1}=0.5,\ \chi_{2}=0$. Early in the inspiral both curves have low errors,
  with SEOBNR errors being lower due to the alignment of the NR data to EOB waveforms as discussed in Sec.~\ref{sec:methods:Ej}. 
  As the inspiral proceeds, the error in the reconstructed curves remains lower, staying 
  below $4\times10^{-4}$ for $x\approx 0.3$.
  }
  \label{fig:NR_reconstruction_SO}
\end{figure}

The fit is done in two steps: first, each $E_{SO}$ curve is fitted to Eq.~\eqref{eq:ESO_Q} and then
all coefficients are fitted across parameter space. To ensure a good interpolation across
parameter space, we use $L_{2}$ regularization with the regularization parameter of $1\times10^{-6}$ (we exclude cases with $\chi_{\rm mw}<0.15$ since the SO component is not as well determined as for the other cases). 
The results are summarized in Figure~\ref{fig:NR_reconstruction_SO} which shows the same data as in Figure \ref{fig:ESO_comparison_all} except that it includes 
equal-mass, equal-spin cases from Table~\ref{tbl:equal_mass}. Figure~\ref{fig:NR_reconstruction_SO} shows that in the early inspiral the phenomenological curves show
excellent agreement, comparable with SEOBNR. The error increases with frequency; in the late inspiral ($x=0.3$) it reaches $\sim 4\times 10^{-4}$, about a factor 8-10 smaller than SEOBNR.

The coefficients for the spin orbit terms are given by
\begin{widetext}
\begin{align}
 a_{3} & =  -37.684  -35.577\chi_{\phi}^{2} +  14.793\chi_{\phi}^{4} + 136.37\nu \\
 a_{4} & =  267.96+  437.71\chi_{\phi}^{2} -199.08\chi_{\phi}^{4} -821.08\nu \\
 a_{5} & = -271.96  -722.11\chi_{\phi}^{2} +  392.16\chi_{\phi}^{4}  + 1208.6\nu \\
 b_{1} & = -10.404  -0.52252\chi_{\phi}^{2}  -1.9139\chi_{\phi}^{4} +   3.9269  \nu\\
 b_{2} & = 30.308 + 15.715\chi_{\phi}^{2} +   2.7542\chi_{\phi}^{4}  -6.7698\nu
\end{align}
\end{widetext}

\section{Summary and conclusion}
\label{sec:conclusion}

In this article, we presented a detailed study of the energetics 
of aligned-spin binary black hole systems. 
This allows direct testing of the conservative dynamics in the strong field regime, see e.g.~\cite{Damour:2011fu,LeTiec:2011dp,Nagar:2015xqa}. 

We discussed in detail 
how the ambiguous offset caused by the initial burst of junk radiation can be 
removed and studied in great detail possible numerical errors/uncertainties. 
Based on the constructed curves, we compared results obtained from numerical relativity simulations 
with PN~\cite{Blanchet:2013haa} and EOB approximants~\cite{Nagar:2015xqa,Taracchini:2013rva,Bohe:2016gbl}, 
as well as for the non-spinning case with an NR surrogate model~\cite{Blackman:2015pia}. 
We find overall a very good performance of EOB approximants as well as
for the surrogate model, while the PN approximant is not capable of 
predicting the conservative dynamics accurately within the last orbits before 
merger. 

For the non-spinning binaries, the obtained fractional differences between the NR data and 
the considered EOB approximants (SEOBNR) is 
$\approx 0.2\%$ during the inspiral and $\approx 1\%$ at merger. 
The NR surrogate achieves fractional errors always on the order of the 
NR uncertainty; i.e., $\approx 0.2\%$.
In addition to these comparisons, we presented a phenomenological fit 
to non-spinning systems with mass ratios up to $q=10$ allowing an accurate representation of the binding energy (differences $\lesssim 0.1\%$). 
We outlined how the obtained fitting parameters, which effectively describe higher order PN orders, can be used for further waveform development. 

For spinning, equal-mass systems, we extracted for the first time 
the individual spin contributions to the binding energy for black hole binaries. 
In particular we studied the spin-orbit, spin-spin, and cubic-in-spin terms. 
We find that while the spin-orbit interaction is accurately modeled with the EOB approximant, 
the spin-spin interaction of the SEOBNR model~\cite{Bohe:2016gbl} as extracted by our method does not agree well with the NR results.
in contrast to the uncalibrated EOB model. This is likely due to the structure of the spin-spin calibration parameters in the SEOBNR
model that incorporate contributions from higher-order terms. 
For further development, a calibration 
with respect to the binding energy may improve the agreement further and allow an efficient way of decoupling
conservative from dissipative dynamics. 

Finally, considering a large set of unequal mass systems with aligned spin 
we again verified the good agreement of the full NR simulations
and the EOB model. Extraction spin-orbit and spin-spin contributions permit analysis of the effect of the particular choice of the binaries effective
spin parameter. We found best performance for the phase effective spin, $\chi_{\phi}$ compared
to the purely mass-weighted effective spin or the effective spin 
as it appears at leading order in the Hamiltonian. 
Additionally, we also presented 
a phenomenological representation of the spin-orbit contribution 
to the binding energy. For the considered cases, the error of the 
phenomenological fit is on the order of $1\%$ during the inspiral 
and below $6\%$ at merger. 

\appendix

\begin{figure*}[t]
  \includegraphics[width=0.495\linewidth]{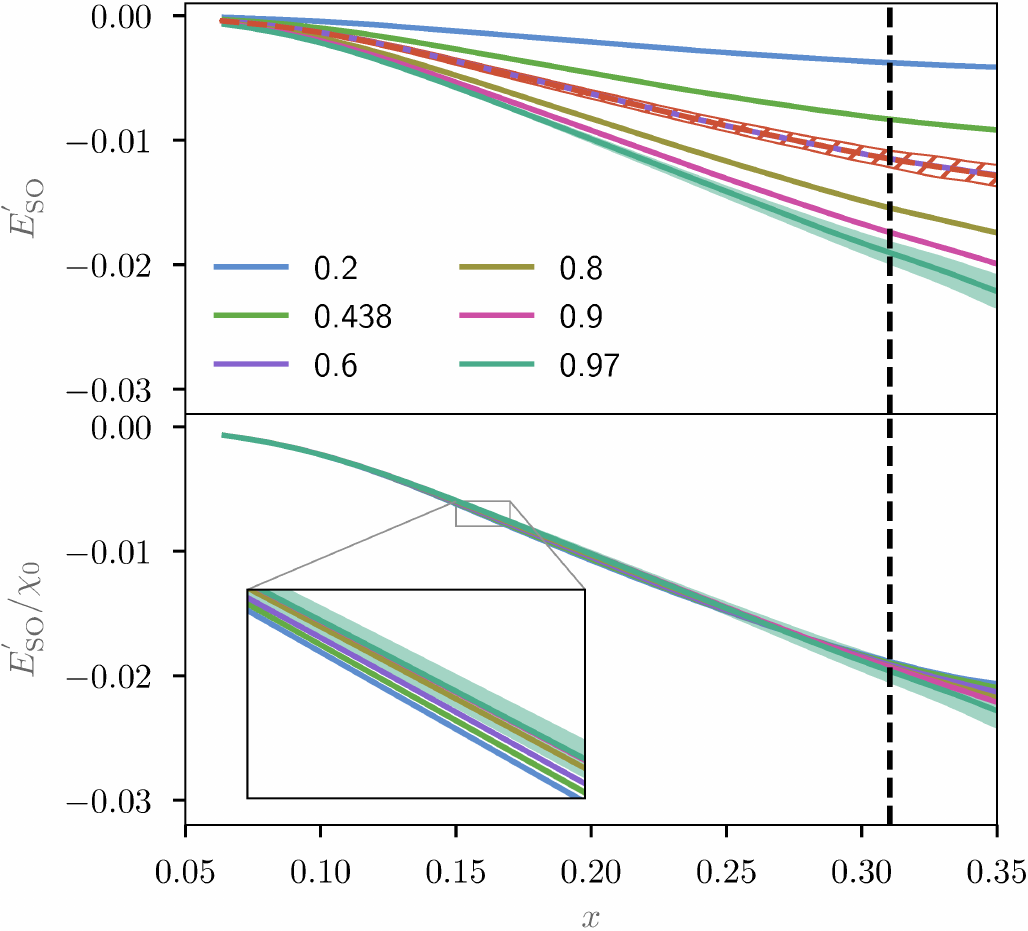}
  \includegraphics[width=0.495\linewidth]{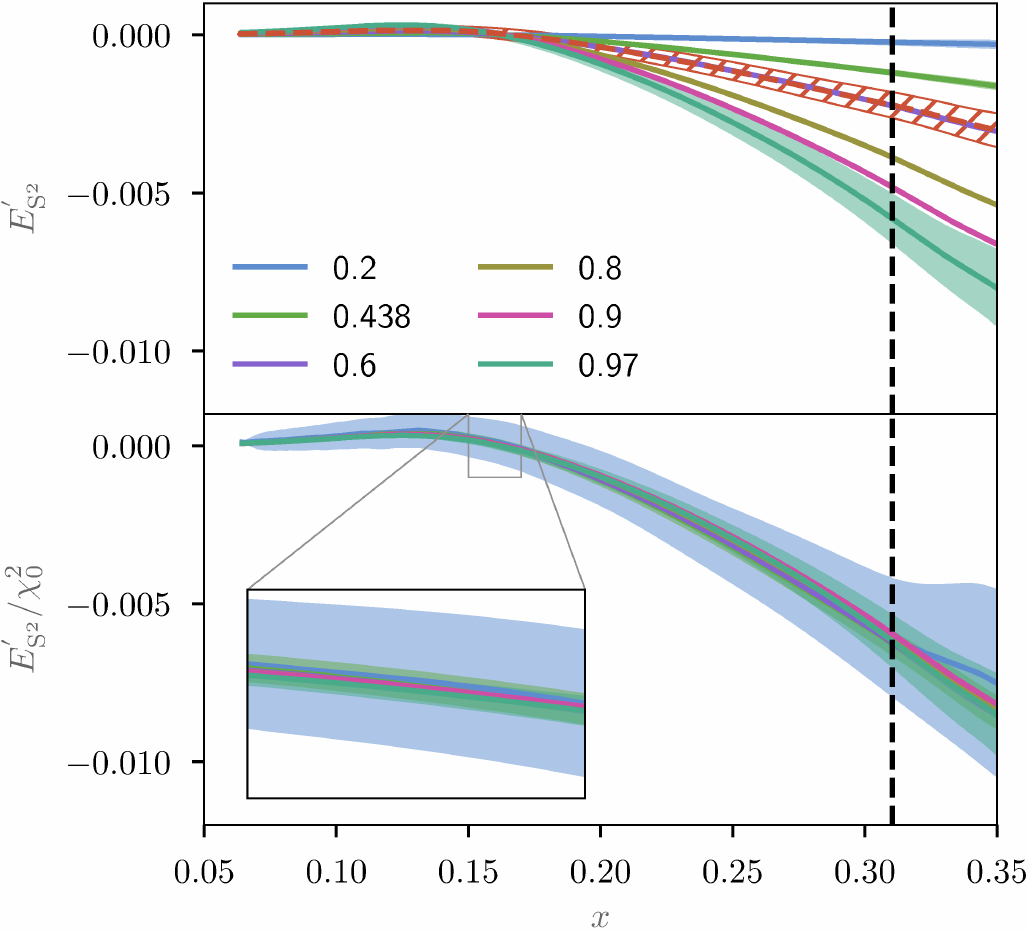}
  \caption{Left: the approximate SO contribution to the binding energy,
    $E^{'}_{SO}$, for all equal-spin cases described in Table~\ref{tbl:equal_mass}. The top panel shows
    the raw curves, and the bottom panel shows the results rescaled by $\chi$. Right:
    the combination of spin-squared contributions, $E^{'}_{S^{2}}$. The panels
    are the same as in the left plot, but rescaling is done by $\chi^{2}$. Note
    that in both panels the rescaling causes the curves to lie very close to
    each other, indicating that the procedure is successful in isolating the
    corresponding contributions to the binding energy. The hatched area shows the
    error in the corresponding combinations as shown in
    Fig.~\ref{fig:e_components_q1}, demonstrating that the simpler combinations
    yield smaller errors.}
  \label{fig:q_scale}
\end{figure*}
%%%%%%%%%%%%%%%%%%%%%%%%%%%%%%%%%%%%%%%%%%%%%%%%%%%%%%%%%%%%%%%%%%%%%%%%%%%%%%%%%%%%%
\section{Extracting spin contributions from a small number of simulations}
\label{sec:Ej:nonprecBBH:chiflexible}

Extracting the contributions to binding energy as described in Sec.~\ref{sec:Ej:constant_spin:extraction}
requires a set of five configurations to determine the value for one spin magnitude.
This amount of simulations is not available for all spin magnitudes, and performing these simulations is computationally expensive.
Therefore, we show that for a wide range of spins, alternative
expressions can be used to derive the SO and SS contributions 
using only two simulations per spin magnitude.
In particular, consider
\begin{align}
E_{SO}^{'}&=&\frac{1}{4}\left(E(\nu,\chi_1,\chi_2)-E(\nu,-\chi_1,-\chi_2)\right),  \label{eq:Ej:nonprecBBH:chiflexible1}  \\ 
E_{S^{2}}^{'}&=&\frac{1}{4}\left(E(\nu,\chi_1,\chi_2)+E(\nu,-\chi_1,-\chi_2)\right)- \nonumber \\ 
  &&\frac{1}{2}E(\nu,0,0). \label{eq:Ej:nonprecBBH:chiflexible2}
\end{align}

Since only two (three) simulations are necessary to extract the linear (quadratic) in spin contributions, 
in contrast to Eqs.~(\ref{eq:e_comps}), results are less noisy. 

To test Eq.~\eqref{eq:Ej:nonprecBBH:chiflexible1} and 
Eq.~\eqref{eq:Ej:nonprecBBH:chiflexible2} we show in Figure~\ref{fig:q_scale} 
the SO and SS contributions for a variety of equal mass cases. If the spin-orbit (spin-squared) contributions
were extracted correctly, then they should scale linearly (quadratically) with the
spin magnitude. This is indeed the case, as the rescaled curves lie virtually on top
of each other, as shown in the bottom  panels of Figure~\ref{fig:q_scale}. The hatched
red area shows the NR error from Figure ~\ref{fig:e_components_q1} which is larger than the estimate of the NR error
as computed from Eqs.~\eqref{eq:Ej:nonprecBBH:chiflexible1},\eqref{eq:Ej:nonprecBBH:chiflexible2}. Thus we see that this method performs as
expected across a wider variety of spin-magnitudes and allows the approximate extraction of linear and
quadratic in spin terms.
%%%%%%%%%%%%%%%%%%%%%%%%%%%%%%%%%%%%%%%%%%%%%%%%%%%%%%%%%%%%%%%%%%%%%%%%%%%%%%%%%%%%%
%\FloatBarrier

%\newpage

\section{Configurations}
\label{sec:config}
\begin{table}[h!]
\begin{ruledtabular}
\def\arraystretch{1.1}
\begin{tabular}{lc||lc}
SXS ID &     $q$ & SXS ID &     $q$ \\
\hline
  0180 &   1.0 & 0007 &   1.5 \\
  0169 &   2.0 & 0259 &   2.5 \\
  0168 &   3.0 & 0167 &   4.0 \\
  0295 &   4.5 & 0056 &   5.0 \\
  0296 &   5.5 & 0297 &   6.5 \\
  0298 &   7.0 & 0299 &   7.5 \\
  0063 &   8.0 & 0300 &   8.5 \\
  0301 &   9.0 & 0302 &   9.5 \\
  0303 &  10.0 & -    & - \\
\hline
  0201 &  2.32 & 0166 &  6.0  \\
  0199 &  8.73 & 0189 &  9.17 \\
\end{tabular}
\end{ruledtabular}
\caption{All non-spinning configurations used in this work. The top half of the table was used to construct the fit,
while the bottom provided validation cases for the phenomenological fit.}
\label{tbl:non_spinning}
\end{table}

\begin{table}[h!]
\begin{ruledtabular}
\def\arraystretch{1.1}
\begin{tabular}{lccc||lccc}
SXS ID &     $q$ &  $\chi_{1}$ &  $\chi_{2}$ & SXS ID &     $q$ &  $\chi_{1}$ &  $\chi_{2}$  \\
\hline
1137 &  1.0 &      -0.970 &      -0.970 & 0159 &  1.0 &      -0.900 &      -0.900 \\
0212 &  1.0 &      -0.800 &      -0.800 & 0215 &  1.0 &      -0.600 &      -0.600 \\
0148 &  1.0 &      -0.438 &      -0.438 & 0149 &  1.0 &      -0.200 &      -0.200 \\
0180 &  1.0 &       0.000 &       0.000 & 0150 &  1.0 &       0.200 &       0.200 \\
1122 &  1.0 &       0.438 &       0.438 & 0228 &  1.0 &       0.600 &       0.600 \\
0230 &  1.0 &       0.800 &       0.800 & 0160 &  1.0 &       0.900 &       0.900 \\
0158 &  1.0 &       0.970 &       0.970 &  -   &    - &         -   &         -   \\
\hline
0214 &  1.0 &      -0.625 &       -0.25 & 0219 &  1.0 &      -0.500 &        0.90 \\
0221 &  1.0 &      -0.400 &        0.80 & 0226 &  1.0 &       0.500 &       -0.90 \\
0229 &  1.0 &       0.650 &        0.25 & 0232 &  1.0 &       0.900 &        0.50 \\
\end{tabular}
\end{ruledtabular}
\caption{Equal mass spinning cases used in this work. Configurations in the bottom half
of the table are only used for validation.}
\label{tbl:equal_mass}
\end{table}

\begin{table}[h!]
\def\arraystretch{1.1}
\begin{ruledtabular}\begin{tabular}{lccc||lccc}
SXS ID &     $q$ &  $\chi_{1}$ &  $\chi_{2}$& SXS ID &     $q$ &  $\chi_{1}$ &  $\chi_{2}$\\
\hline
  0233 &  2.00 &      -0.871 &       0.850 & 
  0234 &  2.00 &      -0.850 &      -0.850 \\
  0236 &  2.00 &      -0.600 &       0.000 &
  0237 &  2.00 &      -0.600 &       0.600 \\
  0235 &  2.00 &      -0.600 &      -0.600 &
  0238 &  2.00 &      -0.500 &      -0.500 \\
  0239 &  2.00 &      -0.371 &       0.850 &
  0240 &  2.00 &      -0.300 &      -0.300 \\
  0241 &  2.00 &      -0.300 &       0.000 &
  0242 &  2.00 &      -0.300 &       0.300 \\
  0243 &  2.00 &      -0.129 &      -0.850 &
  0247 &  2.00 &       0.000 &       0.600 \\
  0246 &  2.00 &       0.000 &       0.300 &
  0245 &  2.00 &       0.000 &      -0.300 \\
  0244 &  2.00 &       0.000 &      -0.600 &
  0248 &  2.00 &       0.129 &       0.850 \\
  0251 &  2.00 &       0.300 &       0.300 &
  0250 &  2.00 &       0.300 &       0.000 \\
  0249 &  2.00 &       0.300 &      -0.300 &
  0252 &  2.00 &       0.371 &      -0.850 \\
  0253 &  2.00 &       0.500 &       0.500 &
  0256 &  2.00 &       0.600 &       0.600 \\
  0255 &  2.00 &       0.600 &       0.000 &
  0254 &  2.00 &       0.600 &      -0.600 \\
  0257 &  2.00 &       0.850 &       0.850 &
  0258 &  2.00 &       0.871 &      -0.850 \\
  0260 &  3.00 &      -0.850 &      -0.850 & 
  0261 &  3.00 &      -0.731 &       0.850 \\ 
  0264 &  3.00 &      -0.600 &      -0.600 & 
  0265 &  3.00 &      -0.600 &      -0.400 \\ 
  0262 &  3.00 &      -0.600 &       0.000 & 
  0266 &  3.00 &      -0.600 &       0.400 \\ 
  0263 &  3.00 &      -0.600 &       0.600 & 
  0267 &  3.00 &      -0.500 &      -0.500 \\ 
  0105 &  3.00 &      -0.500 &       0.000 & 
  0268 &  3.00 &      -0.400 &      -0.600 \\
  0269 &  3.00 &      -0.400 &       0.600 & 
  0270 &  3.00 &      -0.300 &      -0.300 \\ 
  0271 &  3.00 &      -0.300 &       0.000 & 
  0272 &  3.00 &      -0.300 &       0.300 \\ 
  0273 &  3.00 &      -0.269 &      -0.850 & 
  0274 &  3.00 &      -0.231 &       0.850 \\
  0278 &  3.00 &       0.000 &       0.600 & 
  0275 &  3.00 &       0.000 &      -0.600 \\
  0277 &  3.00 &       0.000 &       0.300 &
  0276 &  3.00 &       0.000 &      -0.300 \\
  0279 &  3.00 &       0.231 &      -0.850 & 
  0280 &  3.00 &       0.269 &       0.850 \\ 
  0281 &  3.00 &       0.300 &      -0.300 & 
  0282 &  3.00 &       0.300 &       0.000 \\ 
  0283 &  3.00 &       0.300 &       0.300 & 
  0284 &  3.00 &       0.400 &      -0.600 \\
  0285 &  3.00 &       0.400 &       0.600 & 
  0174 &  3.00 &       0.500 &       0.000 \\
  0286 &  3.00 &       0.500 &       0.500 & 
  0291 &  3.00 &       0.600 &       0.600 \\ 
  0288 &  3.00 &       0.600 &      -0.400 & 
  0287 &  3.00 &       0.600 &      -0.600 \\ 
  0289 &  3.00 &       0.600 &       0.000 & 
  0290 &  3.00 &       0.600 &       0.400 \\ 
  0292 &  3.00 &       0.731 &      -0.850 & 
  0293 &  3.00 &       0.850 &       0.850 \\ 
  0109 &  5.00 &      -0.500 &       0.000 &
  0110 &  5.00 &       0.500 &       0.000 \\
  0207 &  7.00 &      -0.600 &       0.000 &
  0206 &  7.00 &      -0.400 &       0.000 \\
  0204 &  7.00 &       0.400 &       0.000 &
  0202 &  7.00 &       0.600 &       0.000 \\
  0064 &  8.00 &      -0.500 &       0.000 &
  0065 &  8.00 &       0.500 &       0.000 \\
\end{tabular}\end{ruledtabular}
\caption{All unequal mass, spinning binaries used in this work. Cases with an asterisc were only used for validation.}
\label{tbl:unequal_mass}
\end{table}

\newpage
\begin{acknowledgments}

It is a pleasure to thank Sebastiano Bernuzzi, Alessandra Buonanno, 
Roberto Cotesta, Ian Hinder, Sylvain Marsat, 
Maarten van de Meent, Alessandro Nagar, Noah Sennett, Jan Steinhoff and Justin Vines
for helpful discussions. 
We thank Alessandro Nagar for providing the E-j curves based on the
EOB code presented in~\cite{Nagar:2015xqa}.  
We acknowledge the support of the SXS collaboration and emphasize the 
importance of the public catalog~\cite{SXSCatalog}.   
Our work at Cal State Fullerton was supported in part by grants NSF PHY-1307489,
PHY-1606522, and PHY-1654359. The computations were done on ORCA, which is supported
in part by Cal State Fullerton, the Research Corporation for Science Advancement, and NSF PHY-1429873.
\end{acknowledgments}

\FloatBarrier

%%______________________________________________________________

\bibliography{paper20171218.bbl}

\begin{thebibliography}{69}%
\makeatletter
\providecommand \@ifxundefined [1]{%
 \@ifx{#1\undefined}
}%
\providecommand \@ifnum [1]{%
 \ifnum #1\expandafter \@firstoftwo
 \else \expandafter \@secondoftwo
 \fi
}%
\providecommand \@ifx [1]{%
 \ifx #1\expandafter \@firstoftwo
 \else \expandafter \@secondoftwo
 \fi
}%
\providecommand \natexlab [1]{#1}%
\providecommand \enquote  [1]{``#1''}%
\providecommand \bibnamefont  [1]{#1}%
\providecommand \bibfnamefont [1]{#1}%
\providecommand \citenamefont [1]{#1}%
\providecommand \href@noop [0]{\@secondoftwo}%
\providecommand \href [0]{\begingroup \@sanitize@url \@href}%
\providecommand \@href[1]{\@@startlink{#1}\@@href}%
\providecommand \@@href[1]{\endgroup#1\@@endlink}%
\providecommand \@sanitize@url [0]{\catcode `\\12\catcode `\$12\catcode
  `\&12\catcode `\#12\catcode `\^12\catcode `\_12\catcode `\%12\relax}%
\providecommand \@@startlink[1]{}%
\providecommand \@@endlink[0]{}%
\providecommand \url  [0]{\begingroup\@sanitize@url \@url }%
\providecommand \@url [1]{\endgroup\@href {#1}{\urlprefix }}%
\providecommand \urlprefix  [0]{URL }%
\providecommand \Eprint [0]{\href }%
\providecommand \doibase [0]{http://dx.doi.org/}%
\providecommand \selectlanguage [0]{\@gobble}%
\providecommand \bibinfo  [0]{\@secondoftwo}%
\providecommand \bibfield  [0]{\@secondoftwo}%
\providecommand \translation [1]{[#1]}%
\providecommand \BibitemOpen [0]{}%
\providecommand \bibitemStop [0]{}%
\providecommand \bibitemNoStop [0]{.\EOS\space}%
\providecommand \EOS [0]{\spacefactor3000\relax}%
\providecommand \BibitemShut  [1]{\csname bibitem#1\endcsname}%
\let\auto@bib@innerbib\@empty
%</preamble>
\bibitem [{\citenamefont {Abbott}\ \emph
  {et~al.}(2016{\natexlab{a}})\citenamefont {Abbott} \emph
  {et~al.}}]{Abbott:2016blz}%
  \BibitemOpen
  \bibfield  {author} {\bibinfo {author} {\bibfnamefont {B.~P.}\ \bibnamefont
  {Abbott}} \emph {et~al.} (\bibinfo {collaboration} {Virgo, LIGO
  Scientific}),\ }\href {\doibase 10.1103/PhysRevLett.116.061102} {\bibfield
  {journal} {\bibinfo  {journal} {Phys. Rev. Lett.}\ }\textbf {\bibinfo
  {volume} {116}},\ \bibinfo {pages} {061102} (\bibinfo {year}
  {2016}{\natexlab{a}})},\ \Eprint {http://arxiv.org/abs/1602.03837}
  {arXiv:1602.03837 [gr-qc]} \BibitemShut {NoStop}%
%%CITATION = ARXIV:1602.03837;%%
\bibitem [{\citenamefont {Abbott}\ \emph
  {et~al.}(2016{\natexlab{b}})\citenamefont {Abbott} \emph
  {et~al.}}]{Abbott:2016nmj}%
  \BibitemOpen
  \bibfield  {author} {\bibinfo {author} {\bibfnamefont {B.~P.}\ \bibnamefont
  {Abbott}} \emph {et~al.} (\bibinfo {collaboration} {Virgo, LIGO
  Scientific}),\ }\href {\doibase 10.1103/PhysRevLett.116.241103} {\bibfield
  {journal} {\bibinfo  {journal} {Phys. Rev. Lett.}\ }\textbf {\bibinfo
  {volume} {116}},\ \bibinfo {pages} {241103} (\bibinfo {year}
  {2016}{\natexlab{b}})},\ \Eprint {http://arxiv.org/abs/1606.04855}
  {arXiv:1606.04855 [gr-qc]} \BibitemShut {NoStop}%
%%CITATION = ARXIV:1606.04855;%%
\bibitem [{\citenamefont {Abbott}\ \emph
  {et~al.}(2017{\natexlab{a}})\citenamefont {Abbott} \emph
  {et~al.}}]{Abbott:2017vtc}%
  \BibitemOpen
  \bibfield  {author} {\bibinfo {author} {\bibfnamefont {B.~P.}\ \bibnamefont
  {Abbott}} \emph {et~al.} (\bibinfo {collaboration} {VIRGO, LIGO
  Scientific}),\ }\href {\doibase 10.1103/PhysRevLett.118.221101} {\bibfield
  {journal} {\bibinfo  {journal} {Phys. Rev. Lett.}\ }\textbf {\bibinfo
  {volume} {118}},\ \bibinfo {pages} {221101} (\bibinfo {year}
  {2017}{\natexlab{a}})},\ \Eprint {http://arxiv.org/abs/1706.01812}
  {arXiv:1706.01812 [gr-qc]} \BibitemShut {NoStop}%
%%CITATION = ARXIV:1706.01812;%%
\bibitem [{\citenamefont {Abbott}\ \emph
  {et~al.}(2017{\natexlab{b}})\citenamefont {Abbott} \emph
  {et~al.}}]{Abbott:2017oio}%
  \BibitemOpen
  \bibfield  {author} {\bibinfo {author} {\bibfnamefont {B.~P.}\ \bibnamefont
  {Abbott}} \emph {et~al.} (\bibinfo {collaboration} {Virgo, LIGO
  Scientific}),\ }\href {\doibase 10.1103/PhysRevLett.119.141101} {\bibfield
  {journal} {\bibinfo  {journal} {Phys. Rev. Lett.}\ }\textbf {\bibinfo
  {volume} {119}},\ \bibinfo {pages} {141101} (\bibinfo {year}
  {2017}{\natexlab{b}})},\ \Eprint {http://arxiv.org/abs/1709.09660}
  {arXiv:1709.09660 [gr-qc]} \BibitemShut {NoStop}%
%%CITATION = ARXIV:1709.09660;%%
\bibitem [{\citenamefont {Abbott}\ \emph
  {et~al.}(2017{\natexlab{c}})\citenamefont {Abbott} \emph
  {et~al.}}]{Abbott:2017gyy}%
  \BibitemOpen
  \bibfield  {author} {\bibinfo {author} {\bibfnamefont {B.~P.}\ \bibnamefont
  {Abbott}} \emph {et~al.} (\bibinfo {collaboration} {Virgo, LIGO
  Scientific}),\ }\href@noop {} {\  (\bibinfo {year} {2017}{\natexlab{c}})},\
  \Eprint {http://arxiv.org/abs/1711.05578} {arXiv:1711.05578 [astro-ph.HE]}
  \BibitemShut {NoStop}%
%%CITATION = ARXIV:1711.05578;%%
\bibitem [{\citenamefont {Abbott}\ \emph
  {et~al.}(2017{\natexlab{d}})\citenamefont {Abbott} \emph
  {et~al.}}]{TheLIGOScientific:2017qsa}%
  \BibitemOpen
  \bibfield  {author} {\bibinfo {author} {\bibfnamefont {B.~P.}\ \bibnamefont
  {Abbott}} \emph {et~al.} (\bibinfo {collaboration} {Virgo, LIGO
  Scientific}),\ }\href {\doibase 10.1103/PhysRevLett.119.161101} {\bibfield
  {journal} {\bibinfo  {journal} {Phys. Rev. Lett.}\ }\textbf {\bibinfo
  {volume} {119}},\ \bibinfo {pages} {161101} (\bibinfo {year}
  {2017}{\natexlab{d}})},\ \Eprint {http://arxiv.org/abs/1710.05832}
  {arXiv:1710.05832 [gr-qc]} \BibitemShut {NoStop}%
%%CITATION = ARXIV:1710.05832;%%
\bibitem [{\citenamefont {Abbott}\ \emph
  {et~al.}(2016{\natexlab{c}})\citenamefont {Abbott} \emph
  {et~al.}}]{TheLIGOScientific:2016pea}%
  \BibitemOpen
  \bibfield  {author} {\bibinfo {author} {\bibfnamefont {B.~P.}\ \bibnamefont
  {Abbott}} \emph {et~al.} (\bibinfo {collaboration} {Virgo, LIGO
  Scientific}),\ }\href {\doibase 10.1103/PhysRevX.6.041015} {\bibfield
  {journal} {\bibinfo  {journal} {Phys. Rev.}\ }\textbf {\bibinfo {volume}
  {X6}},\ \bibinfo {pages} {041015} (\bibinfo {year} {2016}{\natexlab{c}})},\
  \Eprint {http://arxiv.org/abs/1606.04856} {arXiv:1606.04856 [gr-qc]}
  \BibitemShut {NoStop}%
%%CITATION = ARXIV:1606.04856;%%
\bibitem [{\citenamefont {Blackman}\ \emph {et~al.}(2015)\citenamefont
  {Blackman}, \citenamefont {Field}, \citenamefont {Galley}, \citenamefont
  {Szilágyi}, \citenamefont {Scheel}, \citenamefont {Tiglio},\ and\
  \citenamefont {Hemberger}}]{Blackman:2015pia}%
  \BibitemOpen
  \bibfield  {author} {\bibinfo {author} {\bibfnamefont {J.}~\bibnamefont
  {Blackman}}, \bibinfo {author} {\bibfnamefont {S.~E.}\ \bibnamefont {Field}},
  \bibinfo {author} {\bibfnamefont {C.~R.}\ \bibnamefont {Galley}}, \bibinfo
  {author} {\bibfnamefont {B.}~\bibnamefont {Szilágyi}}, \bibinfo {author}
  {\bibfnamefont {M.~A.}\ \bibnamefont {Scheel}}, \bibinfo {author}
  {\bibfnamefont {M.}~\bibnamefont {Tiglio}}, \ and\ \bibinfo {author}
  {\bibfnamefont {D.~A.}\ \bibnamefont {Hemberger}},\ }\href {\doibase
  10.1103/PhysRevLett.115.121102} {\bibfield  {journal} {\bibinfo  {journal}
  {Phys. Rev. Lett.}\ }\textbf {\bibinfo {volume} {115}},\ \bibinfo {pages}
  {121102} (\bibinfo {year} {2015})},\ \Eprint
  {http://arxiv.org/abs/1502.07758} {arXiv:1502.07758 [gr-qc]} \BibitemShut
  {NoStop}%
%%CITATION = ARXIV:1502.07758;%%
\bibitem [{\citenamefont {Blackman}\ \emph {et~al.}(2017)\citenamefont
  {Blackman}, \citenamefont {Field}, \citenamefont {Scheel}, \citenamefont
  {Galley}, \citenamefont {Ott}, \citenamefont {Boyle}, \citenamefont {Kidder},
  \citenamefont {Pfeiffer},\ and\ \citenamefont
  {Szilágyi}}]{Blackman:2017pcm}%
  \BibitemOpen
  \bibfield  {author} {\bibinfo {author} {\bibfnamefont {J.}~\bibnamefont
  {Blackman}}, \bibinfo {author} {\bibfnamefont {S.~E.}\ \bibnamefont {Field}},
  \bibinfo {author} {\bibfnamefont {M.~A.}\ \bibnamefont {Scheel}}, \bibinfo
  {author} {\bibfnamefont {C.~R.}\ \bibnamefont {Galley}}, \bibinfo {author}
  {\bibfnamefont {C.~D.}\ \bibnamefont {Ott}}, \bibinfo {author} {\bibfnamefont
  {M.}~\bibnamefont {Boyle}}, \bibinfo {author} {\bibfnamefont {L.~E.}\
  \bibnamefont {Kidder}}, \bibinfo {author} {\bibfnamefont {H.~P.}\
  \bibnamefont {Pfeiffer}}, \ and\ \bibinfo {author} {\bibfnamefont
  {B.}~\bibnamefont {Szilágyi}},\ }\href@noop {} {\  (\bibinfo {year}
  {2017})},\ \Eprint {http://arxiv.org/abs/1705.07089} {arXiv:1705.07089
  [gr-qc]} \BibitemShut {NoStop}%
%%CITATION = ARXIV:1705.07089;%%
\bibitem [{\citenamefont {Ajith}\ \emph {et~al.}(2007)\citenamefont {Ajith}
  \emph {et~al.}}]{Ajith:2007qp}%
  \BibitemOpen
  \bibfield  {author} {\bibinfo {author} {\bibfnamefont {P.}~\bibnamefont
  {Ajith}} \emph {et~al.},\ }\bibfield  {booktitle} {\emph {\bibinfo
  {booktitle} {{Gravitational wave data analysis. Proceedings: 11th Workshop,
  GWDAW-11, Potsdam, Germany, Dec 18-21, 2006}}},\ }\href {\doibase
  10.1088/0264-9381/24/19/S31} {\bibfield  {journal} {\bibinfo  {journal}
  {Class. Quant. Grav.}\ }\textbf {\bibinfo {volume} {24}},\ \bibinfo {pages}
  {S689} (\bibinfo {year} {2007})},\ \Eprint {http://arxiv.org/abs/0704.3764}
  {arXiv:0704.3764 [gr-qc]} \BibitemShut {NoStop}%
%%CITATION = ARXIV:0704.3764;%%
\bibitem [{\citenamefont {Ajith}\ \emph {et~al.}(2011)\citenamefont {Ajith}
  \emph {et~al.}}]{Ajith:2009bn}%
  \BibitemOpen
  \bibfield  {author} {\bibinfo {author} {\bibfnamefont {P.}~\bibnamefont
  {Ajith}} \emph {et~al.},\ }\href {\doibase 10.1103/PhysRevLett.106.241101}
  {\bibfield  {journal} {\bibinfo  {journal} {Phys. Rev. Lett.}\ }\textbf
  {\bibinfo {volume} {106}},\ \bibinfo {pages} {241101} (\bibinfo {year}
  {2011})},\ \Eprint {http://arxiv.org/abs/0909.2867} {arXiv:0909.2867 [gr-qc]}
  \BibitemShut {NoStop}%
%%CITATION = ARXIV:0909.2867;%%
\bibitem [{\citenamefont {Hannam}\ \emph {et~al.}(2014)\citenamefont {Hannam},
  \citenamefont {Schmidt}, \citenamefont {Bohé}, \citenamefont {Haegel},
  \citenamefont {Husa}, \citenamefont {Ohme}, \citenamefont {Pratten},\ and\
  \citenamefont {Pürrer}}]{Hannam:2013oca}%
  \BibitemOpen
  \bibfield  {author} {\bibinfo {author} {\bibfnamefont {M.}~\bibnamefont
  {Hannam}}, \bibinfo {author} {\bibfnamefont {P.}~\bibnamefont {Schmidt}},
  \bibinfo {author} {\bibfnamefont {A.}~\bibnamefont {Bohé}}, \bibinfo
  {author} {\bibfnamefont {L.}~\bibnamefont {Haegel}}, \bibinfo {author}
  {\bibfnamefont {S.}~\bibnamefont {Husa}}, \bibinfo {author} {\bibfnamefont
  {F.}~\bibnamefont {Ohme}}, \bibinfo {author} {\bibfnamefont {G.}~\bibnamefont
  {Pratten}}, \ and\ \bibinfo {author} {\bibfnamefont {M.}~\bibnamefont
  {Pürrer}},\ }\href {\doibase 10.1103/PhysRevLett.113.151101} {\bibfield
  {journal} {\bibinfo  {journal} {Phys. Rev. Lett.}\ }\textbf {\bibinfo
  {volume} {113}},\ \bibinfo {pages} {151101} (\bibinfo {year} {2014})},\
  \Eprint {http://arxiv.org/abs/1308.3271} {arXiv:1308.3271 [gr-qc]}
  \BibitemShut {NoStop}%
%%CITATION = ARXIV:1308.3271;%%
\bibitem [{\citenamefont {Blanchet}(2014)}]{Blanchet:2013haa}%
  \BibitemOpen
  \bibfield  {author} {\bibinfo {author} {\bibfnamefont {L.}~\bibnamefont
  {Blanchet}},\ }\href {\doibase 10.12942/lrr-2014-2} {\bibfield  {journal}
  {\bibinfo  {journal} {Living Rev. Rel.}\ }\textbf {\bibinfo {volume} {17}},\
  \bibinfo {pages} {2} (\bibinfo {year} {2014})},\ \Eprint
  {http://arxiv.org/abs/1310.1528} {arXiv:1310.1528 [gr-qc]} \BibitemShut
  {NoStop}%
%%CITATION = ARXIV:1310.1528;%%
\bibitem [{\citenamefont {Buonanno}\ and\ \citenamefont
  {Damour}(1999)}]{Buonanno:1998gg}%
  \BibitemOpen
  \bibfield  {author} {\bibinfo {author} {\bibfnamefont {A.}~\bibnamefont
  {Buonanno}}\ and\ \bibinfo {author} {\bibfnamefont {T.}~\bibnamefont
  {Damour}},\ }\href {\doibase 10.1103/PhysRevD.59.084006} {\bibfield
  {journal} {\bibinfo  {journal} {Phys. Rev.}\ }\textbf {\bibinfo {volume}
  {D59}},\ \bibinfo {pages} {084006} (\bibinfo {year} {1999})},\ \Eprint
  {http://arxiv.org/abs/gr-qc/9811091} {arXiv:gr-qc/9811091 [gr-qc]}
  \BibitemShut {NoStop}%
%%CITATION = GR-QC/9811091;%%
\bibitem [{\citenamefont {Damour}\ \emph {et~al.}(2000)\citenamefont {Damour},
  \citenamefont {Jaranowski},\ and\ \citenamefont {Schaefer}}]{Damour:2000we}%
  \BibitemOpen
  \bibfield  {author} {\bibinfo {author} {\bibfnamefont {T.}~\bibnamefont
  {Damour}}, \bibinfo {author} {\bibfnamefont {P.}~\bibnamefont {Jaranowski}},
  \ and\ \bibinfo {author} {\bibfnamefont {G.}~\bibnamefont {Schaefer}},\
  }\href {\doibase 10.1103/PhysRevD.62.084011} {\bibfield  {journal} {\bibinfo
  {journal} {Phys. Rev.}\ }\textbf {\bibinfo {volume} {D62}},\ \bibinfo {pages}
  {084011} (\bibinfo {year} {2000})},\ \Eprint
  {http://arxiv.org/abs/gr-qc/0005034} {arXiv:gr-qc/0005034 [gr-qc]}
  \BibitemShut {NoStop}%
%%CITATION = GR-QC/0005034;%%
\bibitem [{\citenamefont {Damour}(2001)}]{Damour:2001tu}%
  \BibitemOpen
  \bibfield  {author} {\bibinfo {author} {\bibfnamefont {T.}~\bibnamefont
  {Damour}},\ }\href {\doibase 10.1103/PhysRevD.64.124013} {\bibfield
  {journal} {\bibinfo  {journal} {Phys. Rev.}\ }\textbf {\bibinfo {volume}
  {D64}},\ \bibinfo {pages} {124013} (\bibinfo {year} {2001})},\ \Eprint
  {http://arxiv.org/abs/gr-qc/0103018} {arXiv:gr-qc/0103018 [gr-qc]}
  \BibitemShut {NoStop}%
%%CITATION = GR-QC/0103018;%%
\bibitem [{\citenamefont {Buonanno}\ \emph {et~al.}(2006)\citenamefont
  {Buonanno}, \citenamefont {Chen},\ and\ \citenamefont
  {Damour}}]{Buonanno:2005xu}%
  \BibitemOpen
  \bibfield  {author} {\bibinfo {author} {\bibfnamefont {A.}~\bibnamefont
  {Buonanno}}, \bibinfo {author} {\bibfnamefont {Y.}~\bibnamefont {Chen}}, \
  and\ \bibinfo {author} {\bibfnamefont {T.}~\bibnamefont {Damour}},\ }\href
  {\doibase 10.1103/PhysRevD.74.104005} {\bibfield  {journal} {\bibinfo
  {journal} {Phys. Rev.}\ }\textbf {\bibinfo {volume} {D74}},\ \bibinfo {pages}
  {104005} (\bibinfo {year} {2006})},\ \Eprint
  {http://arxiv.org/abs/gr-qc/0508067} {arXiv:gr-qc/0508067 [gr-qc]}
  \BibitemShut {NoStop}%
%%CITATION = GR-QC/0508067;%%
\bibitem [{\citenamefont {Damour}\ \emph {et~al.}(2008)\citenamefont {Damour},
  \citenamefont {Jaranowski},\ and\ \citenamefont {Schaefer}}]{Damour:2008qf}%
  \BibitemOpen
  \bibfield  {author} {\bibinfo {author} {\bibfnamefont {T.}~\bibnamefont
  {Damour}}, \bibinfo {author} {\bibfnamefont {P.}~\bibnamefont {Jaranowski}},
  \ and\ \bibinfo {author} {\bibfnamefont {G.}~\bibnamefont {Schaefer}},\
  }\href {\doibase 10.1103/PhysRevD.78.024009} {\bibfield  {journal} {\bibinfo
  {journal} {Phys. Rev.}\ }\textbf {\bibinfo {volume} {D78}},\ \bibinfo {pages}
  {024009} (\bibinfo {year} {2008})},\ \Eprint {http://arxiv.org/abs/0803.0915}
  {arXiv:0803.0915 [gr-qc]} \BibitemShut {NoStop}%
%%CITATION = ARXIV:0803.0915;%%
\bibitem [{\citenamefont {Damour}\ \emph {et~al.}(2009)\citenamefont {Damour},
  \citenamefont {Iyer},\ and\ \citenamefont {Nagar}}]{Damour:2008gu}%
  \BibitemOpen
  \bibfield  {author} {\bibinfo {author} {\bibfnamefont {T.}~\bibnamefont
  {Damour}}, \bibinfo {author} {\bibfnamefont {B.~R.}\ \bibnamefont {Iyer}}, \
  and\ \bibinfo {author} {\bibfnamefont {A.}~\bibnamefont {Nagar}},\ }\href
  {\doibase 10.1103/PhysRevD.79.064004} {\bibfield  {journal} {\bibinfo
  {journal} {Phys. Rev.}\ }\textbf {\bibinfo {volume} {D79}},\ \bibinfo {pages}
  {064004} (\bibinfo {year} {2009})},\ \Eprint {http://arxiv.org/abs/0811.2069}
  {arXiv:0811.2069 [gr-qc]} \BibitemShut {NoStop}%
%%CITATION = ARXIV:0811.2069;%%
\bibitem [{\citenamefont {Damour}\ and\ \citenamefont
  {Nagar}(2009)}]{Damour:2009kr}%
  \BibitemOpen
  \bibfield  {author} {\bibinfo {author} {\bibfnamefont {T.}~\bibnamefont
  {Damour}}\ and\ \bibinfo {author} {\bibfnamefont {A.}~\bibnamefont {Nagar}},\
  }\href {\doibase 10.1103/PhysRevD.79.081503} {\bibfield  {journal} {\bibinfo
  {journal} {Phys. Rev.}\ }\textbf {\bibinfo {volume} {D79}},\ \bibinfo {pages}
  {081503} (\bibinfo {year} {2009})},\ \Eprint {http://arxiv.org/abs/0902.0136}
  {arXiv:0902.0136 [gr-qc]} \BibitemShut {NoStop}%
%%CITATION = ARXIV:0902.0136;%%
\bibitem [{\citenamefont {Barausse}\ \emph {et~al.}(2009)\citenamefont
  {Barausse}, \citenamefont {Racine},\ and\ \citenamefont
  {Buonanno}}]{Barausse:2009aa}%
  \BibitemOpen
  \bibfield  {author} {\bibinfo {author} {\bibfnamefont {E.}~\bibnamefont
  {Barausse}}, \bibinfo {author} {\bibfnamefont {E.}~\bibnamefont {Racine}}, \
  and\ \bibinfo {author} {\bibfnamefont {A.}~\bibnamefont {Buonanno}},\ }\href
  {\doibase 10.1103/PhysRevD.85.069904, 10.1103/PhysRevD.80.104025} {\bibfield
  {journal} {\bibinfo  {journal} {Phys. Rev.}\ }\textbf {\bibinfo {volume}
  {D80}},\ \bibinfo {pages} {104025} (\bibinfo {year} {2009})},\ \bibinfo
  {note} {[Erratum: Phys. Rev.D85,069904(2012)]},\ \Eprint
  {http://arxiv.org/abs/0907.4745} {arXiv:0907.4745 [gr-qc]} \BibitemShut
  {NoStop}%
%%CITATION = ARXIV:0907.4745;%%
\bibitem [{\citenamefont {Barausse}\ and\ \citenamefont
  {Buonanno}(2010)}]{Barausse:2009xi}%
  \BibitemOpen
  \bibfield  {author} {\bibinfo {author} {\bibfnamefont {E.}~\bibnamefont
  {Barausse}}\ and\ \bibinfo {author} {\bibfnamefont {A.}~\bibnamefont
  {Buonanno}},\ }\href {\doibase 10.1103/PhysRevD.81.084024} {\bibfield
  {journal} {\bibinfo  {journal} {Phys. Rev.}\ }\textbf {\bibinfo {volume}
  {D81}},\ \bibinfo {pages} {084024} (\bibinfo {year} {2010})},\ \Eprint
  {http://arxiv.org/abs/0912.3517} {arXiv:0912.3517 [gr-qc]} \BibitemShut
  {NoStop}%
%%CITATION = ARXIV:0912.3517;%%
\bibitem [{\citenamefont {Pan}\ \emph {et~al.}(2011{\natexlab{a}})\citenamefont
  {Pan}, \citenamefont {Buonanno}, \citenamefont {Fujita}, \citenamefont
  {Racine},\ and\ \citenamefont {Tagoshi}}]{Pan:2010hz}%
  \BibitemOpen
  \bibfield  {author} {\bibinfo {author} {\bibfnamefont {Y.}~\bibnamefont
  {Pan}}, \bibinfo {author} {\bibfnamefont {A.}~\bibnamefont {Buonanno}},
  \bibinfo {author} {\bibfnamefont {R.}~\bibnamefont {Fujita}}, \bibinfo
  {author} {\bibfnamefont {E.}~\bibnamefont {Racine}}, \ and\ \bibinfo {author}
  {\bibfnamefont {H.}~\bibnamefont {Tagoshi}},\ }\href {\doibase
  10.1103/PhysRevD.83.064003, 10.1103/PhysRevD.87.109901} {\bibfield  {journal}
  {\bibinfo  {journal} {Phys. Rev.}\ }\textbf {\bibinfo {volume} {D83}},\
  \bibinfo {pages} {064003} (\bibinfo {year} {2011}{\natexlab{a}})},\ \bibinfo
  {note} {[Erratum: Phys. Rev.D87,no.10,109901(2013)]},\ \Eprint
  {http://arxiv.org/abs/1006.0431} {arXiv:1006.0431 [gr-qc]} \BibitemShut
  {NoStop}%
%%CITATION = ARXIV:1006.0431;%%
\bibitem [{\citenamefont {Barausse}\ \emph {et~al.}(2012)\citenamefont
  {Barausse}, \citenamefont {Buonanno},\ and\ \citenamefont
  {Le~Tiec}}]{Barausse:2011dq}%
  \BibitemOpen
  \bibfield  {author} {\bibinfo {author} {\bibfnamefont {E.}~\bibnamefont
  {Barausse}}, \bibinfo {author} {\bibfnamefont {A.}~\bibnamefont {Buonanno}},
  \ and\ \bibinfo {author} {\bibfnamefont {A.}~\bibnamefont {Le~Tiec}},\ }\href
  {\doibase 10.1103/PhysRevD.85.064010} {\bibfield  {journal} {\bibinfo
  {journal} {Phys. Rev.}\ }\textbf {\bibinfo {volume} {D85}},\ \bibinfo {pages}
  {064010} (\bibinfo {year} {2012})},\ \Eprint {http://arxiv.org/abs/1111.5610}
  {arXiv:1111.5610 [gr-qc]} \BibitemShut {NoStop}%
%%CITATION = ARXIV:1111.5610;%%
\bibitem [{\citenamefont {Damour}\ \emph {et~al.}(2013)\citenamefont {Damour},
  \citenamefont {Nagar},\ and\ \citenamefont {Bernuzzi}}]{Damour:2012ky}%
  \BibitemOpen
  \bibfield  {author} {\bibinfo {author} {\bibfnamefont {T.}~\bibnamefont
  {Damour}}, \bibinfo {author} {\bibfnamefont {A.}~\bibnamefont {Nagar}}, \
  and\ \bibinfo {author} {\bibfnamefont {S.}~\bibnamefont {Bernuzzi}},\ }\href
  {\doibase 10.1103/PhysRevD.87.084035} {\bibfield  {journal} {\bibinfo
  {journal} {Phys. Rev.}\ }\textbf {\bibinfo {volume} {D87}},\ \bibinfo {pages}
  {084035} (\bibinfo {year} {2013})},\ \Eprint {http://arxiv.org/abs/1212.4357}
  {arXiv:1212.4357 [gr-qc]} \BibitemShut {NoStop}%
%%CITATION = ARXIV:1212.4357;%%
\bibitem [{\citenamefont {Pan}\ \emph {et~al.}(2011{\natexlab{b}})\citenamefont
  {Pan}, \citenamefont {Buonanno}, \citenamefont {Boyle}, \citenamefont
  {Buchman}, \citenamefont {Kidder}, \citenamefont {Pfeiffer},\ and\
  \citenamefont {Scheel}}]{Pan:2011gk}%
  \BibitemOpen
  \bibfield  {author} {\bibinfo {author} {\bibfnamefont {Y.}~\bibnamefont
  {Pan}}, \bibinfo {author} {\bibfnamefont {A.}~\bibnamefont {Buonanno}},
  \bibinfo {author} {\bibfnamefont {M.}~\bibnamefont {Boyle}}, \bibinfo
  {author} {\bibfnamefont {L.~T.}\ \bibnamefont {Buchman}}, \bibinfo {author}
  {\bibfnamefont {L.~E.}\ \bibnamefont {Kidder}}, \bibinfo {author}
  {\bibfnamefont {H.~P.}\ \bibnamefont {Pfeiffer}}, \ and\ \bibinfo {author}
  {\bibfnamefont {M.~A.}\ \bibnamefont {Scheel}},\ }\href {\doibase
  10.1103/PhysRevD.84.124052} {\bibfield  {journal} {\bibinfo  {journal} {Phys.
  Rev.}\ }\textbf {\bibinfo {volume} {D84}},\ \bibinfo {pages} {124052}
  (\bibinfo {year} {2011}{\natexlab{b}})},\ \Eprint
  {http://arxiv.org/abs/1106.1021} {arXiv:1106.1021 [gr-qc]} \BibitemShut
  {NoStop}%
%%CITATION = ARXIV:1106.1021;%%
\bibitem [{\citenamefont {Pan}\ \emph {et~al.}(2014)\citenamefont {Pan},
  \citenamefont {Buonanno}, \citenamefont {Taracchini}, \citenamefont {Kidder},
  \citenamefont {Mroué}, \citenamefont {Pfeiffer}, \citenamefont {Scheel},\
  and\ \citenamefont {Szilágyi}}]{Pan:2013rra}%
  \BibitemOpen
  \bibfield  {author} {\bibinfo {author} {\bibfnamefont {Y.}~\bibnamefont
  {Pan}}, \bibinfo {author} {\bibfnamefont {A.}~\bibnamefont {Buonanno}},
  \bibinfo {author} {\bibfnamefont {A.}~\bibnamefont {Taracchini}}, \bibinfo
  {author} {\bibfnamefont {L.~E.}\ \bibnamefont {Kidder}}, \bibinfo {author}
  {\bibfnamefont {A.~H.}\ \bibnamefont {Mroué}}, \bibinfo {author}
  {\bibfnamefont {H.~P.}\ \bibnamefont {Pfeiffer}}, \bibinfo {author}
  {\bibfnamefont {M.~A.}\ \bibnamefont {Scheel}}, \ and\ \bibinfo {author}
  {\bibfnamefont {B.}~\bibnamefont {Szilágyi}},\ }\href {\doibase
  10.1103/PhysRevD.89.084006} {\bibfield  {journal} {\bibinfo  {journal} {Phys.
  Rev.}\ }\textbf {\bibinfo {volume} {D89}},\ \bibinfo {pages} {084006}
  (\bibinfo {year} {2014})},\ \Eprint {http://arxiv.org/abs/1307.6232}
  {arXiv:1307.6232 [gr-qc]} \BibitemShut {NoStop}%
%%CITATION = ARXIV:1307.6232;%%
\bibitem [{\citenamefont {Taracchini}\ \emph {et~al.}(2014)\citenamefont
  {Taracchini} \emph {et~al.}}]{Taracchini:2013rva}%
  \BibitemOpen
  \bibfield  {author} {\bibinfo {author} {\bibfnamefont {A.}~\bibnamefont
  {Taracchini}} \emph {et~al.},\ }\href {\doibase 10.1103/PhysRevD.89.061502}
  {\bibfield  {journal} {\bibinfo  {journal} {Phys. Rev.}\ }\textbf {\bibinfo
  {volume} {D89}},\ \bibinfo {pages} {061502} (\bibinfo {year} {2014})},\
  \Eprint {http://arxiv.org/abs/1311.2544} {arXiv:1311.2544 [gr-qc]}
  \BibitemShut {NoStop}%
%%CITATION = ARXIV:1311.2544;%%
\bibitem [{\citenamefont {Damour}\ and\ \citenamefont
  {Nagar}(2014)}]{Damour:2014sva}%
  \BibitemOpen
  \bibfield  {author} {\bibinfo {author} {\bibfnamefont {T.}~\bibnamefont
  {Damour}}\ and\ \bibinfo {author} {\bibfnamefont {A.}~\bibnamefont {Nagar}},\
  }\href {\doibase 10.1103/PhysRevD.90.044018} {\bibfield  {journal} {\bibinfo
  {journal} {Phys. Rev.}\ }\textbf {\bibinfo {volume} {D90}},\ \bibinfo {pages}
  {044018} (\bibinfo {year} {2014})},\ \Eprint {http://arxiv.org/abs/1406.6913}
  {arXiv:1406.6913 [gr-qc]} \BibitemShut {NoStop}%
%%CITATION = ARXIV:1406.6913;%%
\bibitem [{\citenamefont {Nagar}\ \emph {et~al.}(2016)\citenamefont {Nagar},
  \citenamefont {Damour}, \citenamefont {Reisswig},\ and\ \citenamefont
  {Pollney}}]{Nagar:2015xqa}%
  \BibitemOpen
  \bibfield  {author} {\bibinfo {author} {\bibfnamefont {A.}~\bibnamefont
  {Nagar}}, \bibinfo {author} {\bibfnamefont {T.}~\bibnamefont {Damour}},
  \bibinfo {author} {\bibfnamefont {C.}~\bibnamefont {Reisswig}}, \ and\
  \bibinfo {author} {\bibfnamefont {D.}~\bibnamefont {Pollney}},\ }\href
  {\doibase 10.1103/PhysRevD.93.044046} {\bibfield  {journal} {\bibinfo
  {journal} {Phys. Rev.}\ }\textbf {\bibinfo {volume} {D93}},\ \bibinfo {pages}
  {044046} (\bibinfo {year} {2016})},\ \Eprint
  {http://arxiv.org/abs/1506.08457} {arXiv:1506.08457 [gr-qc]} \BibitemShut
  {NoStop}%
%%CITATION = ARXIV:1506.08457;%%
\bibitem [{\citenamefont {Nagar}\ and\ \citenamefont
  {Shah}(2016)}]{Nagar:2016ayt}%
  \BibitemOpen
  \bibfield  {author} {\bibinfo {author} {\bibfnamefont {A.}~\bibnamefont
  {Nagar}}\ and\ \bibinfo {author} {\bibfnamefont {A.}~\bibnamefont {Shah}},\
  }\href {\doibase 10.1103/PhysRevD.94.104017} {\bibfield  {journal} {\bibinfo
  {journal} {Phys. Rev.}\ }\textbf {\bibinfo {volume} {D94}},\ \bibinfo {pages}
  {104017} (\bibinfo {year} {2016})},\ \Eprint
  {http://arxiv.org/abs/1606.00207} {arXiv:1606.00207 [gr-qc]} \BibitemShut
  {NoStop}%
%%CITATION = ARXIV:1606.00207;%%
\bibitem [{\citenamefont {Bini}\ and\ \citenamefont
  {Damour}(2016)}]{Bini:2016cje}%
  \BibitemOpen
  \bibfield  {author} {\bibinfo {author} {\bibfnamefont {D.}~\bibnamefont
  {Bini}}\ and\ \bibinfo {author} {\bibfnamefont {T.}~\bibnamefont {Damour}},\
  }\href {\doibase 10.1103/PhysRevD.93.104040} {\bibfield  {journal} {\bibinfo
  {journal} {Phys. Rev.}\ }\textbf {\bibinfo {volume} {D93}},\ \bibinfo {pages}
  {104040} (\bibinfo {year} {2016})},\ \Eprint
  {http://arxiv.org/abs/1603.09175} {arXiv:1603.09175 [gr-qc]} \BibitemShut
  {NoStop}%
%%CITATION = ARXIV:1603.09175;%%
\bibitem [{\citenamefont {Babak}\ \emph {et~al.}(2017)\citenamefont {Babak},
  \citenamefont {Taracchini},\ and\ \citenamefont {Buonanno}}]{Babak:2016tgq}%
  \BibitemOpen
  \bibfield  {author} {\bibinfo {author} {\bibfnamefont {S.}~\bibnamefont
  {Babak}}, \bibinfo {author} {\bibfnamefont {A.}~\bibnamefont {Taracchini}}, \
  and\ \bibinfo {author} {\bibfnamefont {A.}~\bibnamefont {Buonanno}},\ }\href
  {\doibase 10.1103/PhysRevD.95.024010} {\bibfield  {journal} {\bibinfo
  {journal} {Phys. Rev.}\ }\textbf {\bibinfo {volume} {D95}},\ \bibinfo {pages}
  {024010} (\bibinfo {year} {2017})},\ \Eprint
  {http://arxiv.org/abs/1607.05661} {arXiv:1607.05661 [gr-qc]} \BibitemShut
  {NoStop}%
%%CITATION = ARXIV:1607.05661;%%
\bibitem [{\citenamefont {Bohé}\ \emph {et~al.}(2017)\citenamefont {Bohé}
  \emph {et~al.}}]{Bohe:2016gbl}%
  \BibitemOpen
  \bibfield  {author} {\bibinfo {author} {\bibfnamefont {A.}~\bibnamefont
  {Bohé}} \emph {et~al.},\ }\href {\doibase 10.1103/PhysRevD.95.044028}
  {\bibfield  {journal} {\bibinfo  {journal} {Phys. Rev.}\ }\textbf {\bibinfo
  {volume} {D95}},\ \bibinfo {pages} {044028} (\bibinfo {year} {2017})},\
  \Eprint {http://arxiv.org/abs/1611.03703} {arXiv:1611.03703 [gr-qc]}
  \BibitemShut {NoStop}%
%%CITATION = ARXIV:1611.03703;%%
\bibitem [{\citenamefont {Damour}\ \emph {et~al.}(2012)\citenamefont {Damour},
  \citenamefont {Nagar}, \citenamefont {Pollney},\ and\ \citenamefont
  {Reisswig}}]{Damour:2011fu}%
  \BibitemOpen
  \bibfield  {author} {\bibinfo {author} {\bibfnamefont {T.}~\bibnamefont
  {Damour}}, \bibinfo {author} {\bibfnamefont {A.}~\bibnamefont {Nagar}},
  \bibinfo {author} {\bibfnamefont {D.}~\bibnamefont {Pollney}}, \ and\
  \bibinfo {author} {\bibfnamefont {C.}~\bibnamefont {Reisswig}},\ }\href
  {\doibase 10.1103/PhysRevLett.108.131101} {\bibfield  {journal} {\bibinfo
  {journal} {Phys. Rev. Lett.}\ }\textbf {\bibinfo {volume} {108}},\ \bibinfo
  {pages} {131101} (\bibinfo {year} {2012})},\ \Eprint
  {http://arxiv.org/abs/1110.2938} {arXiv:1110.2938 [gr-qc]} \BibitemShut
  {NoStop}%
%%CITATION = ARXIV:1110.2938;%%
\bibitem [{SpE()}]{SpECwebsite}%
  \BibitemOpen
  \href@noop {} {}\bibinfo {howpublished}
  {\url{http://www.black-holes.org/SpEC.html}}\BibitemShut {NoStop}%
\bibitem [{\citenamefont {Scheel}\ \emph {et~al.}(2006)\citenamefont {Scheel},
  \citenamefont {Pfeiffer}, \citenamefont {Lindblom}, \citenamefont {Kidder},
  \citenamefont {Rinne},\ and\ \citenamefont {Teukolsky}}]{Scheel:2006gg}%
  \BibitemOpen
  \bibfield  {author} {\bibinfo {author} {\bibfnamefont {M.~A.}\ \bibnamefont
  {Scheel}}, \bibinfo {author} {\bibfnamefont {H.~P.}\ \bibnamefont
  {Pfeiffer}}, \bibinfo {author} {\bibfnamefont {L.}~\bibnamefont {Lindblom}},
  \bibinfo {author} {\bibfnamefont {L.~E.}\ \bibnamefont {Kidder}}, \bibinfo
  {author} {\bibfnamefont {O.}~\bibnamefont {Rinne}}, \ and\ \bibinfo {author}
  {\bibfnamefont {S.~A.}\ \bibnamefont {Teukolsky}},\ }\href {\doibase
  10.1103/PhysRevD.74.104006} {\bibfield  {journal} {\bibinfo  {journal} {Phys.
  Rev.}\ }\textbf {\bibinfo {volume} {D74}},\ \bibinfo {pages} {104006}
  (\bibinfo {year} {2006})},\ \Eprint {http://arxiv.org/abs/gr-qc/0607056}
  {arXiv:gr-qc/0607056 [gr-qc]} \BibitemShut {NoStop}%
%%CITATION = GR-QC/0607056;%%
\bibitem [{\citenamefont {Szilagyi}\ \emph {et~al.}(2009)\citenamefont
  {Szilagyi}, \citenamefont {Lindblom},\ and\ \citenamefont
  {Scheel}}]{Szilagyi:2009qz}%
  \BibitemOpen
  \bibfield  {author} {\bibinfo {author} {\bibfnamefont {B.}~\bibnamefont
  {Szilagyi}}, \bibinfo {author} {\bibfnamefont {L.}~\bibnamefont {Lindblom}},
  \ and\ \bibinfo {author} {\bibfnamefont {M.~A.}\ \bibnamefont {Scheel}},\
  }\href {\doibase 10.1103/PhysRevD.80.124010} {\bibfield  {journal} {\bibinfo
  {journal} {Phys. Rev.}\ }\textbf {\bibinfo {volume} {D80}},\ \bibinfo {pages}
  {124010} (\bibinfo {year} {2009})},\ \Eprint {http://arxiv.org/abs/0909.3557}
  {arXiv:0909.3557 [gr-qc]} \BibitemShut {NoStop}%
%%CITATION = ARXIV:0909.3557;%%
\bibitem [{\citenamefont {Buchman}\ \emph {et~al.}(2012)\citenamefont
  {Buchman}, \citenamefont {Pfeiffer}, \citenamefont {Scheel},\ and\
  \citenamefont {Szilagyi}}]{Buchman:2012dw}%
  \BibitemOpen
  \bibfield  {author} {\bibinfo {author} {\bibfnamefont {L.~T.}\ \bibnamefont
  {Buchman}}, \bibinfo {author} {\bibfnamefont {H.~P.}\ \bibnamefont
  {Pfeiffer}}, \bibinfo {author} {\bibfnamefont {M.~A.}\ \bibnamefont
  {Scheel}}, \ and\ \bibinfo {author} {\bibfnamefont {B.}~\bibnamefont
  {Szilagyi}},\ }\href {\doibase 10.1103/PhysRevD.86.084033} {\bibfield
  {journal} {\bibinfo  {journal} {Phys. Rev.}\ }\textbf {\bibinfo {volume}
  {D86}},\ \bibinfo {pages} {084033} (\bibinfo {year} {2012})},\ \Eprint
  {http://arxiv.org/abs/1206.3015} {arXiv:1206.3015 [gr-qc]} \BibitemShut
  {NoStop}%
%%CITATION = ARXIV:1206.3015;%%
\bibitem [{SXS()}]{SXSCatalog}%
  \BibitemOpen
  \href@noop {} {}\bibinfo {howpublished}
  {\url{http://www.black-holes.org/waveforms}}\BibitemShut {NoStop}%
\bibitem [{\citenamefont {Mroue}\ \emph {et~al.}(2013)\citenamefont {Mroue}
  \emph {et~al.}}]{Mroue:2013xna}%
  \BibitemOpen
  \bibfield  {author} {\bibinfo {author} {\bibfnamefont {A.~H.}\ \bibnamefont
  {Mroue}} \emph {et~al.},\ }\href {\doibase 10.1103/PhysRevLett.111.241104}
  {\bibfield  {journal} {\bibinfo  {journal} {Phys. Rev. Lett.}\ }\textbf
  {\bibinfo {volume} {111}},\ \bibinfo {pages} {241104} (\bibinfo {year}
  {2013})},\ \Eprint {http://arxiv.org/abs/1304.6077} {arXiv:1304.6077 [gr-qc]}
  \BibitemShut {NoStop}%
%%CITATION = ARXIV:1304.6077;%%
\bibitem [{\citenamefont {Lovelace}\ \emph {et~al.}(2008)\citenamefont
  {Lovelace}, \citenamefont {Owen}, \citenamefont {Pfeiffer},\ and\
  \citenamefont {Chu}}]{Lovelace:2008tw}%
  \BibitemOpen
  \bibfield  {author} {\bibinfo {author} {\bibfnamefont {G.}~\bibnamefont
  {Lovelace}}, \bibinfo {author} {\bibfnamefont {R.}~\bibnamefont {Owen}},
  \bibinfo {author} {\bibfnamefont {H.~P.}\ \bibnamefont {Pfeiffer}}, \ and\
  \bibinfo {author} {\bibfnamefont {T.}~\bibnamefont {Chu}},\ }\href {\doibase
  10.1103/PhysRevD.78.084017} {\bibfield  {journal} {\bibinfo  {journal} {Phys.
  Rev.}\ }\textbf {\bibinfo {volume} {D78}},\ \bibinfo {pages} {084017}
  (\bibinfo {year} {2008})},\ \Eprint {http://arxiv.org/abs/0805.4192}
  {arXiv:0805.4192 [gr-qc]} \BibitemShut {NoStop}%
%%CITATION = ARXIV:0805.4192;%%
\bibitem [{\citenamefont {Pfeiffer}\ and\ \citenamefont
  {York}(2003)}]{Pfeiffer:2002iy}%
  \BibitemOpen
  \bibfield  {author} {\bibinfo {author} {\bibfnamefont {H.~P.}\ \bibnamefont
  {Pfeiffer}}\ and\ \bibinfo {author} {\bibfnamefont {J.~W.}\ \bibnamefont
  {York}, \bibfnamefont {Jr.}},\ }\href {\doibase 10.1103/PhysRevD.67.044022}
  {\bibfield  {journal} {\bibinfo  {journal} {Phys. Rev.}\ }\textbf {\bibinfo
  {volume} {D67}},\ \bibinfo {pages} {044022} (\bibinfo {year} {2003})},\
  \Eprint {http://arxiv.org/abs/gr-qc/0207095} {arXiv:gr-qc/0207095 [gr-qc]}
  \BibitemShut {NoStop}%
%%CITATION = GR-QC/0207095;%%
\bibitem [{\citenamefont {Pfeiffer}\ \emph {et~al.}(2003)\citenamefont
  {Pfeiffer}, \citenamefont {Kidder}, \citenamefont {Scheel},\ and\
  \citenamefont {Teukolsky}}]{Pfeiffer:2002wt}%
  \BibitemOpen
  \bibfield  {author} {\bibinfo {author} {\bibfnamefont {H.~P.}\ \bibnamefont
  {Pfeiffer}}, \bibinfo {author} {\bibfnamefont {L.~E.}\ \bibnamefont
  {Kidder}}, \bibinfo {author} {\bibfnamefont {M.~A.}\ \bibnamefont {Scheel}},
  \ and\ \bibinfo {author} {\bibfnamefont {S.~A.}\ \bibnamefont {Teukolsky}},\
  }\href {\doibase 10.1016/S0010-4655(02)00847-0} {\bibfield  {journal}
  {\bibinfo  {journal} {Comput. Phys. Commun.}\ }\textbf {\bibinfo {volume}
  {152}},\ \bibinfo {pages} {253} (\bibinfo {year} {2003})},\ \Eprint
  {http://arxiv.org/abs/gr-qc/0202096} {arXiv:gr-qc/0202096 [gr-qc]}
  \BibitemShut {NoStop}%
%%CITATION = GR-QC/0202096;%%
\bibitem [{\citenamefont {Lindblom}\ \emph {et~al.}(2006)\citenamefont
  {Lindblom}, \citenamefont {Scheel}, \citenamefont {Kidder}, \citenamefont
  {Owen},\ and\ \citenamefont {Rinne}}]{Lindblom:2005qh}%
  \BibitemOpen
  \bibfield  {author} {\bibinfo {author} {\bibfnamefont {L.}~\bibnamefont
  {Lindblom}}, \bibinfo {author} {\bibfnamefont {M.~A.}\ \bibnamefont
  {Scheel}}, \bibinfo {author} {\bibfnamefont {L.~E.}\ \bibnamefont {Kidder}},
  \bibinfo {author} {\bibfnamefont {R.}~\bibnamefont {Owen}}, \ and\ \bibinfo
  {author} {\bibfnamefont {O.}~\bibnamefont {Rinne}},\ }\href {\doibase
  10.1088/0264-9381/23/16/S09} {\bibfield  {journal} {\bibinfo  {journal}
  {Class. Quant. Grav.}\ }\textbf {\bibinfo {volume} {23}},\ \bibinfo {pages}
  {S447} (\bibinfo {year} {2006})},\ \Eprint
  {http://arxiv.org/abs/gr-qc/0512093} {arXiv:gr-qc/0512093 [gr-qc]}
  \BibitemShut {NoStop}%
%%CITATION = GR-QC/0512093;%%
\bibitem [{\citenamefont {Pretorius}(2005)}]{Pretorius:2005gq}%
  \BibitemOpen
  \bibfield  {author} {\bibinfo {author} {\bibfnamefont {F.}~\bibnamefont
  {Pretorius}},\ }\href {\doibase 10.1103/PhysRevLett.95.121101} {\bibfield
  {journal} {\bibinfo  {journal} {Phys. Rev. Lett.}\ }\textbf {\bibinfo
  {volume} {95}},\ \bibinfo {pages} {121101} (\bibinfo {year} {2005})},\
  \Eprint {http://arxiv.org/abs/gr-qc/0507014} {arXiv:gr-qc/0507014 [gr-qc]}
  \BibitemShut {NoStop}%
%%CITATION = GR-QC/0507014;%%
\bibitem [{\citenamefont {Friedrich}(1985)}]{Friedrich1985}%
  \BibitemOpen
  \bibfield  {author} {\bibinfo {author} {\bibfnamefont {H.}~\bibnamefont
  {Friedrich}},\ }\href {\doibase 10.1007/BF01217728} {\bibfield  {journal}
  {\bibinfo  {journal} {Comm. Math. Phys.}\ }\textbf {\bibinfo {volume}
  {100}},\ \bibinfo {pages} {525} (\bibinfo {year} {1985})}\BibitemShut
  {NoStop}%
\bibitem [{\citenamefont {Lindblom}\ and\ \citenamefont
  {Szilagyi}(2009)}]{Lindblom:2009tu}%
  \BibitemOpen
  \bibfield  {author} {\bibinfo {author} {\bibfnamefont {L.}~\bibnamefont
  {Lindblom}}\ and\ \bibinfo {author} {\bibfnamefont {B.}~\bibnamefont
  {Szilagyi}},\ }\href {\doibase 10.1103/PhysRevD.80.084019} {\bibfield
  {journal} {\bibinfo  {journal} {Phys. Rev.}\ }\textbf {\bibinfo {volume}
  {D80}},\ \bibinfo {pages} {084019} (\bibinfo {year} {2009})},\ \Eprint
  {http://arxiv.org/abs/0904.4873} {arXiv:0904.4873 [gr-qc]} \BibitemShut
  {NoStop}%
%%CITATION = ARXIV:0904.4873;%%
\bibitem [{\citenamefont {Hemberger}\ \emph {et~al.}(2013)\citenamefont
  {Hemberger}, \citenamefont {Scheel}, \citenamefont {Kidder}, \citenamefont
  {Szilágyi}, \citenamefont {Lovelace}, \citenamefont {Taylor},\ and\
  \citenamefont {Teukolsky}}]{Hemberger:2012jz}%
  \BibitemOpen
  \bibfield  {author} {\bibinfo {author} {\bibfnamefont {D.~A.}\ \bibnamefont
  {Hemberger}}, \bibinfo {author} {\bibfnamefont {M.~A.}\ \bibnamefont
  {Scheel}}, \bibinfo {author} {\bibfnamefont {L.~E.}\ \bibnamefont {Kidder}},
  \bibinfo {author} {\bibfnamefont {B.}~\bibnamefont {Szilágyi}}, \bibinfo
  {author} {\bibfnamefont {G.}~\bibnamefont {Lovelace}}, \bibinfo {author}
  {\bibfnamefont {N.~W.}\ \bibnamefont {Taylor}}, \ and\ \bibinfo {author}
  {\bibfnamefont {S.~A.}\ \bibnamefont {Teukolsky}},\ }\href {\doibase
  10.1088/0264-9381/30/11/115001} {\bibfield  {journal} {\bibinfo  {journal}
  {Class. Quant. Grav.}\ }\textbf {\bibinfo {volume} {30}},\ \bibinfo {pages}
  {115001} (\bibinfo {year} {2013})},\ \Eprint {http://arxiv.org/abs/1211.6079}
  {arXiv:1211.6079 [gr-qc]} \BibitemShut {NoStop}%
%%CITATION = ARXIV:1211.6079;%%
\bibitem [{\citenamefont {Scheel}\ \emph {et~al.}(2015)\citenamefont {Scheel},
  \citenamefont {Giesler}, \citenamefont {Hemberger}, \citenamefont {Lovelace},
  \citenamefont {Kuper}, \citenamefont {Boyle}, \citenamefont {Szilágyi},\
  and\ \citenamefont {Kidder}}]{Scheel:2014ina}%
  \BibitemOpen
  \bibfield  {author} {\bibinfo {author} {\bibfnamefont {M.~A.}\ \bibnamefont
  {Scheel}}, \bibinfo {author} {\bibfnamefont {M.}~\bibnamefont {Giesler}},
  \bibinfo {author} {\bibfnamefont {D.~A.}\ \bibnamefont {Hemberger}}, \bibinfo
  {author} {\bibfnamefont {G.}~\bibnamefont {Lovelace}}, \bibinfo {author}
  {\bibfnamefont {K.}~\bibnamefont {Kuper}}, \bibinfo {author} {\bibfnamefont
  {M.}~\bibnamefont {Boyle}}, \bibinfo {author} {\bibfnamefont
  {B.}~\bibnamefont {Szilágyi}}, \ and\ \bibinfo {author} {\bibfnamefont
  {L.~E.}\ \bibnamefont {Kidder}},\ }\href {\doibase
  10.1088/0264-9381/32/10/105009} {\bibfield  {journal} {\bibinfo  {journal}
  {Class. Quant. Grav.}\ }\textbf {\bibinfo {volume} {32}},\ \bibinfo {pages}
  {105009} (\bibinfo {year} {2015})},\ \Eprint {http://arxiv.org/abs/1412.1803}
  {arXiv:1412.1803 [gr-qc]} \BibitemShut {NoStop}%
%%CITATION = ARXIV:1412.1803;%%
\bibitem [{\citenamefont {Szilágyi}(2014)}]{Szilagyi:2014fna}%
  \BibitemOpen
  \bibfield  {author} {\bibinfo {author} {\bibfnamefont {B.}~\bibnamefont
  {Szilágyi}},\ }\href {\doibase 10.1142/S0218271814300146} {\bibfield
  {journal} {\bibinfo  {journal} {Int. J. Mod. Phys.}\ }\textbf {\bibinfo
  {volume} {D23}},\ \bibinfo {pages} {1430014} (\bibinfo {year} {2014})},\
  \Eprint {http://arxiv.org/abs/1405.3693} {arXiv:1405.3693 [gr-qc]}
  \BibitemShut {NoStop}%
%%CITATION = ARXIV:1405.3693;%%
\bibitem [{\citenamefont {Szilágyi}\ \emph {et~al.}(2015)\citenamefont
  {Szilágyi}, \citenamefont {Blackman}, \citenamefont {Buonanno},
  \citenamefont {Taracchini}, \citenamefont {Pfeiffer}, \citenamefont {Scheel},
  \citenamefont {Chu}, \citenamefont {Kidder},\ and\ \citenamefont
  {Pan}}]{Szilagyi:2015rwa}%
  \BibitemOpen
  \bibfield  {author} {\bibinfo {author} {\bibfnamefont {B.}~\bibnamefont
  {Szilágyi}}, \bibinfo {author} {\bibfnamefont {J.}~\bibnamefont {Blackman}},
  \bibinfo {author} {\bibfnamefont {A.}~\bibnamefont {Buonanno}}, \bibinfo
  {author} {\bibfnamefont {A.}~\bibnamefont {Taracchini}}, \bibinfo {author}
  {\bibfnamefont {H.~P.}\ \bibnamefont {Pfeiffer}}, \bibinfo {author}
  {\bibfnamefont {M.~A.}\ \bibnamefont {Scheel}}, \bibinfo {author}
  {\bibfnamefont {T.}~\bibnamefont {Chu}}, \bibinfo {author} {\bibfnamefont
  {L.~E.}\ \bibnamefont {Kidder}}, \ and\ \bibinfo {author} {\bibfnamefont
  {Y.}~\bibnamefont {Pan}},\ }\href {\doibase 10.1103/PhysRevLett.115.031102}
  {\bibfield  {journal} {\bibinfo  {journal} {Phys. Rev. Lett.}\ }\textbf
  {\bibinfo {volume} {115}},\ \bibinfo {pages} {031102} (\bibinfo {year}
  {2015})},\ \Eprint {http://arxiv.org/abs/1502.04953} {arXiv:1502.04953
  [gr-qc]} \BibitemShut {NoStop}%
%%CITATION = ARXIV:1502.04953;%%
\bibitem [{\citenamefont {Barkett}\ \emph {et~al.}(2016)\citenamefont {Barkett}
  \emph {et~al.}}]{Barkett:2015wia}%
  \BibitemOpen
  \bibfield  {author} {\bibinfo {author} {\bibfnamefont {K.}~\bibnamefont
  {Barkett}} \emph {et~al.},\ }\href {\doibase 10.1103/PhysRevD.93.044064}
  {\bibfield  {journal} {\bibinfo  {journal} {Phys. Rev.}\ }\textbf {\bibinfo
  {volume} {D93}},\ \bibinfo {pages} {044064} (\bibinfo {year} {2016})},\
  \Eprint {http://arxiv.org/abs/1509.05782} {arXiv:1509.05782 [gr-qc]}
  \BibitemShut {NoStop}%
%%CITATION = ARXIV:1509.05782;%%
\bibitem [{\citenamefont {Lovelace}\ \emph {et~al.}(2015)\citenamefont
  {Lovelace} \emph {et~al.}}]{Lovelace:2014twa}%
  \BibitemOpen
  \bibfield  {author} {\bibinfo {author} {\bibfnamefont {G.}~\bibnamefont
  {Lovelace}} \emph {et~al.},\ }\href {\doibase 10.1088/0264-9381/32/6/065007}
  {\bibfield  {journal} {\bibinfo  {journal} {Class. Quant. Grav.}\ }\textbf
  {\bibinfo {volume} {32}},\ \bibinfo {pages} {065007} (\bibinfo {year}
  {2015})},\ \Eprint {http://arxiv.org/abs/1411.7297} {arXiv:1411.7297 [gr-qc]}
  \BibitemShut {NoStop}%
%%CITATION = ARXIV:1411.7297;%%
\bibitem [{\citenamefont {Ruiz}\ \emph {et~al.}(2008)\citenamefont {Ruiz},
  \citenamefont {Takahashi}, \citenamefont {Alcubierre},\ and\ \citenamefont
  {Nunez}}]{Ruiz:2007yx}%
  \BibitemOpen
  \bibfield  {author} {\bibinfo {author} {\bibfnamefont {M.}~\bibnamefont
  {Ruiz}}, \bibinfo {author} {\bibfnamefont {R.}~\bibnamefont {Takahashi}},
  \bibinfo {author} {\bibfnamefont {M.}~\bibnamefont {Alcubierre}}, \ and\
  \bibinfo {author} {\bibfnamefont {D.}~\bibnamefont {Nunez}},\ }\href
  {\doibase 10.1007/s10714-007-0570-8, 10.1007/s10714-008-0684-7} {\bibfield
  {journal} {\bibinfo  {journal} {Gen. Rel. Grav.}\ }\textbf {\bibinfo {volume}
  {40}},\ \bibinfo {pages} {2467} (\bibinfo {year} {2008})},\ \Eprint
  {http://arxiv.org/abs/0707.4654} {arXiv:0707.4654 [gr-qc]} \BibitemShut
  {NoStop}%
%%CITATION = ARXIV:0707.4654;%%
\bibitem [{\citenamefont {Bernuzzi}\ \emph {et~al.}(2014)\citenamefont
  {Bernuzzi}, \citenamefont {Dietrich}, \citenamefont {Tichy},\ and\
  \citenamefont {Brügmann}}]{Bernuzzi:2013rza}%
  \BibitemOpen
  \bibfield  {author} {\bibinfo {author} {\bibfnamefont {S.}~\bibnamefont
  {Bernuzzi}}, \bibinfo {author} {\bibfnamefont {T.}~\bibnamefont {Dietrich}},
  \bibinfo {author} {\bibfnamefont {W.}~\bibnamefont {Tichy}}, \ and\ \bibinfo
  {author} {\bibfnamefont {B.}~\bibnamefont {Brügmann}},\ }\href {\doibase
  10.1103/PhysRevD.89.104021} {\bibfield  {journal} {\bibinfo  {journal} {Phys.
  Rev.}\ }\textbf {\bibinfo {volume} {D89}},\ \bibinfo {pages} {104021}
  (\bibinfo {year} {2014})},\ \Eprint {http://arxiv.org/abs/1311.4443}
  {arXiv:1311.4443 [gr-qc]} \BibitemShut {NoStop}%
%%CITATION = ARXIV:1311.4443;%%
\bibitem [{\citenamefont {Dietrich}\ \emph
  {et~al.}(2017{\natexlab{a}})\citenamefont {Dietrich}, \citenamefont
  {Bernuzzi}, \citenamefont {Ujevic},\ and\ \citenamefont
  {Tichy}}]{Dietrich:2016lyp}%
  \BibitemOpen
  \bibfield  {author} {\bibinfo {author} {\bibfnamefont {T.}~\bibnamefont
  {Dietrich}}, \bibinfo {author} {\bibfnamefont {S.}~\bibnamefont {Bernuzzi}},
  \bibinfo {author} {\bibfnamefont {M.}~\bibnamefont {Ujevic}}, \ and\ \bibinfo
  {author} {\bibfnamefont {W.}~\bibnamefont {Tichy}},\ }\href {\doibase
  10.1103/PhysRevD.95.044045} {\bibfield  {journal} {\bibinfo  {journal} {Phys.
  Rev.}\ }\textbf {\bibinfo {volume} {D95}},\ \bibinfo {pages} {044045}
  (\bibinfo {year} {2017}{\natexlab{a}})},\ \Eprint
  {http://arxiv.org/abs/1611.07367} {arXiv:1611.07367 [gr-qc]} \BibitemShut
  {NoStop}%
%%CITATION = ARXIV:1611.07367;%%
\bibitem [{\citenamefont {Dietrich}\ \emph
  {et~al.}(2017{\natexlab{b}})\citenamefont {Dietrich}, \citenamefont
  {Bernuzzi}, \citenamefont {Bruegmann}, \citenamefont {Ujevic},\ and\
  \citenamefont {Tichy}}]{Dietrich:2017xqb}%
  \BibitemOpen
  \bibfield  {author} {\bibinfo {author} {\bibfnamefont {T.}~\bibnamefont
  {Dietrich}}, \bibinfo {author} {\bibfnamefont {S.}~\bibnamefont {Bernuzzi}},
  \bibinfo {author} {\bibfnamefont {B.}~\bibnamefont {Bruegmann}}, \bibinfo
  {author} {\bibfnamefont {M.}~\bibnamefont {Ujevic}}, \ and\ \bibinfo {author}
  {\bibfnamefont {W.}~\bibnamefont {Tichy}},\ }\href@noop {} {\  (\bibinfo
  {year} {2017}{\natexlab{b}})},\ \Eprint {http://arxiv.org/abs/1712.02992}
  {arXiv:1712.02992 [gr-qc]} \BibitemShut {NoStop}%
%%CITATION = ARXIV:1712.02992;%%
\bibitem [{\citenamefont {Levi}\ and\ \citenamefont
  {Steinhoff}(2016)}]{Levi:2016ofk}%
  \BibitemOpen
  \bibfield  {author} {\bibinfo {author} {\bibfnamefont {M.}~\bibnamefont
  {Levi}}\ and\ \bibinfo {author} {\bibfnamefont {J.}~\bibnamefont
  {Steinhoff}},\ }\href@noop {} {\  (\bibinfo {year} {2016})},\ \Eprint
  {http://arxiv.org/abs/1607.04252} {arXiv:1607.04252 [gr-qc]} \BibitemShut
  {NoStop}%
%%CITATION = ARXIV:1607.04252;%%
\bibitem [{\citenamefont {Lovelace}\ \emph {et~al.}(2016)\citenamefont
  {Lovelace} \emph {et~al.}}]{Lovelace:2016uwp}%
  \BibitemOpen
  \bibfield  {author} {\bibinfo {author} {\bibfnamefont {G.}~\bibnamefont
  {Lovelace}} \emph {et~al.},\ }\href {\doibase 10.1088/0264-9381/33/24/244002}
  {\bibfield  {journal} {\bibinfo  {journal} {Class. Quant. Grav.}\ }\textbf
  {\bibinfo {volume} {33}},\ \bibinfo {pages} {244002} (\bibinfo {year}
  {2016})},\ \Eprint {http://arxiv.org/abs/1607.05377} {arXiv:1607.05377
  [gr-qc]} \BibitemShut {NoStop}%
%%CITATION = ARXIV:1607.05377;%%
\bibitem [{\citenamefont {Kumar}\ \emph {et~al.}(2015)\citenamefont {Kumar},
  \citenamefont {Barkett}, \citenamefont {Bhagwat}, \citenamefont {Afshari},
  \citenamefont {Brown}, \citenamefont {Lovelace}, \citenamefont {Scheel},\
  and\ \citenamefont {Szilágyi}}]{Kumar:2015tha}%
  \BibitemOpen
  \bibfield  {author} {\bibinfo {author} {\bibfnamefont {P.}~\bibnamefont
  {Kumar}}, \bibinfo {author} {\bibfnamefont {K.}~\bibnamefont {Barkett}},
  \bibinfo {author} {\bibfnamefont {S.}~\bibnamefont {Bhagwat}}, \bibinfo
  {author} {\bibfnamefont {N.}~\bibnamefont {Afshari}}, \bibinfo {author}
  {\bibfnamefont {D.~A.}\ \bibnamefont {Brown}}, \bibinfo {author}
  {\bibfnamefont {G.}~\bibnamefont {Lovelace}}, \bibinfo {author}
  {\bibfnamefont {M.~A.}\ \bibnamefont {Scheel}}, \ and\ \bibinfo {author}
  {\bibfnamefont {B.}~\bibnamefont {Szilágyi}},\ }\href {\doibase
  10.1103/PhysRevD.92.102001} {\bibfield  {journal} {\bibinfo  {journal} {Phys.
  Rev.}\ }\textbf {\bibinfo {volume} {D92}},\ \bibinfo {pages} {102001}
  (\bibinfo {year} {2015})},\ \Eprint {http://arxiv.org/abs/1507.00103}
  {arXiv:1507.00103 [gr-qc]} \BibitemShut {NoStop}%
%%CITATION = ARXIV:1507.00103;%%
\bibitem [{\citenamefont {Boyle}(2013)}]{Boyle:2013nka}%
  \BibitemOpen
  \bibfield  {author} {\bibinfo {author} {\bibfnamefont {M.}~\bibnamefont
  {Boyle}},\ }\href {\doibase 10.1103/PhysRevD.87.104006} {\bibfield  {journal}
  {\bibinfo  {journal} {Phys. Rev.}\ }\textbf {\bibinfo {volume} {D87}},\
  \bibinfo {pages} {104006} (\bibinfo {year} {2013})},\ \Eprint
  {http://arxiv.org/abs/1302.2919} {arXiv:1302.2919 [gr-qc]} \BibitemShut
  {NoStop}%
%%CITATION = ARXIV:1302.2919;%%
\bibitem [{\citenamefont {Boyle}\ and\ \citenamefont
  {Mroue}(2009)}]{Boyle:2009vi}%
  \BibitemOpen
  \bibfield  {author} {\bibinfo {author} {\bibfnamefont {M.}~\bibnamefont
  {Boyle}}\ and\ \bibinfo {author} {\bibfnamefont {A.~H.}\ \bibnamefont
  {Mroue}},\ }\href {\doibase 10.1103/PhysRevD.80.124045} {\bibfield  {journal}
  {\bibinfo  {journal} {Phys. Rev.}\ }\textbf {\bibinfo {volume} {D80}},\
  \bibinfo {pages} {124045} (\bibinfo {year} {2009})},\ \Eprint
  {http://arxiv.org/abs/0905.3177} {arXiv:0905.3177 [gr-qc]} \BibitemShut
  {NoStop}%
%%CITATION = ARXIV:0905.3177;%%
\bibitem [{\citenamefont {Chu}\ \emph {et~al.}(2016)\citenamefont {Chu},
  \citenamefont {Fong}, \citenamefont {Kumar}, \citenamefont {Pfeiffer},
  \citenamefont {Boyle}, \citenamefont {Hemberger}, \citenamefont {Kidder},
  \citenamefont {Scheel},\ and\ \citenamefont {Szilagyi}}]{Chu:2015kft}%
  \BibitemOpen
  \bibfield  {author} {\bibinfo {author} {\bibfnamefont {T.}~\bibnamefont
  {Chu}}, \bibinfo {author} {\bibfnamefont {H.}~\bibnamefont {Fong}}, \bibinfo
  {author} {\bibfnamefont {P.}~\bibnamefont {Kumar}}, \bibinfo {author}
  {\bibfnamefont {H.~P.}\ \bibnamefont {Pfeiffer}}, \bibinfo {author}
  {\bibfnamefont {M.}~\bibnamefont {Boyle}}, \bibinfo {author} {\bibfnamefont
  {D.~A.}\ \bibnamefont {Hemberger}}, \bibinfo {author} {\bibfnamefont {L.~E.}\
  \bibnamefont {Kidder}}, \bibinfo {author} {\bibfnamefont {M.~A.}\
  \bibnamefont {Scheel}}, \ and\ \bibinfo {author} {\bibfnamefont
  {B.}~\bibnamefont {Szilagyi}},\ }\href {\doibase
  10.1088/0264-9381/33/16/165001} {\bibfield  {journal} {\bibinfo  {journal}
  {Class. Quant. Grav.}\ }\textbf {\bibinfo {volume} {33}},\ \bibinfo {pages}
  {165001} (\bibinfo {year} {2016})},\ \Eprint
  {http://arxiv.org/abs/1512.06800} {arXiv:1512.06800 [gr-qc]} \BibitemShut
  {NoStop}%
%%CITATION = ARXIV:1512.06800;%%
\bibitem [{\citenamefont {Taylor}\ \emph {et~al.}(2013)\citenamefont {Taylor},
  \citenamefont {Boyle}, \citenamefont {Reisswig}, \citenamefont {Scheel},
  \citenamefont {Chu}, \citenamefont {Kidder},\ and\ \citenamefont
  {Szilágyi}}]{Taylor:2013zia}%
  \BibitemOpen
  \bibfield  {author} {\bibinfo {author} {\bibfnamefont {N.~W.}\ \bibnamefont
  {Taylor}}, \bibinfo {author} {\bibfnamefont {M.}~\bibnamefont {Boyle}},
  \bibinfo {author} {\bibfnamefont {C.}~\bibnamefont {Reisswig}}, \bibinfo
  {author} {\bibfnamefont {M.~A.}\ \bibnamefont {Scheel}}, \bibinfo {author}
  {\bibfnamefont {T.}~\bibnamefont {Chu}}, \bibinfo {author} {\bibfnamefont
  {L.~E.}\ \bibnamefont {Kidder}}, \ and\ \bibinfo {author} {\bibfnamefont
  {B.}~\bibnamefont {Szilágyi}},\ }\href {\doibase 10.1103/PhysRevD.88.124010}
  {\bibfield  {journal} {\bibinfo  {journal} {Phys. Rev.}\ }\textbf {\bibinfo
  {volume} {D88}},\ \bibinfo {pages} {124010} (\bibinfo {year} {2013})},\
  \Eprint {http://arxiv.org/abs/1309.3605} {arXiv:1309.3605 [gr-qc]}
  \BibitemShut {NoStop}%
\bibitem [{\citenamefont {Le~Tiec}\ \emph {et~al.}(2012)\citenamefont
  {Le~Tiec}, \citenamefont {Barausse},\ and\ \citenamefont
  {Buonanno}}]{LeTiec:2011dp}%
  \BibitemOpen
  \bibfield  {author} {\bibinfo {author} {\bibfnamefont {A.}~\bibnamefont
  {Le~Tiec}}, \bibinfo {author} {\bibfnamefont {E.}~\bibnamefont {Barausse}}, \
  and\ \bibinfo {author} {\bibfnamefont {A.}~\bibnamefont {Buonanno}},\ }\href
  {\doibase 10.1103/PhysRevLett.108.131103} {\bibfield  {journal} {\bibinfo
  {journal} {Phys. Rev. Lett.}\ }\textbf {\bibinfo {volume} {108}},\ \bibinfo
  {pages} {131103} (\bibinfo {year} {2012})},\ \Eprint
  {http://arxiv.org/abs/1111.5609} {arXiv:1111.5609 [gr-qc]} \BibitemShut
  {NoStop}%
%%CITATION = ARXIV:1111.5609;%%
\bibitem [{\citenamefont {Damour}\ \emph {et~al.}(2014)\citenamefont {Damour},
  \citenamefont {Jaranowski},\ and\ \citenamefont {Schäfer}}]{Damour:2014jta}%
  \BibitemOpen
  \bibfield  {author} {\bibinfo {author} {\bibfnamefont {T.}~\bibnamefont
  {Damour}}, \bibinfo {author} {\bibfnamefont {P.}~\bibnamefont {Jaranowski}},
  \ and\ \bibinfo {author} {\bibfnamefont {G.}~\bibnamefont {Schäfer}},\
  }\href {\doibase 10.1103/PhysRevD.89.064058} {\bibfield  {journal} {\bibinfo
  {journal} {Phys. Rev.}\ }\textbf {\bibinfo {volume} {D89}},\ \bibinfo {pages}
  {064058} (\bibinfo {year} {2014})},\ \Eprint {http://arxiv.org/abs/1401.4548}
  {arXiv:1401.4548 [gr-qc]} \BibitemShut {NoStop}%
%%CITATION = ARXIV:1401.4548;%%
\bibitem [{\citenamefont {Kapadia}\ \emph {et~al.}(2016)\citenamefont
  {Kapadia}, \citenamefont {Johnson-McDaniel},\ and\ \citenamefont
  {Ajith}}]{Kapadia:2015yra}%
  \BibitemOpen
  \bibfield  {author} {\bibinfo {author} {\bibfnamefont {S.~J.}\ \bibnamefont
  {Kapadia}}, \bibinfo {author} {\bibfnamefont {N.~K.}\ \bibnamefont
  {Johnson-McDaniel}}, \ and\ \bibinfo {author} {\bibfnamefont
  {P.}~\bibnamefont {Ajith}},\ }\href {\doibase 10.1103/PhysRevD.93.024006}
  {\bibfield  {journal} {\bibinfo  {journal} {Phys. Rev.}\ }\textbf {\bibinfo
  {volume} {D93}},\ \bibinfo {pages} {024006} (\bibinfo {year} {2016})},\
  \Eprint {http://arxiv.org/abs/1509.06366} {arXiv:1509.06366 [gr-qc]}
  \BibitemShut {NoStop}%
%%CITATION = ARXIV:1509.06366;%%
\bibitem [{\citenamefont {Campanelli}\ \emph {et~al.}(2006)\citenamefont
  {Campanelli}, \citenamefont {Lousto},\ and\ \citenamefont
  {Zlochower}}]{Campanelli:2006uy}%
  \BibitemOpen
  \bibfield  {author} {\bibinfo {author} {\bibfnamefont {M.}~\bibnamefont
  {Campanelli}}, \bibinfo {author} {\bibfnamefont {C.~O.}\ \bibnamefont
  {Lousto}}, \ and\ \bibinfo {author} {\bibfnamefont {Y.}~\bibnamefont
  {Zlochower}},\ }\href {\doibase 10.1103/PhysRevD.74.041501} {\bibfield
  {journal} {\bibinfo  {journal} {Phys. Rev.}\ }\textbf {\bibinfo {volume}
  {D74}},\ \bibinfo {pages} {041501} (\bibinfo {year} {2006})},\ \Eprint
  {http://arxiv.org/abs/gr-qc/0604012} {arXiv:gr-qc/0604012 [gr-qc]}
  \BibitemShut {NoStop}%
\end{thebibliography}%

\end{document}